\documentclass[aps, prl, twocolumn,amsmath,amssymb,floatfix,reprint,longbibliography]{revtex4-2}
\usepackage{graphicx}
\usepackage{physics}
\usepackage{times}
\usepackage{dcolumn}
\usepackage[dvipsnames]{xcolor}
\usepackage[normalem]{ulem}  
\usepackage{hyperref} % For hyperlinks in the PDF
\hypersetup{colorlinks, allcolors=blue}
\allowdisplaybreaks
\let\oldsection\section
\renewcommand{\section}[1]{\hspace*{1em}\textit{#1}---}

\begin{document}

\title{Mechanism for Nodal Topological Superconductivity on PtBi$_2$ Surface}

\author{Kristian M{\ae}land}
\email[Contact author: ]{kristian.maeland@uni-wuerzburg.de}
\author{Giorgio Sangiovanni}
\author{Bj{\"o}rn Trauzettel}
\affiliation{Institute for Theoretical Physics and Astrophysics, University of W{\"u}rzburg, D-97074 W{\"u}rzburg, Germany}
\affiliation{W{\"u}rzburg-Dresden Cluster of Excellence ctd.qmat, D-97074 W{\"u}rzburg, Germany}

\begin{abstract}
Experiments show that the Weyl semimetal PtBi$_2$ hosts unconventional superconductivity in its topological surface states. Hence, the material is a candidate for intrinsic topological superconductivity. Measurements indicate nodal gaps in the center of the Fermi arcs. We derive that anisotropic electron-phonon coupling on Weyl semimetal surfaces, combined with statically screened Coulomb repulsion, is a microscopic mechanism for this nodal pairing. The dominant solution of the linearized gap equation shows nodal gaps when the surface state bandwidth is comparable to the maximum phonon energy, as is the case in PtBi$_2$. We further predict that if the screening of Coulomb interaction on the surface is enhanced by Coulomb engineering, the superconducting gap becomes nodeless, and the critical temperature increases.
\end{abstract}

\maketitle

%------------------------------------------INTRODUCTION--------------------------------------------------------------------------
\section{Introduction}%
Trigonal PtBi$_2$ is a Weyl semimetal \cite{Yan2017WeylRev, Armitage2018WeylRev, Shipunov2020ExpWeylSC, Veyrat2023ExpWeyl2DSC, OLeary2025PtBi2, Vocaturo2024PtBi2Effective, Palumbo2025DFT, Kuibarov2024FermiArcSCNat, Kuibarov2025ARPES} with a dominant surface superconductivity measured with both scanning tunneling microscopy and angle resolved photoemission spectroscopy (ARPES) \cite{Kuibarov2024FermiArcSCNat, Changdar2025iwave, Kuibarov2025ARPES, Kuibarov2025SepHighTc, Schimmel2023ExpWeylSCSTM, Hoffmann2024PtBi2STM, Huang2025PtBi2STM, Moreno2025PtBi2vortex}. The material is a platform for intrinsic topological superconductivity \cite{Kuibarov2024FermiArcSCNat, Changdar2025iwave, Huang2025PtBi2STM, Maeland2025Jun} and has prerequisites for high critical temperature $T_c$ \cite{Kuibarov2025SepHighTc}. Hence, PtBi$_2$ holds promise for future technological applications in, e.g., topological quantum computation \cite{TopoQuantumCompRevModPhys, Leijnse2012TSCrev, Bernevig2013, TopoSCrevSato}. 
ARPES measurements of the gap amplitude combined with symmetry analysis indicate twelve nodes in the superconducting gap function consistent with $i$-wave pairing \cite{Changdar2025iwave}. It is of high interest to find a microscopic mechanism for this highly unconventional superconducting state. Understanding the mechanism can help find ways to enhance $T_c$ \cite{Kuibarov2025SepHighTc}.

ARPES measures the absolute value of the superconducting gap and, due to experimental resolution, cannot conclusively prove the presence of nodes.
Like with the $d$-wave high-$T_c$ superconductors, phase sensitive measurements could confirm the nodes \cite{vanHarlingen1993phase}.
Here, we take a theoretical perspective and ask if we can find a microscopic mechanism that can explain the presence of the nodes. Unlike high-$T_c$ superconductors, there are no indications of strong correlations in PtBi$_2$ \cite{Changdar2025iwave}, 
motivating a search for weak-coupling mechanisms.

The electronic states are most surface localized in the center of the Fermi arcs \cite{Min2019FA}. Hence, if surface superconductivity dominates, a maximum gap in the center of the arcs is expected \cite{Trama2024TRSWeylSM_SC}.
At the same time, Coulomb repulsion is strongest in the center of the Fermi arcs where the electronic states are most compressed in real space. Hence, the inclusion of Coulomb interactions holds promise for nodal pairing.

In PtBi$_2$, measurements indicate that the bandwidth of the surface state is comparable to the maximum phonon energy \cite{Kuibarov2024FermiArcSCNat, Changdar2025iwave, Kuibarov2025ARPES, Kuibarov2025SepHighTc, Bashlakov2022PtBi2phononExp}. For such a situation, we find that a direct sum of phonon-mediated electron-electron attraction and statically screened Coulomb repulsion gives nodal pairing in the Fermi arcs. That could explain the observed nodal pairing on the surface of PtBi$_2$ \cite{Changdar2025iwave}. 
The Coulomb repulsion between bulk and surface states is negligible due to their separation in real space. Then, phonon and Coulomb interactions have a similar range in terms of momentum. 
Thus, electrons are forced to pair with higher angular momentum to avoid the screened Coulomb repulsion. 
We find that the phonon-mediated electron-electron interaction on the surface of Weyl semimetals has a long-ranged component to support such high-angular-momentum pairing.

The traditional view of how Coulomb interaction modifies phonon-mediated superconductivity is through quantitative changes.
When the electron bandwidth is much larger than the maximum phonon energy, the effective repulsion is reduced, as described by the Morel-Anderson pseudopotential \cite{Bogoliubov1958, Morel1962Anderson}. 
This situation changes when the electron bandwidth is comparable to the maximum phonon energy. 
The mechanism predicted by us differs from others based on electronic correlations \cite{Kohn1965Luttinger, Rosner2018Plasmon, Veld2023Plasmon, Veld2025Plasmon, Kobayashi2022Apr, Moriya2003spinfluct,  Kuroki2004Aug, Kuroki2005Jan, Romer2021SROspinfluctuations, Schnell2006dwavephonon}, including plasmon \cite{Rosner2018Plasmon, Veld2023Plasmon, Veld2025Plasmon} and spin-fluctuation mechanisms \cite{Schnell2006dwavephonon} which have been combined with phonons. 
In our case, we include Coulomb interactions in a lowest order, purely repulsive, weak-coupling approach.

\begin{figure}
    \centering
    \includegraphics[width=0.9\linewidth]{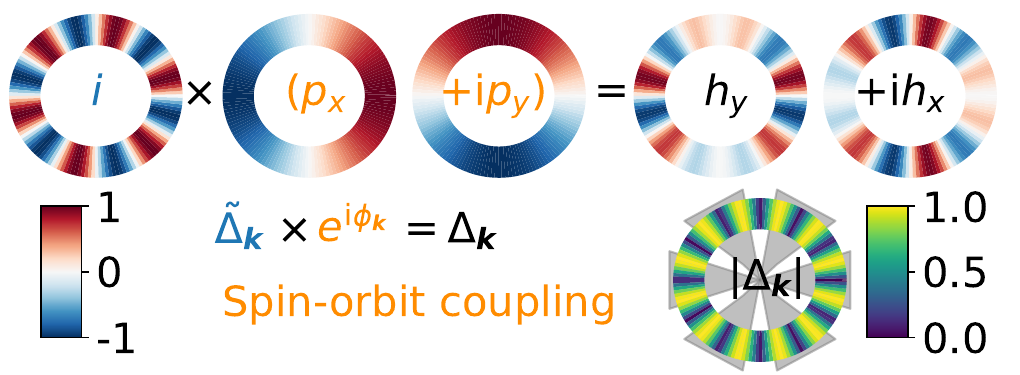}
    \caption{
    Illustration of the gap $\Tilde{\Delta}_{\boldsymbol{k}}e^{\mathrm{i}\phi_{\boldsymbol{k}}} = \Delta_{\boldsymbol{k}}$ on a circular Fermi surface around the $\boldsymbol{\Gamma}$ point. An even-parity $i$-wave function $\Tilde{\Delta}_{\boldsymbol{k}}$ multiplied by $p_x+\mathrm{i} p_y$-wave momentum dependence from spin-orbit coupling yields an odd-parity gap function $\Delta_{\boldsymbol{k}}$.
    The angular extent of the Fermi arcs is indicated in gray for the absolute value of the gap $|\Delta_{\boldsymbol{k}}|$. The absolute value of the gap reproduces the experimentally observed gap profile in the Fermi arcs of PtBi$_2$ \cite{Changdar2025iwave}.}
    \label{fig:gapsym}
\end{figure}

We consider Cooper pairing of the long-lived excitations on the surface. The Fermi arcs are nondegenerate, yielding a single gap function that must be odd in momentum due to Pauli exchange statistics. 
We single out
an odd-parity factor $e^{\mathrm{i}\phi_{\boldsymbol{k}}}$
in the odd-parity gap function $\Delta_{\boldsymbol{k}} = \Tilde{\Delta}_{\boldsymbol{k}}e^{\mathrm{i}\phi_{\boldsymbol{k}}}$ \cite{Scheurer2016NCSTSC}.
The function $\Tilde{\Delta}_{\boldsymbol{k}}$ is the even-parity 
part of the gap function. 
We find that
$e^{\mathrm{i}\phi_{\boldsymbol{k}}}$ has a $p_x+\mathrm{i}p_y$-wave momentum dependence that originates with spin-orbit coupling (SOC) via the electronic eigenstates \cite{Maeland2025Jun}. With zero Coulomb repulsion we find fully gapped $p_x+\mathrm{i}p_y$-wave pairing, which we interpret as anisotropic $s$-wave pairing from phonons multiplied by $e^{\mathrm{i}\phi_{\boldsymbol{k}}}$.
Upon including Coulomb interactions, we find dominant $i$-wave pairing, such that the gap $\Delta_{\boldsymbol{k}}$ is odd-parity $i\times(p_x+\mathrm{i} p_y)$-wave, as illustrated in Fig.~\ref{fig:gapsym}. 
This gap has nodes in the same location as $\Tilde{\Delta}_{\boldsymbol{k}}$, which is $i$-wave. 
As noted in Fig.~\ref{fig:gapsym}, the gap can be considered as $h_y+\mathrm{i}h_x$-wave. Since it is nodal, we prefer to call it $i\times(p_x+\mathrm{i} p_y)$-wave or $i$-wave for short. We spend the rest of the Letter explaining our model and show a parameter range where $i$-wave pairing is dominant on the surface. We argue that this parameter range is plausible in PtBi$_2$. In Supplemental Material (SM) \cite{Suppl}, we explore the parameter space to demonstrate that the $i$-wave pairing is robust.

%------------------------------MODEL-------------------------------------------------------

\section{Electrons}% 
For the Weyl semimetal normal state, we adopt the effective model from Ref.~\cite{Vocaturo2024PtBi2Effective}, designed to capture the fermiology of PtBi$_2$. 
Trigonal PtBi$_2$ is in the P31m space group, with broken inversion symmetry and retained time-reversal symmetry \cite{Shipunov2020ExpWeylSC}. 
The effective model considers a hexagonal crystal with two orbitals per site and the same symmetries as PtBi$_2$, yielding the same number and general location of Weyl nodes and Fermi arcs.
The Hamiltonian is
\begin{equation}
    H_{\text{el}} = -\sum_{i\ell\sigma} (\mu+\mu_\ell)c_{i\ell\sigma}^\dagger c_{i\ell\sigma} + H_{\text{hop}} + H_{\text{SOC}} + H_{\gamma}. 
\end{equation}
The electron operator $c_{i\ell\sigma}^{(\dagger)}$ destroys (creates) an electron at site $i$ in orbital $\ell= A,B$ with spin $\sigma$. $\mu$ is the chemical potential and $\mu_A = -\mu_o, \mu_B = \mu_o$ are on-site energies depending on the orbital. 
The hopping term includes nearest neighbor (NN) hopping with $t$ ($t_o$) being the in-plane intraorbital (interorbital) hopping and $\beta$ ($\beta_o$) being the out-of-plane intraorbital (interorbital) hopping. 
The SOC term is of Rashba type with a chiral $p$-wave momentum dependence and strength $\alpha$. The last term, $H_{\gamma}$, breaks inversion symmetry, parametrized by $\gamma$. 
See End Matter (EM) for the expressions.

We consider a slab geometry with $L$ layers and introduce a partial Fourier transform (FT) in the plane. We find the $4L$ normal state electron bands $\epsilon_{\boldsymbol{k}n}$ by diagonalization \cite{Suppl}. We also introduce the layer dependent weight of their eigenstates, defined as
$
    W_{\boldsymbol{k}n} = \sum_{z_i = 1}^{L} (z_i-1)|\psi_{\boldsymbol{k}n,z_i}|^2/(L-1),
$
where $|\psi_{\boldsymbol{k}n,z_i}|^2$ is the sum of squares of the 4 entries in the eigenvector of band $n$ associated with layer $z_i$. 
We use lattice constant $a=1$, reduced Planck's constant $\hbar = 1$, roman font for imaginary unit $\mathrm{i}$, a bar over 3D vectors [$\Bar{\boldsymbol{r}}_i = (x_i,y_i,z_i)$], and no bar over 2D vectors [$\boldsymbol{r}_i = (x_i, y_i)$] throughout.

\section{Coulomb interaction}%
To model a screened Coulomb interaction we include on-site and NN repulsion, $H_C = H_{U} + H_{V}$, see EM. For the on-site term we use a two-orbital Hund's rule model with parameters $U$ and $J < U/3$ \cite{Georges2013Hund, Budich2013Hund, Amaricci2015Hund}. 
We model the NN repulsion as a density-density interaction with strength $V$. 
We perform a partial FT and transform to the band basis using $c_{\boldsymbol{k} z_i \ell\sigma} = \sum_n v_{\boldsymbol{k} n z_i \ell\sigma}d_{\boldsymbol{k}n}$ to obtain $H_C = \sum_{\boldsymbol{kk}'\boldsymbol{q}n_1 n_2 n_3 n_4} V_{\boldsymbol{k}+\boldsymbol{q}, \boldsymbol{k}'-\boldsymbol{q}, \boldsymbol{k}', \boldsymbol{k}}^{n_1 n_2 n_3 n_4} d_{\boldsymbol{k}+\boldsymbol{q},n_1}^\dagger d_{\boldsymbol{k}'-\boldsymbol{q},n_2}^\dagger d_{\boldsymbol{k}'n_3}d_{\boldsymbol{k}n_4}$, where $V_{\boldsymbol{k}+\boldsymbol{q}, \boldsymbol{k}'-\boldsymbol{q}, \boldsymbol{k}', \boldsymbol{k}}^{n_1 n_2 n_3 n_4}$ is defined in SM \cite{Suppl}. The main momentum dependence of the Coulomb interaction is through the electronic eigenstates. From in-plane NN repulsion, there is an extra momentum dependence $\gamma(\boldsymbol{q}) = 2\cos q_y + 4 \cos(\sqrt{3}q_x/2)\cos(q_y/2)$ which is largest at small momentum transfers, giving a more peaked Coulomb repulsion in the center of the Fermi arcs.

\begin{figure}
    \centering
    \includegraphics[width=\linewidth]{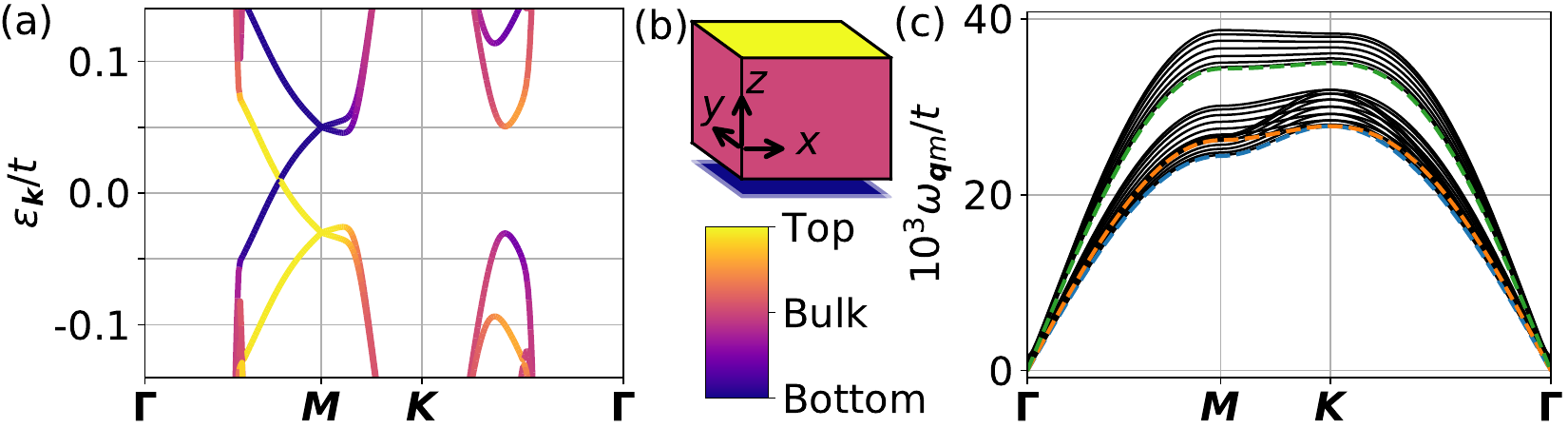}
    \caption{
    (a) Electron bands with a focus on surface states close to the Fermi level in slab geometry shown along a path between high symmetry points in the first Brillouin zone.
    The bands are colored by the location of their eigenstate $W_{\boldsymbol{k}n}$ along the $z$ direction, as indicated in (b). In panel (c), the colored dashed lines show the three bulk acoustic phonon modes, while the black lines illustrate the phonon spectrum in slab geometry. The parameters are $t_o/t = 0.24$, $\beta/t = -1.5$, $\beta_o = -0.4$, $\mu/t = -0.01, \mu_o/t = -0.7, \alpha/t = -0.03$, $\gamma/t = -0.04$, $\gamma_1 = -(0.01t)^2$, $\gamma_3 = 0.43\gamma_1$, $\gamma_4 = 1.5\gamma_1$, $\gamma_5 = 0.432\gamma_1$, $\gamma_6 = 0.6 \gamma_1$, (a) $L=30$, and (b) $L = 7$.} 
    \label{fig:bands}
\end{figure}

\section{Phonons}%
We derive the phonon spectrum using a force constant approach \cite{BruusFlensberg, Klogetvedt2023, Syljuasen2024, Thingstad2020Jun, Leraand2025Feb, Maeland2025Jun, Leraand2026Mar}. We determine the phonon modes for the single-atomic basis hexagonal crystal employed for the electron model, assuming the symmetries of the P31m space group. This approach results in a phenomenological model with five free force constants, $\gamma_1, \gamma_3, \gamma_4, \gamma_5,$ and $\gamma_6$, which we tune to get a spectrum with a similar energy range as reported for PtBi$_2$ \cite{Bashlakov2022PtBi2phononExp}. 
The phonon description is quantized in terms of phonon destruction (creation) operators $a_{\boldsymbol{q}m}^{(\dagger)}$, and the phonon energy spectrum $\omega_{\boldsymbol{q}m}$ is found by diagonalizing the dynamical matrix. The eigenvectors $\hat{e}_{\boldsymbol{q}m}$ contain information about the phonon modes. See EM for details. There are three acoustic phonon modes in the bulk. In slab geometry, they are projected on the in-plane momentum yielding $3L$ modes \cite{Lucas1968PhononOBC, Benedek2010PhononOBC}. Three of these modes are acoustic, while the remaining $3L-3$ modes have a nonzero energy at $\boldsymbol{q} = 0$. 
Even though these $3L-3$ modes also relate to the bulk acoustic modes, we name them optical phonons to separate phonon modes with zero energy and nonzero energy at zero in-plane momentum. 
Figure \ref{fig:bands} shows electron bands and phonon spectrum.

\section{Electron-phonon coupling}%
We imagine that the ions move by small distances away from their equilibrium positions and perform a Taylor expansion of the hopping term. This yields terms that couple electron hopping to ion displacements. By quantizing the ion displacements in terms of phonon operators we get an expression for the electron-phonon coupling (EPC) \cite{Thingstad2020Jun, Leraand2025Feb, Maeland2025Jun, Leraand2026Mar}, see EM. After transforming to the electron band basis,
the EPC Hamiltonian reads
$
    H_{\text{EPC}} = \sum_{\boldsymbol{k}\boldsymbol{q}m} g_{\boldsymbol{k}+\boldsymbol{q}, \boldsymbol{k}}^{m} (a_{-\boldsymbol{q},m}^\dagger+a_{\boldsymbol{q}m})d_{\boldsymbol{k}+\boldsymbol{q}}^\dagger d_{\boldsymbol{k}}.
$
We focus on the band that has a Fermi surface (FS) and drop the band index.
The EPC $g$ factor is
$g_{\boldsymbol{k} \boldsymbol{k}'}^{m} = \sum_{\ell\ell'\sigma z_i \delta_z} g_{\boldsymbol{k} \boldsymbol{k}' m}^{\ell\ell' z_i \delta_z}v_{\boldsymbol{k},z_i+\delta_z,\ell,\sigma}^* v_{\boldsymbol{k}' z_i \ell'\sigma},$
where
\begin{align}
\label{eq:gEPC}
    g_{\boldsymbol{k} \boldsymbol{k}' m}^{\ell\ell' z_i \delta_z} &= \sum_{\boldsymbol{\delta}} \frac{\chi t_{\ell\ell'}(\Bar{\boldsymbol{\delta}})}{\sqrt{2N_L M\omega_{\boldsymbol{k}-\boldsymbol{k}',m}}}(e^{-\mathrm{i}\boldsymbol{k}'\cdot \boldsymbol{\delta}}\Bar{\boldsymbol{e}}_{\boldsymbol{k}-\boldsymbol{k}', m}^{z_i+\delta_z}\nonumber\\
    &-e^{-\mathrm{i}\boldsymbol{k}\cdot\boldsymbol{\delta}}\Bar{\boldsymbol{e}}_{\boldsymbol{k}-\boldsymbol{k}',m}^{z_i})\cdot \Bar{\boldsymbol{\delta}}.
\end{align}
Here, $\chi$ is a dimensionless number related to the spread of atomic orbitals.
The hopping parameter $t_{\ell\ell'}(\Bar{\boldsymbol{\delta}})$ depends on orbital indices and the direction of the NN vector. 
$N_L$ is the number of sites per layer, and $M$ is the ion mass.
$\Bar{\boldsymbol{e}}_{\boldsymbol{q} m}^{z_i}$ is the part of the phonon eigenstate related to layer $z_i$. If $\delta_z = 0$, the in-plane EPC strength $g_{\boldsymbol{k} \boldsymbol{k}' m}^{\ell\ell' z_i, 0} \to 0$ if $\boldsymbol{k}' \to \boldsymbol{k}$. That is a lattice version of the known behavior from the jellium model \cite{BruusFlensberg}. However, if $\delta_z = \pm 1$, the phonon eigenstates in Eq.~\eqref{eq:gEPC} are generally different. Then, the out-of-plane EPC strength  $g_{\boldsymbol{k} \boldsymbol{k}' m}^{\ell\ell' z_i, \pm 1}$ does not necessarily go to zero for zero momentum transfer. If the optical modes then have a low energy at $\boldsymbol{q} = 0$, we see that the coupling can be large. Hence, the out-of-plane EPC does not behave according to the jellium model, which has major consequences for the superconducting pairing for surface states. The three acoustic modes in slab geometry have layer-independent phonon eigenstates at $\boldsymbol{q} = 0$. Hence, they behave similarly to the jellium model also for out-of-plane EPC.

\section{Superconductivity}%
We employ a generalization of the Bardeen-Cooper-Schrieffer (BCS) theory of superconductivity \cite{BCS, Sigrist}. Focusing on zero-momentum Cooper pairing, we consider the interaction $H_{\text{BCS}} = \sum_{\boldsymbol{kk}'} V_{\boldsymbol{k}'\boldsymbol{k}} d_{\boldsymbol{k}'}^\dagger d_{-\boldsymbol{k}'}^\dagger d_{-\boldsymbol{k}} d_{\boldsymbol{k}}$ written in the normal state electron band basis. Here, $V_{\boldsymbol{k}'\boldsymbol{k}} = V_{\boldsymbol{k}'\boldsymbol{k}}^{C} + V_{\boldsymbol{k}'\boldsymbol{k}}^{\text{ph}}$.
From the Coulomb repulsion $H_C$, we choose the band with a FS and insert the assumed zero-momentum pairing to obtain $V_{\boldsymbol{k}'\boldsymbol{k}}^{C}$. 
From the EPC, we derive an effective electron-electron interaction mediated by the phonons. The phonon contribution in the static limit is $V_{\boldsymbol{k}'\boldsymbol{k}}^{\text{ph}} = -\sum_m g_{\boldsymbol{k}'\boldsymbol{k}}^m g_{-\boldsymbol{k}',-\boldsymbol{k}}^m/\omega_{\boldsymbol{k}'-\boldsymbol{k},m}$ \cite{Suppl}.

The linearized gap equation, valid close to $T_c$, is $\Delta_{\boldsymbol{k}} = -\sum_{\boldsymbol{k}'} \bar{V}_{\boldsymbol{k}\boldsymbol{k}'} \Delta_{\boldsymbol{k}'} \chi_{\boldsymbol{k}'}(T_c).$
The symmetrized interaction $\bar{V}_{\boldsymbol{k}\boldsymbol{k}'} = (V_{\boldsymbol{k}\boldsymbol{k}'} - V_{\boldsymbol{k},-\boldsymbol{k}'}- V_{-\boldsymbol{k},\boldsymbol{k}'}+ V_{-\boldsymbol{k},-\boldsymbol{k}'})/2$ is the only part that contributes in the gap equation. The sum over $\sum_{\boldsymbol{k}'}$ runs over the full first Brillouin zone (1BZ), while the main contributions are expected from the region close to the FS where $\chi_{\boldsymbol{k}'}(T_c) = \tanh(|\epsilon_{\boldsymbol{k}'}|/2k_B T_c)/2|\epsilon_{\boldsymbol{k}'}|$ is peaked. Hence, a solution of the gap equation requires turning the sum into an integral, and then approximating the integral with adaptive integration involving a denser grid of points close to the FS.

In general, one expects that the Coulomb repulsion is active in the entire electron bandwidth, 
while the main contribution from the phonon-mediated attraction occurs in a small energy window around the FS given by the maximum phonon energy $\omega_D$.
Then, the Coulomb repulsion has only a weak effect on the critical temperature, quantified by the Morel-Anderson pseudopotential \cite{Bogoliubov1958, Morel1962Anderson}. 
This applies to isotropic $s$-wave pairing.
However, the general idea survives more complicated interactions and unconventional, momentum-dependent gaps. The gap from phonons alone is $p_x+\mathrm{i}p_y$-wave in our case \cite{Maeland2025Jun}. By solving the gap equation including Coulomb interaction in the full 1BZ, we find that the gap is $p_x+\mathrm{i}p_y$-wave close to the FS, while at momenta outside $|\epsilon_{\boldsymbol{k}}| < \omega_D$, the gap is $-p_x-\mathrm{i}p_y$-wave. Hence, the gap takes advantage of the Coulomb repulsion between regions close to the FS and regions farther from the FS by changing sign radially \cite{Suppl}.

If the surface state bandwidth $W$ is reduced to be comparable to the maximum phonon energy $\omega_D$, the radial sign change of the gap becomes less effective. Then, we find that a gap which is $i\times (p_x +\mathrm{i}p_y)$-wave within the range of phonons ($|\epsilon_{\boldsymbol{k}}| < \omega_D$), and zero at other momenta, becomes dominant over the fully gapped pairing with a radial sign change. We demonstrate in SM that the exact limit where the electron bandwidth is the same as the maximum phonon energy is not needed \cite{Suppl}. In this Letter, for simplicity, we focus on the case where the surface state bandwidth is approximately equal to $\omega_D$, as illustrated in Fig.~\ref{fig:bands}. 
Then, the phonon-mediated interaction is active in the entire energy range of the surface state. Meanwhile, bulk and surface states decouple both for the phonon-mediated interaction and for Coulomb repulsion due to negligible overlap of the eigenstates. Thus, phonon and electronic correlations have the same range. 
Comparable surface state electron bandwidth and $\omega_D$ indeed appears to be the case in PtBi$_2$, where ARPES measurements \cite{Kuibarov2024FermiArcSCNat, Kuibarov2025ARPES, Changdar2025iwave, Kuibarov2025SepHighTc} indicate that the surface state extends about $10$ to $20$~meV below the FS, and Yanson point-contact spectroscopy shows that the maximum phonon energy is approximately $20$~meV \cite{Bashlakov2022PtBi2phononExp}.

\begin{figure*}
    \centering
    \includegraphics[width=0.95\linewidth]{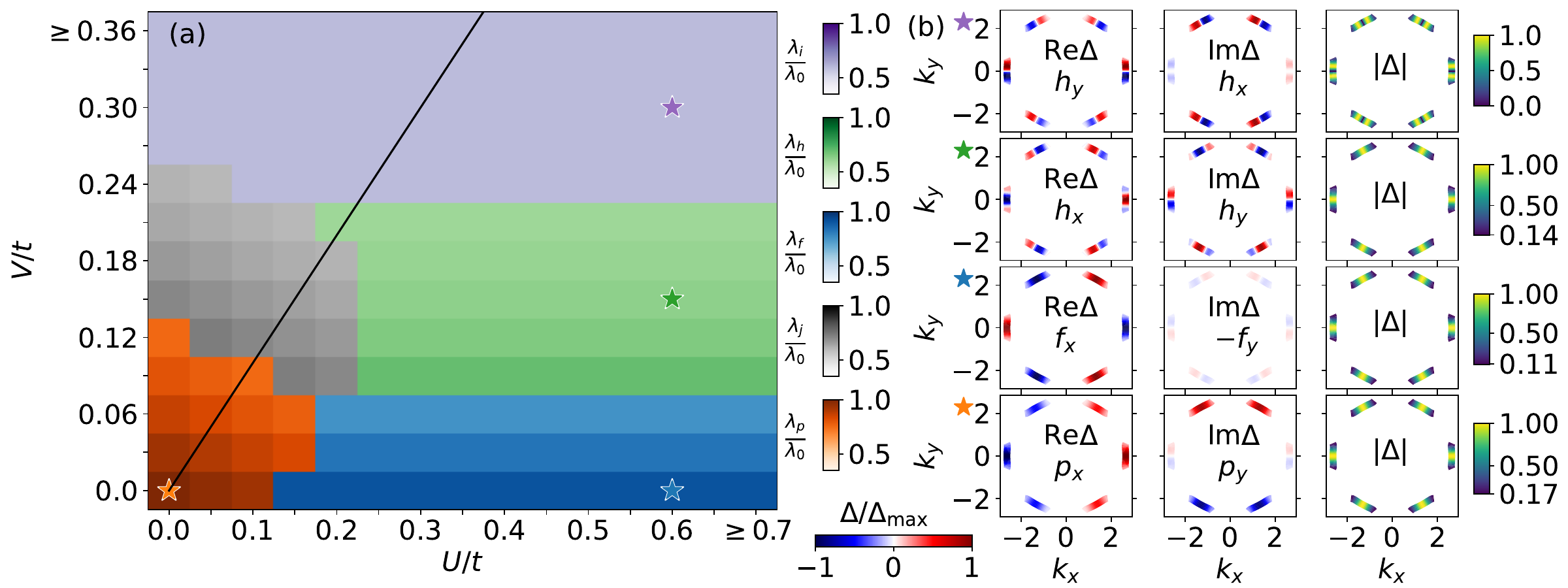}
    \caption{(a) Dimensionless coupling $\lambda$ as a function of on-site $U$ and nearest-neighbor $V$ Coulomb repulsion, scaled by $\lambda_0$ which is $\lambda$ at $U=V = 0$. Colors indicate the symmetry of the superconducting gap with the largest critical temperature. The black line shows $V = U$ and the region below that line is most realistic. Nodal $i\times (p_x+\mathrm{i}p_y)$-wave pairing dominates the parameter space. Stars indicate the value of $U$ and $V$ used for the rows in (b), where the real, imaginary, and absolute value of the gap is shown on the bottom surface Fermi arc. The gaps are shown in units of their own largest absolute value. The parameters are $J = 0.2U$, $\chi = 30$, $Mt = 24350$, $L=30$, $N_{\text{samp}} = 126$, and otherwise the same as Fig.~\ref{fig:bands}.}
    \label{fig:PD}
\end{figure*}

The FS averaged gap equation for the case where the surface state bandwidth is comparable to $\omega_D$ is an eigenvalue problem $\lambda \Delta_{k_\parallel} =  -\sum_{k'_\parallel } N_{k'_\parallel} \bar{V}_{k_\parallel k'_\parallel}^{\text{FS}}\Delta_{k'_\parallel }$, where $k_\parallel$ is the component of the momentum parallel to the FS, $N_{k'_\parallel}$ is a momentum dependent density of states (DOS) factor, and $\bar{V}_{k_\parallel k'_\parallel}^{\text{FS}}$ is the sum of phonon and Coulomb interactions on the FS. The largest eigenvalue $\lambda$ is the dimensionless coupling. Its corresponding eigenvector gives the momentum dependence of the dominant gap function. 
BCS theory provides an estimate of $T_c$ through $k_B T_c \approx 1.13 \omega_D e^{-1/\lambda}$, valid for weak coupling with $\lambda \ll 1$ \cite{BCS, SFsuperconductivity}. The main strength of the generalized BCS theory is to predict the dominant gap symmetry, which is our focus.

Figure~\ref{fig:PD} shows the dimensionless coupling as a function of on-site $U$ and NN $V$ Coulomb repulsion and indicates which superconducting pairing has the highest $T_c$ at each point. We label the dimensionless coupling in each region by the symmetry of the gap, $\lambda_p, \lambda_j, \lambda_f, \lambda_h$, and $\lambda_i$. Fully gapped $p_x+\mathrm{i}p_y$-wave dominates at weak Coulomb repulsion, while there is a closely related $j_x-\mathrm{i}j_y$-wave state with accidental nodes that mostly exists for unrealistic values of $V/U$. We discuss this state in detail in SM \cite{Suppl}. 
Fully gapped $f_x-\mathrm{i}f_y$-wave and $h_x+\mathrm{i}h_y$-wave states exist at intermediate Coulomb repulsions. Nodal $i\times(p_x+\mathrm{i}p_y)$-wave is preferred for $V/t \geq 0.24$ when $U/t \geq 0.1$. Hence, the nodal pairing that matches the ARPES experiment \cite{Changdar2025iwave} dominates the parameter space with combined phonon and Coulomb mechanisms. 

Note that $\lambda_p$ and $\lambda_j$ decrease both with $U$ and with $V$. However, $\lambda_f$ and $\lambda_h$ are independent of $U$ and decrease only with $V$. Remarkably, $\lambda_i$ is independent of both $U$ and $V$, which is why it ends up dominating the parameter space. 
To understand this behavior, we consider the Cooper pairing in real space \cite{Suppl}. 
The fully gapped $p$-wave and closely related nodal $j$-wave state involves on-site and NN pairing of electrons. Therefore, $\lambda_p$ and $\lambda_j$ depend on both on-site and NN repulsion. Meanwhile, the fully gapped $f$- and $h$- wave states have no on-site pairing but do contain NN Cooper pairing.  Thus $\lambda_f$ and $\lambda_h$ depend on $V$ but not on $U$. The $i\times(p_x+\mathrm{i}p_y)$-wave pairing is purely longer ranged than NN in real space, so $\lambda_i$ is independent of both $U$ and $V$. 

From the absolute value of the gaps in Fig.~\ref{fig:PD}(b), we see that all gap symmetries have a maximum gap in the center of the Fermi arc except for the $i$-wave pairing. This is because the phonon-mediated attraction is strongest in the center of the arc.

All five superconducting states are topologically nontrivial of different types. The $p_x+\mathrm{i}p_y$-wave state is an intrinsic realization of the Fu and Kane model \cite{Fu2008Kane}. Time-reversal symmetry is preserved \cite{Scheurer2016NCSTSC}, but a small out-of-plane magnetic field yields Majorana bound states in the core of vortices and potential chiral edge states on the 2D surface \cite{Fukui2010FuKaneTopoStability}.
The fully gapped $f$- and $h$-wave states are similar states with higher winding \cite{Venditti2026d+id}. Interestingly, these latter states spontaneously break time-reversal symmetry 
and can thus have chiral edge states without a magnetic field \cite{Andersen2024TRSB, Sigrist, Suppl}.
The nodal $i$-wave and $j$-wave states are weak topological superconductors \cite{TopoSCrevSato}. There is no full gap, but superconducting nodes on a nondegenerate FS are Majorana fermions. This results in Majorana zero energy flat bands on the hinges of the material \cite{TopoSCrevSato, Changdar2025iwave}. Investigations of potential applications of these edge states include Refs.~\cite{Lapp2022nodalMajorana, Lapp2025nodalMajorana}.

We find that $\lambda_i \approx 0.302$ while $\lambda_0 \approx 0.506$. With parameters relevant to PtBi$_2$, we find that the nodal state can still have a $T_c$ on the order of 10 K \cite{Suppl}.
The main point is that the critical temperature of the nodal $i$-wave pairing need not be less than 25\% of the nodeless pairing at zero Coulomb interaction.

At the same time, our results in Fig.~\ref{fig:PD} indicate that if we can better screen the Coulomb interaction on the surface, we should expect the gap to become nodeless and $T_c$ to increase. Coulomb engineering \cite{Steinke2020CoulombEngineering, vanLoon2023CoulombEngineering, Veld2023Plasmon, Rosner2018Plasmon, Veld2025Plasmon} is possible by placing another material close to the surface of our Weyl semimetal to act as a dielectric environment. 
Alternatively, doping the material to tune the bulk and surface DOS \cite{Kuibarov2025SepHighTc, Rosner2018Plasmon} could affect the screening of the Coulomb interaction.

We ensure low energy of the optical phonon modes at $\boldsymbol{q} = 0$. 
The interpretation of this parameter choice is that the bulk modes have a weak dispersion along $q_z$ for $q_x = q_y = 0$. 
In these modes the layers only move relative to each other while all ions in a layer move equally.
Given the van der Waals coupling between the trilayers in PtBi$_2$ \cite{Veyrat2023ExpWeyl2DSC} it is reasonable that these modes cost little energy to excite compared with modes where ions within layers move relative to each other. 
While the electron model and parameter values are chosen to match PtBi$_2$, the physics that we describe in this Letter could also apply to other Weyl semimetals.

The situation $W \approx \omega_D$ suggests a nonadiabatic regime where Migdal's theorem is not valid \cite{Cappelluti2023nonadiabatic, migdal1958interaction}. While the surface state electrons have a comparable bandwidth to the phonons, they have a much larger slope such that Migdal's theorem should still apply \cite{Roy2014Migdal2D, Suppl}. We expect that studies of superconductivity beyond the adiabatic limit, including vertex corrections, would give modifications to the prediction of $T_c$ and the nature of the phase transition \cite{Marsiglio2020ElishbergRev, Chubukov2020Eliashberg, Schrodi2020beyondMigdal, Cappelluti2023nonadiabatic, Miserev2025range, Veyrat2023ExpWeyl2DSC} but not to the symmetry of the gap.

%------------------------------CONCLUSION-------------------------------------------------
\section{Conclusion}%
The Fermi arc surface states in PtBi$_2$ have a small bandwidth and demonstrate nodal $i$-wave superconductivity in ARPES experiments. We show that a phonon-mediated interaction combined with statically screened Coulomb repulsion can explain the nodal pairing. 
This is due to a surface state bandwidth that is comparable to the maximum phonon energy such that Cooper pairing must be longer ranged in real space to avoid the short range Coulomb repulsion. Thus, gaps with larger angular momentum are preferred. Electron-phonon coupling for topological surface states has a long-ranged component to support such pairing.
We predict that better screening of Coulomb interactions will lead to nodeless gaps and higher critical temperature.

\section{Note added}%
Two preprints discussing mechanisms for nodal superconductivity in Fermi arcs appeared recently \cite{Dsouza2026KL, Buccheri2026ph}.

%--------------------------------ACKNOWLEDGEMENTS-------------------------------------------
\section{Acknowledgments}%
We thank Jeroen van den Brink, Kristoffer Leraand, Asle Sudb{\o}, Pavlo Sukhachov, Even Thingstad, Carsten Timm, and Ludovica Zullo for useful discussions.
This work was supported by the Deutsche Forschungsgemeinschaft (DFG, German Research Foundation) through SFB 1170 (project ID 258499086) and DFG through the W{\"u}rzburg-Dresden Cluster of Excellence ctd.qmat (EXC 2147, project ID 390858490).

%----------------------------BIBLIOGRAPHY---------------------------------------
\bibliography{main.bbl}

\begin{thebibliography}{96}%
\makeatletter
\providecommand \@ifxundefined [1]{%
 \@ifx{#1\undefined}
}%
\providecommand \@ifnum [1]{%
 \ifnum #1\expandafter \@firstoftwo
 \else \expandafter \@secondoftwo
 \fi
}%
\providecommand \@ifx [1]{%
 \ifx #1\expandafter \@firstoftwo
 \else \expandafter \@secondoftwo
 \fi
}%
\providecommand \natexlab [1]{#1}%
\providecommand \enquote  [1]{``#1''}%
\providecommand \bibnamefont  [1]{#1}%
\providecommand \bibfnamefont [1]{#1}%
\providecommand \citenamefont [1]{#1}%
\providecommand \href@noop [0]{\@secondoftwo}%
\providecommand \href [0]{\begingroup \@sanitize@url \@href}%
\providecommand \@href[1]{\@@startlink{#1}\@@href}%
\providecommand \@@href[1]{\endgroup#1\@@endlink}%
\providecommand \@sanitize@url [0]{\catcode `\\12\catcode `\$12\catcode `\&12\catcode `\#12\catcode `\^12\catcode `\_12\catcode `\%12\relax}%
\providecommand \@@startlink[1]{}%
\providecommand \@@endlink[0]{}%
\providecommand \url  [0]{\begingroup\@sanitize@url \@url }%
\providecommand \@url [1]{\endgroup\@href {#1}{\urlprefix }}%
\providecommand \urlprefix  [0]{URL }%
\providecommand \Eprint [0]{\href }%
\providecommand \doibase [0]{https://doi.org/}%
\providecommand \selectlanguage [0]{\@gobble}%
\providecommand \bibinfo  [0]{\@secondoftwo}%
\providecommand \bibfield  [0]{\@secondoftwo}%
\providecommand \translation [1]{[#1]}%
\providecommand \BibitemOpen [0]{}%
\providecommand \bibitemStop [0]{}%
\providecommand \bibitemNoStop [0]{.\EOS\space}%
\providecommand \EOS [0]{\spacefactor3000\relax}%
\providecommand \BibitemShut  [1]{\csname bibitem#1\endcsname}%
\let\auto@bib@innerbib\@empty
%</preamble>
\bibitem [{\citenamefont {Yan}\ and\ \citenamefont {Felser}(2017)}]{Yan2017WeylRev}%
  \BibitemOpen
  \bibfield  {author} {\bibinfo {author} {\bibfnamefont {B.}~\bibnamefont {Yan}}\ and\ \bibinfo {author} {\bibfnamefont {C.}~\bibnamefont {Felser}},\ }\bibfield  {title} {\bibinfo {title} {{Topological Materials: Weyl Semimetals}},\ }\href {https://doi.org/10.1146/annurev-conmatphys-031016-025458} {\bibfield  {journal} {\bibinfo  {journal} {Annu. Rev. Condens. Matter Phys.}\ }\textbf {\bibinfo {volume} {8}},\ \bibinfo {pages} {337} (\bibinfo {year} {2017})}\BibitemShut {NoStop}%
\bibitem [{\citenamefont {Armitage}\ \emph {et~al.}(2018)\citenamefont {Armitage}, \citenamefont {Mele},\ and\ \citenamefont {Vishwanath}}]{Armitage2018WeylRev}%
  \BibitemOpen
  \bibfield  {author} {\bibinfo {author} {\bibfnamefont {N.~P.}\ \bibnamefont {Armitage}}, \bibinfo {author} {\bibfnamefont {E.~J.}\ \bibnamefont {Mele}},\ and\ \bibinfo {author} {\bibfnamefont {A.}~\bibnamefont {Vishwanath}},\ }\bibfield  {title} {\bibinfo {title} {{Weyl and Dirac semimetals in three-dimensional solids}},\ }\href {https://doi.org/10.1103/RevModPhys.90.015001} {\bibfield  {journal} {\bibinfo  {journal} {Rev. Mod. Phys.}\ }\textbf {\bibinfo {volume} {90}},\ \bibinfo {pages} {015001} (\bibinfo {year} {2018})}\BibitemShut {NoStop}%
\bibitem [{\citenamefont {Shipunov}\ \emph {et~al.}(2020)\citenamefont {Shipunov}, \citenamefont {Kovalchuk}, \citenamefont {Piening}, \citenamefont {Labracherie}, \citenamefont {Veyrat}, \citenamefont {Wolf}, \citenamefont {Lubk}, \citenamefont {Subakti}, \citenamefont {Giraud}, \citenamefont {Dufouleur}, \citenamefont {Shokri}, \citenamefont {Caglieris}, \citenamefont {Hess}, \citenamefont {Efremov}, \citenamefont {B{\ifmmode\ddot{u}\else\"{u}\fi}chner},\ and\ \citenamefont {Aswartham}}]{Shipunov2020ExpWeylSC}%
  \BibitemOpen
  \bibfield  {author} {\bibinfo {author} {\bibfnamefont {G.}~\bibnamefont {Shipunov}}, \bibinfo {author} {\bibfnamefont {I.}~\bibnamefont {Kovalchuk}}, \bibinfo {author} {\bibfnamefont {B.~R.}\ \bibnamefont {Piening}}, \bibinfo {author} {\bibfnamefont {V.}~\bibnamefont {Labracherie}}, \bibinfo {author} {\bibfnamefont {A.}~\bibnamefont {Veyrat}}, \bibinfo {author} {\bibfnamefont {D.}~\bibnamefont {Wolf}}, \bibinfo {author} {\bibfnamefont {A.}~\bibnamefont {Lubk}}, \bibinfo {author} {\bibfnamefont {S.}~\bibnamefont {Subakti}}, \bibinfo {author} {\bibfnamefont {R.}~\bibnamefont {Giraud}}, \bibinfo {author} {\bibfnamefont {J.}~\bibnamefont {Dufouleur}}, \bibinfo {author} {\bibfnamefont {S.}~\bibnamefont {Shokri}}, \bibinfo {author} {\bibfnamefont {F.}~\bibnamefont {Caglieris}}, \bibinfo {author} {\bibfnamefont {C.}~\bibnamefont {Hess}}, \bibinfo {author} {\bibfnamefont {D.~V.}\ \bibnamefont {Efremov}}, \bibinfo {author} {\bibfnamefont {B.}~\bibnamefont {B{\ifmmode\ddot{u}\else\"{u}\fi}chner}},\ and\ \bibinfo
  {author} {\bibfnamefont {S.}~\bibnamefont {Aswartham}},\ }\bibfield  {title} {\bibinfo {title} {{Polymorphic ${\mathrm{PtBi}}_{2}$: Growth, structure, and superconducting properties}},\ }\href {https://doi.org/10.1103/PhysRevMaterials.4.124202} {\bibfield  {journal} {\bibinfo  {journal} {Phys. Rev. Mater.}\ }\textbf {\bibinfo {volume} {4}},\ \bibinfo {pages} {124202} (\bibinfo {year} {2020})}\BibitemShut {NoStop}%
\bibitem [{\citenamefont {Veyrat}\ \emph {et~al.}(2023)\citenamefont {Veyrat}, \citenamefont {Labracherie}, \citenamefont {Bashlakov}, \citenamefont {Caglieris}, \citenamefont {Facio}, \citenamefont {Shipunov}, \citenamefont {Charvin}, \citenamefont {Acharya}, \citenamefont {Naidyuk}, \citenamefont {Giraud}, \citenamefont {van~den Brink}, \citenamefont {B{\ifmmode\ddot{u}\else\"{u}\fi}chner}, \citenamefont {Hess}, \citenamefont {Aswartham},\ and\ \citenamefont {Dufouleur}}]{Veyrat2023ExpWeyl2DSC}%
  \BibitemOpen
  \bibfield  {author} {\bibinfo {author} {\bibfnamefont {A.}~\bibnamefont {Veyrat}}, \bibinfo {author} {\bibfnamefont {V.}~\bibnamefont {Labracherie}}, \bibinfo {author} {\bibfnamefont {D.~L.}\ \bibnamefont {Bashlakov}}, \bibinfo {author} {\bibfnamefont {F.}~\bibnamefont {Caglieris}}, \bibinfo {author} {\bibfnamefont {J.~I.}\ \bibnamefont {Facio}}, \bibinfo {author} {\bibfnamefont {G.}~\bibnamefont {Shipunov}}, \bibinfo {author} {\bibfnamefont {T.}~\bibnamefont {Charvin}}, \bibinfo {author} {\bibfnamefont {R.}~\bibnamefont {Acharya}}, \bibinfo {author} {\bibfnamefont {Y.}~\bibnamefont {Naidyuk}}, \bibinfo {author} {\bibfnamefont {R.}~\bibnamefont {Giraud}}, \bibinfo {author} {\bibfnamefont {J.}~\bibnamefont {van~den Brink}}, \bibinfo {author} {\bibfnamefont {B.}~\bibnamefont {B{\ifmmode\ddot{u}\else\"{u}\fi}chner}}, \bibinfo {author} {\bibfnamefont {C.}~\bibnamefont {Hess}}, \bibinfo {author} {\bibfnamefont {S.}~\bibnamefont {Aswartham}},\ and\ \bibinfo {author} {\bibfnamefont {J.}~\bibnamefont {Dufouleur}},\
  }\bibfield  {title} {\bibinfo {title} {{Berezinskii{\textendash}Kosterlitz{\textendash}Thouless Transition in the Type-I Weyl Semimetal PtBi$_2$}},\ }\href {https://doi.org/10.1021/acs.nanolett.2c04297} {\bibfield  {journal} {\bibinfo  {journal} {Nano Lett.}\ }\textbf {\bibinfo {volume} {23}},\ \bibinfo {pages} {1229} (\bibinfo {year} {2023})}\BibitemShut {NoStop}%
\bibitem [{\citenamefont {O'Leary}\ \emph {et~al.}(2025)\citenamefont {O'Leary}, \citenamefont {Li}, \citenamefont {Wang}, \citenamefont {Schrunk}, \citenamefont {Eaton}, \citenamefont {Canfield},\ and\ \citenamefont {Kaminski}}]{OLeary2025PtBi2}%
  \BibitemOpen
  \bibfield  {author} {\bibinfo {author} {\bibfnamefont {E.}~\bibnamefont {O'Leary}}, \bibinfo {author} {\bibfnamefont {Z.}~\bibnamefont {Li}}, \bibinfo {author} {\bibfnamefont {L.-L.}\ \bibnamefont {Wang}}, \bibinfo {author} {\bibfnamefont {B.}~\bibnamefont {Schrunk}}, \bibinfo {author} {\bibfnamefont {A.}~\bibnamefont {Eaton}}, \bibinfo {author} {\bibfnamefont {P.~C.}\ \bibnamefont {Canfield}},\ and\ \bibinfo {author} {\bibfnamefont {A.}~\bibnamefont {Kaminski}},\ }\bibfield  {title} {\bibinfo {title} {{Topography of Fermi arcs in $t\text{\ensuremath{-}}{\mathrm{PtBi}}_{2}$ using high-resolution angle-resolved photoemission spectroscopy}},\ }\href {https://doi.org/10.1103/n5pz-j2sl} {\bibfield  {journal} {\bibinfo  {journal} {Phys. Rev. B}\ }\textbf {\bibinfo {volume} {112}},\ \bibinfo {pages} {085154} (\bibinfo {year} {2025})}\BibitemShut {NoStop}%
\bibitem [{\citenamefont {Vocaturo}\ \emph {et~al.}(2024)\citenamefont {Vocaturo}, \citenamefont {Koepernik}, \citenamefont {Facio}, \citenamefont {Timm}, \citenamefont {Fulga}, \citenamefont {Janson},\ and\ \citenamefont {van~den Brink}}]{Vocaturo2024PtBi2Effective}%
  \BibitemOpen
  \bibfield  {author} {\bibinfo {author} {\bibfnamefont {R.}~\bibnamefont {Vocaturo}}, \bibinfo {author} {\bibfnamefont {K.}~\bibnamefont {Koepernik}}, \bibinfo {author} {\bibfnamefont {J.~I.}\ \bibnamefont {Facio}}, \bibinfo {author} {\bibfnamefont {C.}~\bibnamefont {Timm}}, \bibinfo {author} {\bibfnamefont {I.~C.}\ \bibnamefont {Fulga}}, \bibinfo {author} {\bibfnamefont {O.}~\bibnamefont {Janson}},\ and\ \bibinfo {author} {\bibfnamefont {J.}~\bibnamefont {van~den Brink}},\ }\bibfield  {title} {\bibinfo {title} {{Electronic structure of the surface-superconducting Weyl semimetal ${\mathrm{PtBi}}_{2}$}},\ }\href {https://doi.org/10.1103/PhysRevB.110.054504} {\bibfield  {journal} {\bibinfo  {journal} {Phys. Rev. B}\ }\textbf {\bibinfo {volume} {110}},\ \bibinfo {pages} {054504} (\bibinfo {year} {2024})}\BibitemShut {NoStop}%
\bibitem [{\citenamefont {Palumbo}\ \emph {et~al.}(2025)\citenamefont {Palumbo}, \citenamefont {Cornaglia},\ and\ \citenamefont {Facio}}]{Palumbo2025DFT}%
  \BibitemOpen
  \bibfield  {author} {\bibinfo {author} {\bibfnamefont {S.}~\bibnamefont {Palumbo}}, \bibinfo {author} {\bibfnamefont {P.~S.}\ \bibnamefont {Cornaglia}},\ and\ \bibinfo {author} {\bibfnamefont {J.~I.}\ \bibnamefont {Facio}},\ }\bibfield  {title} {\bibinfo {title} {{Interplay between inversion and translation symmetries in trigonal ${\mathrm{PtBi}}_{2}$}},\ }\href {https://doi.org/10.1103/bb1q-4qt5} {\bibfield  {journal} {\bibinfo  {journal} {Phys. Rev. B}\ }\textbf {\bibinfo {volume} {112}},\ \bibinfo {pages} {205125} (\bibinfo {year} {2025})}\BibitemShut {NoStop}%
\bibitem [{\citenamefont {Kuibarov}\ \emph {et~al.}(2024)\citenamefont {Kuibarov}, \citenamefont {Suvorov}, \citenamefont {Vocaturo}, \citenamefont {Fedorov}, \citenamefont {Lou}, \citenamefont {Merkwitz}, \citenamefont {Voroshnin}, \citenamefont {Facio}, \citenamefont {Koepernik}, \citenamefont {Yaresko}, \citenamefont {Shipunov}, \citenamefont {Aswartham}, \citenamefont {Brink}, \citenamefont {B{\ifmmode\ddot{u}\else\"{u}\fi}chner},\ and\ \citenamefont {Borisenko}}]{Kuibarov2024FermiArcSCNat}%
  \BibitemOpen
  \bibfield  {author} {\bibinfo {author} {\bibfnamefont {A.}~\bibnamefont {Kuibarov}}, \bibinfo {author} {\bibfnamefont {O.}~\bibnamefont {Suvorov}}, \bibinfo {author} {\bibfnamefont {R.}~\bibnamefont {Vocaturo}}, \bibinfo {author} {\bibfnamefont {A.}~\bibnamefont {Fedorov}}, \bibinfo {author} {\bibfnamefont {R.}~\bibnamefont {Lou}}, \bibinfo {author} {\bibfnamefont {L.}~\bibnamefont {Merkwitz}}, \bibinfo {author} {\bibfnamefont {V.}~\bibnamefont {Voroshnin}}, \bibinfo {author} {\bibfnamefont {J.~I.}\ \bibnamefont {Facio}}, \bibinfo {author} {\bibfnamefont {K.}~\bibnamefont {Koepernik}}, \bibinfo {author} {\bibfnamefont {A.}~\bibnamefont {Yaresko}}, \bibinfo {author} {\bibfnamefont {G.}~\bibnamefont {Shipunov}}, \bibinfo {author} {\bibfnamefont {S.}~\bibnamefont {Aswartham}}, \bibinfo {author} {\bibfnamefont {J.~v.~d.}\ \bibnamefont {Brink}}, \bibinfo {author} {\bibfnamefont {B.}~\bibnamefont {B{\ifmmode\ddot{u}\else\"{u}\fi}chner}},\ and\ \bibinfo {author} {\bibfnamefont {S.}~\bibnamefont {Borisenko}},\
  }\bibfield  {title} {\bibinfo {title} {{Evidence of superconducting Fermi arcs}},\ }\href {https://doi.org/10.1038/s41586-023-06977-7} {\bibfield  {journal} {\bibinfo  {journal} {Nature}\ }\textbf {\bibinfo {volume} {626}},\ \bibinfo {pages} {294} (\bibinfo {year} {2024})}\BibitemShut {NoStop}%
\bibitem [{\citenamefont {Kuibarov}\ \emph {et~al.}(2025{\natexlab{a}})\citenamefont {Kuibarov}, \citenamefont {Changdar}, \citenamefont {Fedorov}, \citenamefont {Lou}, \citenamefont {Suvorov}, \citenamefont {Misheneva}, \citenamefont {Harnagea}, \citenamefont {Kovalchuk}, \citenamefont {Wurmehl}, \citenamefont {B{\ifmmode\ddot{u}\else\"{u}\fi}chner},\ and\ \citenamefont {Borisenko}}]{Kuibarov2025ARPES}%
  \BibitemOpen
  \bibfield  {author} {\bibinfo {author} {\bibfnamefont {A.}~\bibnamefont {Kuibarov}}, \bibinfo {author} {\bibfnamefont {S.}~\bibnamefont {Changdar}}, \bibinfo {author} {\bibfnamefont {A.}~\bibnamefont {Fedorov}}, \bibinfo {author} {\bibfnamefont {R.}~\bibnamefont {Lou}}, \bibinfo {author} {\bibfnamefont {O.}~\bibnamefont {Suvorov}}, \bibinfo {author} {\bibfnamefont {V.}~\bibnamefont {Misheneva}}, \bibinfo {author} {\bibfnamefont {L.}~\bibnamefont {Harnagea}}, \bibinfo {author} {\bibfnamefont {I.}~\bibnamefont {Kovalchuk}}, \bibinfo {author} {\bibfnamefont {S.}~\bibnamefont {Wurmehl}}, \bibinfo {author} {\bibfnamefont {B.}~\bibnamefont {B{\ifmmode\ddot{u}\else\"{u}\fi}chner}},\ and\ \bibinfo {author} {\bibfnamefont {S.}~\bibnamefont {Borisenko}},\ }\bibfield  {title} {\bibinfo {title} {{Measuring superconducting arcs by angle-resolved photoemission spectroscopy}},\ }\href {https://doi.org/10.1103/5q36-wgl9} {\bibfield  {journal} {\bibinfo  {journal} {Phys. Rev. B}\ }\textbf {\bibinfo {volume} {112}},\
  \bibinfo {pages} {144518} (\bibinfo {year} {2025}{\natexlab{a}})}\BibitemShut {NoStop}%
\bibitem [{\citenamefont {Changdar}\ \emph {et~al.}(2025)\citenamefont {Changdar}, \citenamefont {Suvorov}, \citenamefont {Kuibarov}, \citenamefont {Thirupathaiah}, \citenamefont {Shipunov}, \citenamefont {Aswartham}, \citenamefont {Wurmehl}, \citenamefont {Kovalchuk}, \citenamefont {Koepernik}, \citenamefont {Timm}, \citenamefont {B{\ifmmode\ddot{u}\else\"{u}\fi}chner}, \citenamefont {Fulga}, \citenamefont {Borisenko},\ and\ \citenamefont {van~den Brink}}]{Changdar2025iwave}%
  \BibitemOpen
  \bibfield  {author} {\bibinfo {author} {\bibfnamefont {S.}~\bibnamefont {Changdar}}, \bibinfo {author} {\bibfnamefont {O.}~\bibnamefont {Suvorov}}, \bibinfo {author} {\bibfnamefont {A.}~\bibnamefont {Kuibarov}}, \bibinfo {author} {\bibfnamefont {S.}~\bibnamefont {Thirupathaiah}}, \bibinfo {author} {\bibfnamefont {G.}~\bibnamefont {Shipunov}}, \bibinfo {author} {\bibfnamefont {S.}~\bibnamefont {Aswartham}}, \bibinfo {author} {\bibfnamefont {S.}~\bibnamefont {Wurmehl}}, \bibinfo {author} {\bibfnamefont {I.}~\bibnamefont {Kovalchuk}}, \bibinfo {author} {\bibfnamefont {K.}~\bibnamefont {Koepernik}}, \bibinfo {author} {\bibfnamefont {C.}~\bibnamefont {Timm}}, \bibinfo {author} {\bibfnamefont {B.}~\bibnamefont {B{\ifmmode\ddot{u}\else\"{u}\fi}chner}}, \bibinfo {author} {\bibfnamefont {I.~C.}\ \bibnamefont {Fulga}}, \bibinfo {author} {\bibfnamefont {S.}~\bibnamefont {Borisenko}},\ and\ \bibinfo {author} {\bibfnamefont {J.}~\bibnamefont {van~den Brink}},\ }\bibfield  {title} {\bibinfo {title} {{Topological nodal
  i-wave superconductivity in PtBi$_2$}},\ }\href {https://doi.org/10.1038/s41586-025-09712-6} {\bibfield  {journal} {\bibinfo  {journal} {Nature}\ }\textbf {\bibinfo {volume} {647}},\ \bibinfo {pages} {613} (\bibinfo {year} {2025})}\BibitemShut {NoStop}%
\bibitem [{\citenamefont {Kuibarov}\ \emph {et~al.}(2025{\natexlab{b}})\citenamefont {Kuibarov}, \citenamefont {Changdar}, \citenamefont {Vocaturo}, \citenamefont {Suvorov}, \citenamefont {Fedorov}, \citenamefont {Lou}, \citenamefont {Krivenkov}, \citenamefont {Harnagea}, \citenamefont {Wurmehl}, \citenamefont {Brink}, \citenamefont {B{\ifmmode\ddot{u}\else\"{u}\fi}chner},\ and\ \citenamefont {Borisenko}}]{Kuibarov2025SepHighTc}%
  \BibitemOpen
  \bibfield  {author} {\bibinfo {author} {\bibfnamefont {A.}~\bibnamefont {Kuibarov}}, \bibinfo {author} {\bibfnamefont {S.}~\bibnamefont {Changdar}}, \bibinfo {author} {\bibfnamefont {R.}~\bibnamefont {Vocaturo}}, \bibinfo {author} {\bibfnamefont {O.}~\bibnamefont {Suvorov}}, \bibinfo {author} {\bibfnamefont {A.}~\bibnamefont {Fedorov}}, \bibinfo {author} {\bibfnamefont {R.}~\bibnamefont {Lou}}, \bibinfo {author} {\bibfnamefont {M.}~\bibnamefont {Krivenkov}}, \bibinfo {author} {\bibfnamefont {L.}~\bibnamefont {Harnagea}}, \bibinfo {author} {\bibfnamefont {S.}~\bibnamefont {Wurmehl}}, \bibinfo {author} {\bibfnamefont {J.~v.~d.}\ \bibnamefont {Brink}}, \bibinfo {author} {\bibfnamefont {B.}~\bibnamefont {B{\ifmmode\ddot{u}\else\"{u}\fi}chner}},\ and\ \bibinfo {author} {\bibfnamefont {S.}~\bibnamefont {Borisenko}},\ }\bibfield  {title} {\bibinfo {title} {{Three prerequisites for high-temperature superconductivity in t-PtBi$_2$}},\ }\href {https://doi.org/10.48550/arXiv.2509.02178} {\bibfield  {journal} {\bibinfo
   {journal} {arXiv:2509.02178}\ } (\bibinfo {year} {2025}{\natexlab{b}})}\BibitemShut {NoStop}%
\bibitem [{\citenamefont {Schimmel}\ \emph {et~al.}(2024)\citenamefont {Schimmel}, \citenamefont {Fasano}, \citenamefont {Hoffmann}, \citenamefont {Besproswanny}, \citenamefont {Corredor~Bohorquez}, \citenamefont {Puig}, \citenamefont {Elshalem}, \citenamefont {Kalisky}, \citenamefont {Shipunov}, \citenamefont {Baumann}, \citenamefont {Aswartham}, \citenamefont {B{\ifmmode\ddot{u}\else\"{u}\fi}chner},\ and\ \citenamefont {Hess}}]{Schimmel2023ExpWeylSCSTM}%
  \BibitemOpen
  \bibfield  {author} {\bibinfo {author} {\bibfnamefont {S.}~\bibnamefont {Schimmel}}, \bibinfo {author} {\bibfnamefont {Y.}~\bibnamefont {Fasano}}, \bibinfo {author} {\bibfnamefont {S.}~\bibnamefont {Hoffmann}}, \bibinfo {author} {\bibfnamefont {J.}~\bibnamefont {Besproswanny}}, \bibinfo {author} {\bibfnamefont {L.~T.}\ \bibnamefont {Corredor~Bohorquez}}, \bibinfo {author} {\bibfnamefont {J.}~\bibnamefont {Puig}}, \bibinfo {author} {\bibfnamefont {B.-C.}\ \bibnamefont {Elshalem}}, \bibinfo {author} {\bibfnamefont {B.}~\bibnamefont {Kalisky}}, \bibinfo {author} {\bibfnamefont {G.}~\bibnamefont {Shipunov}}, \bibinfo {author} {\bibfnamefont {D.}~\bibnamefont {Baumann}}, \bibinfo {author} {\bibfnamefont {S.}~\bibnamefont {Aswartham}}, \bibinfo {author} {\bibfnamefont {B.}~\bibnamefont {B{\ifmmode\ddot{u}\else\"{u}\fi}chner}},\ and\ \bibinfo {author} {\bibfnamefont {C.}~\bibnamefont {Hess}},\ }\bibfield  {title} {\bibinfo {title} {{Surface superconductivity in the topological Weyl semimetal t-PtBi$_2$}},\ }\href
  {https://doi.org/10.1038/s41467-024-54389-6} {\bibfield  {journal} {\bibinfo  {journal} {Nat. Commun.}\ }\textbf {\bibinfo {volume} {15}},\ \bibinfo {pages} {9895} (\bibinfo {year} {2024})}\BibitemShut {NoStop}%
\bibitem [{\citenamefont {Hoffmann}\ \emph {et~al.}(2024)\citenamefont {Hoffmann}, \citenamefont {Schimmel}, \citenamefont {Vocaturo}, \citenamefont {Puig}, \citenamefont {Shipunov}, \citenamefont {Janson}, \citenamefont {Aswartham}, \citenamefont {Baumann}, \citenamefont {B{\ifmmode\ddot{u}\else\"{u}\fi}chner}, \citenamefont {van~den Brink}, \citenamefont {Fasano}, \citenamefont {Facio},\ and\ \citenamefont {Hess}}]{Hoffmann2024PtBi2STM}%
  \BibitemOpen
  \bibfield  {author} {\bibinfo {author} {\bibfnamefont {S.}~\bibnamefont {Hoffmann}}, \bibinfo {author} {\bibfnamefont {S.}~\bibnamefont {Schimmel}}, \bibinfo {author} {\bibfnamefont {R.}~\bibnamefont {Vocaturo}}, \bibinfo {author} {\bibfnamefont {J.}~\bibnamefont {Puig}}, \bibinfo {author} {\bibfnamefont {G.}~\bibnamefont {Shipunov}}, \bibinfo {author} {\bibfnamefont {O.}~\bibnamefont {Janson}}, \bibinfo {author} {\bibfnamefont {S.}~\bibnamefont {Aswartham}}, \bibinfo {author} {\bibfnamefont {D.}~\bibnamefont {Baumann}}, \bibinfo {author} {\bibfnamefont {B.}~\bibnamefont {B{\ifmmode\ddot{u}\else\"{u}\fi}chner}}, \bibinfo {author} {\bibfnamefont {J.}~\bibnamefont {van~den Brink}}, \bibinfo {author} {\bibfnamefont {Y.}~\bibnamefont {Fasano}}, \bibinfo {author} {\bibfnamefont {J.~I.}\ \bibnamefont {Facio}},\ and\ \bibinfo {author} {\bibfnamefont {C.}~\bibnamefont {Hess}},\ }\bibfield  {title} {\bibinfo {title} {{Fermi Arcs Dominating the Electronic Surface Properties of Trigonal PtBi$_2$}},\ }\href
  {https://doi.org/10.1002/apxr.202400150} {\bibfield  {journal} {\bibinfo  {journal} {Adv. Phys. Res.}\ }\textbf {\bibinfo {volume} {4}},\ \bibinfo {pages} {2400150} (\bibinfo {year} {2024})}\BibitemShut {NoStop}%
\bibitem [{\citenamefont {Huang}\ \emph {et~al.}(2025)\citenamefont {Huang}, \citenamefont {Zhao}, \citenamefont {Schimmel}, \citenamefont {Besproswanny}, \citenamefont {H{\ifmmode\ddot{a}\else\"{a}\fi}rtl}, \citenamefont {Hess}, \citenamefont {B{\ifmmode\ddot{u}\else\"{u}\fi}chner},\ and\ \citenamefont {Bode}}]{Huang2025PtBi2STM}%
  \BibitemOpen
  \bibfield  {author} {\bibinfo {author} {\bibfnamefont {X.}~\bibnamefont {Huang}}, \bibinfo {author} {\bibfnamefont {L.}~\bibnamefont {Zhao}}, \bibinfo {author} {\bibfnamefont {S.}~\bibnamefont {Schimmel}}, \bibinfo {author} {\bibfnamefont {J.}~\bibnamefont {Besproswanny}}, \bibinfo {author} {\bibfnamefont {P.}~\bibnamefont {H{\ifmmode\ddot{a}\else\"{a}\fi}rtl}}, \bibinfo {author} {\bibfnamefont {C.}~\bibnamefont {Hess}}, \bibinfo {author} {\bibfnamefont {B.}~\bibnamefont {B{\ifmmode\ddot{u}\else\"{u}\fi}chner}},\ and\ \bibinfo {author} {\bibfnamefont {M.}~\bibnamefont {Bode}},\ }\bibfield  {title} {\bibinfo {title} {{Sizable superconducting gap and anisotropic chiral topological superconductivity in the Weyl semimetal PtBi$_2$}},\ }\href {https://doi.org/10.48550/arXiv.2507.13843} {\bibfield  {journal} {\bibinfo  {journal} {arXiv:2507.13843}\ } (\bibinfo {year} {2025})}\BibitemShut {NoStop}%
\bibitem [{\citenamefont {Moreno}\ \emph {et~al.}(2025)\citenamefont {Moreno}, \citenamefont {Talavera}, \citenamefont {Herrera}, \citenamefont {Valle}, \citenamefont {Li}, \citenamefont {Wang}, \citenamefont {Bud'ko}, \citenamefont {Buzdin}, \citenamefont {Guillam{\ifmmode\acute{o}\else\'{o}\fi}n}, \citenamefont {Canfield},\ and\ \citenamefont {Suderow}}]{Moreno2025PtBi2vortex}%
  \BibitemOpen
  \bibfield  {author} {\bibinfo {author} {\bibfnamefont {J.~A.}\ \bibnamefont {Moreno}}, \bibinfo {author} {\bibfnamefont {P.~G.}\ \bibnamefont {Talavera}}, \bibinfo {author} {\bibfnamefont {E.}~\bibnamefont {Herrera}}, \bibinfo {author} {\bibfnamefont {S.~L.}\ \bibnamefont {Valle}}, \bibinfo {author} {\bibfnamefont {Z.}~\bibnamefont {Li}}, \bibinfo {author} {\bibfnamefont {L.-L.}\ \bibnamefont {Wang}}, \bibinfo {author} {\bibfnamefont {S.}~\bibnamefont {Bud'ko}}, \bibinfo {author} {\bibfnamefont {A.~I.}\ \bibnamefont {Buzdin}}, \bibinfo {author} {\bibfnamefont {I.}~\bibnamefont {Guillam{\ifmmode\acute{o}\else\'{o}\fi}n}}, \bibinfo {author} {\bibfnamefont {P.~C.}\ \bibnamefont {Canfield}},\ and\ \bibinfo {author} {\bibfnamefont {H.}~\bibnamefont {Suderow}},\ }\bibfield  {title} {\bibinfo {title} {{Robust surface superconductivity and vortex lattice in the Weyl semimetal $\gamma$-PtBi$_2$}},\ }\href {https://doi.org/10.48550/arXiv.2508.04867} {\bibfield  {journal} {\bibinfo  {journal} {arXiv:2508.04867}\ }
  (\bibinfo {year} {2025})}\BibitemShut {NoStop}%
\bibitem [{\citenamefont {M{\ae}land}\ \emph {et~al.}(2025)\citenamefont {M{\ae}land}, \citenamefont {Bahari},\ and\ \citenamefont {Trauzettel}}]{Maeland2025Jun}%
  \BibitemOpen
  \bibfield  {author} {\bibinfo {author} {\bibfnamefont {K.}~\bibnamefont {M{\ae}land}}, \bibinfo {author} {\bibfnamefont {M.}~\bibnamefont {Bahari}},\ and\ \bibinfo {author} {\bibfnamefont {B.}~\bibnamefont {Trauzettel}},\ }\bibfield  {title} {\bibinfo {title} {{Phonon-mediated intrinsic topological superconductivity in Fermi arcs}},\ }\href {https://doi.org/10.1103/47vs-qgzk} {\bibfield  {journal} {\bibinfo  {journal} {Phys. Rev. B}\ }\textbf {\bibinfo {volume} {112}},\ \bibinfo {pages} {104507} (\bibinfo {year} {2025})}\BibitemShut {NoStop}%
\bibitem [{\citenamefont {Nayak}\ \emph {et~al.}(2008)\citenamefont {Nayak}, \citenamefont {Simon}, \citenamefont {Stern}, \citenamefont {Freedman},\ and\ \citenamefont {Das~Sarma}}]{TopoQuantumCompRevModPhys}%
  \BibitemOpen
  \bibfield  {author} {\bibinfo {author} {\bibfnamefont {C.}~\bibnamefont {Nayak}}, \bibinfo {author} {\bibfnamefont {S.~H.}\ \bibnamefont {Simon}}, \bibinfo {author} {\bibfnamefont {A.}~\bibnamefont {Stern}}, \bibinfo {author} {\bibfnamefont {M.}~\bibnamefont {Freedman}},\ and\ \bibinfo {author} {\bibfnamefont {S.}~\bibnamefont {Das~Sarma}},\ }\bibfield  {title} {\bibinfo {title} {{Non-Abelian} anyons and topological quantum computation},\ }\href {https://doi.org/10.1103/RevModPhys.80.1083} {\bibfield  {journal} {\bibinfo  {journal} {Rev. Mod. Phys.}\ }\textbf {\bibinfo {volume} {80}},\ \bibinfo {pages} {1083} (\bibinfo {year} {2008})}\BibitemShut {NoStop}%
\bibitem [{\citenamefont {Leijnse}\ and\ \citenamefont {Flensberg}(2012)}]{Leijnse2012TSCrev}%
  \BibitemOpen
  \bibfield  {author} {\bibinfo {author} {\bibfnamefont {M.}~\bibnamefont {Leijnse}}\ and\ \bibinfo {author} {\bibfnamefont {K.}~\bibnamefont {Flensberg}},\ }\bibfield  {title} {\bibinfo {title} {{Introduction to topological superconductivity and Majorana fermions}},\ }\href {https://doi.org/10.1088/0268-1242/27/12/124003} {\bibfield  {journal} {\bibinfo  {journal} {Semicond. Sci. Technol.}\ }\textbf {\bibinfo {volume} {27}},\ \bibinfo {pages} {124003} (\bibinfo {year} {2012})}\BibitemShut {NoStop}%
\bibitem [{\citenamefont {Bernevig}\ and\ \citenamefont {Hughes}(2013)}]{Bernevig2013}%
  \BibitemOpen
  \bibfield  {author} {\bibinfo {author} {\bibfnamefont {B.~A.}\ \bibnamefont {Bernevig}}\ and\ \bibinfo {author} {\bibfnamefont {T.~L.}\ \bibnamefont {Hughes}},\ }\href@noop {} {\emph {\bibinfo {title} {Topological Insulators and Topological Superconductors}}}\ (\bibinfo  {publisher} {Princeton University Press},\ \bibinfo {address} {Princeton, NJ},\ \bibinfo {year} {2013})\BibitemShut {NoStop}%
\bibitem [{\citenamefont {Sato}\ and\ \citenamefont {Ando}(2017)}]{TopoSCrevSato}%
  \BibitemOpen
  \bibfield  {author} {\bibinfo {author} {\bibfnamefont {M.}~\bibnamefont {Sato}}\ and\ \bibinfo {author} {\bibfnamefont {Y.}~\bibnamefont {Ando}},\ }\bibfield  {title} {\bibinfo {title} {Topological superconductors: a review},\ }\href {https://doi.org/10.1088/1361-6633/aa6ac7} {\bibfield  {journal} {\bibinfo  {journal} {Rep. Prog. Phys.}\ }\textbf {\bibinfo {volume} {80}},\ \bibinfo {pages} {076501} (\bibinfo {year} {2017})}\BibitemShut {NoStop}%
\bibitem [{\citenamefont {Wollman}\ \emph {et~al.}(1993)\citenamefont {Wollman}, \citenamefont {Van~Harlingen}, \citenamefont {Lee}, \citenamefont {Ginsberg},\ and\ \citenamefont {Leggett}}]{vanHarlingen1993phase}%
  \BibitemOpen
  \bibfield  {author} {\bibinfo {author} {\bibfnamefont {D.~A.}\ \bibnamefont {Wollman}}, \bibinfo {author} {\bibfnamefont {D.~J.}\ \bibnamefont {Van~Harlingen}}, \bibinfo {author} {\bibfnamefont {W.~C.}\ \bibnamefont {Lee}}, \bibinfo {author} {\bibfnamefont {D.~M.}\ \bibnamefont {Ginsberg}},\ and\ \bibinfo {author} {\bibfnamefont {A.~J.}\ \bibnamefont {Leggett}},\ }\bibfield  {title} {\bibinfo {title} {{Experimental determination of the superconducting pairing state in YBCO from the phase coherence of YBCO-Pb dc SQUIDs}},\ }\href {https://doi.org/10.1103/PhysRevLett.71.2134} {\bibfield  {journal} {\bibinfo  {journal} {Phys. Rev. Lett.}\ }\textbf {\bibinfo {volume} {71}},\ \bibinfo {pages} {2134} (\bibinfo {year} {1993})}\BibitemShut {NoStop}%
\bibitem [{\citenamefont {Min}\ \emph {et~al.}(2019)\citenamefont {Min}, \citenamefont {Bentmann}, \citenamefont {Neu}, \citenamefont {Eck}, \citenamefont {Moser}, \citenamefont {Figgemeier}, \citenamefont {{\ifmmode\ddot{U}\else\"{U}\fi}nzelmann}, \citenamefont {Kissner}, \citenamefont {Lutz}, \citenamefont {Koch}, \citenamefont {Jozwiak}, \citenamefont {Bostwick}, \citenamefont {Rotenberg}, \citenamefont {Thomale}, \citenamefont {Sangiovanni}, \citenamefont {Siegrist}, \citenamefont {Di~Sante},\ and\ \citenamefont {Reinert}}]{Min2019FA}%
  \BibitemOpen
  \bibfield  {author} {\bibinfo {author} {\bibfnamefont {C.-H.}\ \bibnamefont {Min}}, \bibinfo {author} {\bibfnamefont {H.}~\bibnamefont {Bentmann}}, \bibinfo {author} {\bibfnamefont {J.~N.}\ \bibnamefont {Neu}}, \bibinfo {author} {\bibfnamefont {P.}~\bibnamefont {Eck}}, \bibinfo {author} {\bibfnamefont {S.}~\bibnamefont {Moser}}, \bibinfo {author} {\bibfnamefont {T.}~\bibnamefont {Figgemeier}}, \bibinfo {author} {\bibfnamefont {M.}~\bibnamefont {{\ifmmode\ddot{U}\else\"{U}\fi}nzelmann}}, \bibinfo {author} {\bibfnamefont {K.}~\bibnamefont {Kissner}}, \bibinfo {author} {\bibfnamefont {P.}~\bibnamefont {Lutz}}, \bibinfo {author} {\bibfnamefont {R.~J.}\ \bibnamefont {Koch}}, \bibinfo {author} {\bibfnamefont {C.}~\bibnamefont {Jozwiak}}, \bibinfo {author} {\bibfnamefont {A.}~\bibnamefont {Bostwick}}, \bibinfo {author} {\bibfnamefont {E.}~\bibnamefont {Rotenberg}}, \bibinfo {author} {\bibfnamefont {R.}~\bibnamefont {Thomale}}, \bibinfo {author} {\bibfnamefont {G.}~\bibnamefont {Sangiovanni}}, \bibinfo {author}
  {\bibfnamefont {T.}~\bibnamefont {Siegrist}}, \bibinfo {author} {\bibfnamefont {D.}~\bibnamefont {Di~Sante}},\ and\ \bibinfo {author} {\bibfnamefont {F.}~\bibnamefont {Reinert}},\ }\bibfield  {title} {\bibinfo {title} {{Orbital Fingerprint of Topological Fermi Arcs in the Weyl Semimetal TaP}},\ }\href {https://doi.org/10.1103/PhysRevLett.122.116402} {\bibfield  {journal} {\bibinfo  {journal} {Phys. Rev. Lett.}\ }\textbf {\bibinfo {volume} {122}},\ \bibinfo {pages} {116402} (\bibinfo {year} {2019})}\BibitemShut {NoStop}%
\bibitem [{\citenamefont {Trama}\ \emph {et~al.}(2025)\citenamefont {Trama}, \citenamefont {K{\ifmmode\ddot{o}\else\"{o}\fi}nye}, \citenamefont {Fulga},\ and\ \citenamefont {van~den Brink}}]{Trama2024TRSWeylSM_SC}%
  \BibitemOpen
  \bibfield  {author} {\bibinfo {author} {\bibfnamefont {M.}~\bibnamefont {Trama}}, \bibinfo {author} {\bibfnamefont {V.}~\bibnamefont {K{\ifmmode\ddot{o}\else\"{o}\fi}nye}}, \bibinfo {author} {\bibfnamefont {I.~C.}\ \bibnamefont {Fulga}},\ and\ \bibinfo {author} {\bibfnamefont {J.}~\bibnamefont {van~den Brink}},\ }\bibfield  {title} {\bibinfo {title} {{Self-consistent surface superconductivity in time-reversal symmetric Weyl semimetals}},\ }\href {https://doi.org/10.1103/bdtb-mb8c} {\bibfield  {journal} {\bibinfo  {journal} {Phys. Rev. B}\ }\textbf {\bibinfo {volume} {112}},\ \bibinfo {pages} {064514} (\bibinfo {year} {2025})}\BibitemShut {NoStop}%
\bibitem [{\citenamefont {Bashlakov}\ \emph {et~al.}(2022)\citenamefont {Bashlakov}, \citenamefont {Kvitnitskaya}, \citenamefont {Shipunov}, \citenamefont {Aswartham}, \citenamefont {Feya}, \citenamefont {Efremov}, \citenamefont {B{\ifmmode\ddot{u}\else\"{u}\fi}chner},\ and\ \citenamefont {Naidyuk}}]{Bashlakov2022PtBi2phononExp}%
  \BibitemOpen
  \bibfield  {author} {\bibinfo {author} {\bibfnamefont {D.~L.}\ \bibnamefont {Bashlakov}}, \bibinfo {author} {\bibfnamefont {O.~E.}\ \bibnamefont {Kvitnitskaya}}, \bibinfo {author} {\bibfnamefont {G.}~\bibnamefont {Shipunov}}, \bibinfo {author} {\bibfnamefont {S.}~\bibnamefont {Aswartham}}, \bibinfo {author} {\bibfnamefont {O.~D.}\ \bibnamefont {Feya}}, \bibinfo {author} {\bibfnamefont {D.~V.}\ \bibnamefont {Efremov}}, \bibinfo {author} {\bibfnamefont {B.}~\bibnamefont {B{\ifmmode\ddot{u}\else\"{u}\fi}chner}},\ and\ \bibinfo {author} {\bibfnamefont {{\relax Yu}.~G.}\ \bibnamefont {Naidyuk}},\ }\bibfield  {title} {\bibinfo {title} {{Electron-phonon interaction and point contact enhanced superconductivity in trigonal PtBi$_2$}},\ }\href {https://doi.org/10.1063/10.0014014} {\bibfield  {journal} {\bibinfo  {journal} {Low Temp. Phys.}\ }\textbf {\bibinfo {volume} {48}},\ \bibinfo {pages} {747} (\bibinfo {year} {2022})}\BibitemShut {NoStop}%
\bibitem [{\citenamefont {Bogoljubov}\ \emph {et~al.}(1958)\citenamefont {Bogoljubov}, \citenamefont {Tolmachov},\ and\ \citenamefont {{\ifmmode\check{S}\else\v{S}\fi}irkov}}]{Bogoliubov1958}%
  \BibitemOpen
  \bibfield  {author} {\bibinfo {author} {\bibfnamefont {N.~N.}\ \bibnamefont {Bogoljubov}}, \bibinfo {author} {\bibfnamefont {V.~V.}\ \bibnamefont {Tolmachov}},\ and\ \bibinfo {author} {\bibfnamefont {D.~V.}\ \bibnamefont {{\ifmmode\check{S}\else\v{S}\fi}irkov}},\ }\bibfield  {title} {\bibinfo {title} {{A New Method in the Theory of Superconductivity}},\ }\href {https://doi.org/10.1002/prop.19580061102} {\bibfield  {journal} {\bibinfo  {journal} {Fortschr. Phys.}\ }\textbf {\bibinfo {volume} {6}},\ \bibinfo {pages} {605} (\bibinfo {year} {1958})}\BibitemShut {NoStop}%
\bibitem [{\citenamefont {Morel}\ and\ \citenamefont {Anderson}(1962)}]{Morel1962Anderson}%
  \BibitemOpen
  \bibfield  {author} {\bibinfo {author} {\bibfnamefont {P.}~\bibnamefont {Morel}}\ and\ \bibinfo {author} {\bibfnamefont {P.~W.}\ \bibnamefont {Anderson}},\ }\bibfield  {title} {\bibinfo {title} {{Calculation of the Superconducting State Parameters with Retarded Electron-Phonon Interaction}},\ }\href {https://doi.org/10.1103/PhysRev.125.1263} {\bibfield  {journal} {\bibinfo  {journal} {Phys. Rev.}\ }\textbf {\bibinfo {volume} {125}},\ \bibinfo {pages} {1263} (\bibinfo {year} {1962})}\BibitemShut {NoStop}%
\bibitem [{\citenamefont {Kohn}\ and\ \citenamefont {Luttinger}(1965)}]{Kohn1965Luttinger}%
  \BibitemOpen
  \bibfield  {author} {\bibinfo {author} {\bibfnamefont {W.}~\bibnamefont {Kohn}}\ and\ \bibinfo {author} {\bibfnamefont {J.~M.}\ \bibnamefont {Luttinger}},\ }\bibfield  {title} {\bibinfo {title} {{New Mechanism for Superconductivity}},\ }\href {https://doi.org/10.1103/PhysRevLett.15.524} {\bibfield  {journal} {\bibinfo  {journal} {Phys. Rev. Lett.}\ }\textbf {\bibinfo {volume} {15}},\ \bibinfo {pages} {524} (\bibinfo {year} {1965})}\BibitemShut {NoStop}%
\bibitem [{\citenamefont {R{\ifmmode\ddot{o}\else\"{o}\fi}sner}\ \emph {et~al.}(2018)\citenamefont {R{\ifmmode\ddot{o}\else\"{o}\fi}sner}, \citenamefont {Groenewald}, \citenamefont {Sch{\ifmmode\ddot{o}\else\"{o}\fi}nhoff}, \citenamefont {Berges}, \citenamefont {Haas},\ and\ \citenamefont {Wehling}}]{Rosner2018Plasmon}%
  \BibitemOpen
  \bibfield  {author} {\bibinfo {author} {\bibfnamefont {M.}~\bibnamefont {R{\ifmmode\ddot{o}\else\"{o}\fi}sner}}, \bibinfo {author} {\bibfnamefont {R.~E.}\ \bibnamefont {Groenewald}}, \bibinfo {author} {\bibfnamefont {G.}~\bibnamefont {Sch{\ifmmode\ddot{o}\else\"{o}\fi}nhoff}}, \bibinfo {author} {\bibfnamefont {J.}~\bibnamefont {Berges}}, \bibinfo {author} {\bibfnamefont {S.}~\bibnamefont {Haas}},\ and\ \bibinfo {author} {\bibfnamefont {T.~O.}\ \bibnamefont {Wehling}},\ }\bibfield  {title} {\bibinfo {title} {{Plasmonic Superconductivity in Layered Materials}},\ }\href {https://doi.org/10.48550/arXiv.1803.04576} {\bibfield  {journal} {\bibinfo  {journal} {arXiv:1803.04576}\ } (\bibinfo {year} {2018})}\BibitemShut {NoStop}%
\bibitem [{\citenamefont {Veld}\ \emph {et~al.}(2023)\citenamefont {Veld}, \citenamefont {Katsnelson}, \citenamefont {Millis},\ and\ \citenamefont {R{\ifmmode\ddot{o}\else\"{o}\fi}sner}}]{Veld2023Plasmon}%
  \BibitemOpen
  \bibfield  {author} {\bibinfo {author} {\bibfnamefont {Y.~i.}\ \bibnamefont {Veld}}, \bibinfo {author} {\bibfnamefont {M.~I.}\ \bibnamefont {Katsnelson}}, \bibinfo {author} {\bibfnamefont {A.~J.}\ \bibnamefont {Millis}},\ and\ \bibinfo {author} {\bibfnamefont {M.}~\bibnamefont {R{\ifmmode\ddot{o}\else\"{o}\fi}sner}},\ }\bibfield  {title} {\bibinfo {title} {{Screening induced crossover between phonon- and plasmon-mediated pairing in layered superconductors}},\ }\href {https://doi.org/10.1088/2053-1583/acf944} {\bibfield  {journal} {\bibinfo  {journal} {2D Mater.}\ }\textbf {\bibinfo {volume} {10}},\ \bibinfo {pages} {045031} (\bibinfo {year} {2023})}\BibitemShut {NoStop}%
\bibitem [{\citenamefont {Veld}\ \emph {et~al.}(2025)\citenamefont {Veld}, \citenamefont {Katsnelson}, \citenamefont {Millis},\ and\ \citenamefont {R{\ifmmode\ddot{o}\else\"{o}\fi}sner}}]{Veld2025Plasmon}%
  \BibitemOpen
  \bibfield  {author} {\bibinfo {author} {\bibfnamefont {Y.~i.}\ \bibnamefont {Veld}}, \bibinfo {author} {\bibfnamefont {M.~I.}\ \bibnamefont {Katsnelson}}, \bibinfo {author} {\bibfnamefont {A.~J.}\ \bibnamefont {Millis}},\ and\ \bibinfo {author} {\bibfnamefont {M.}~\bibnamefont {R{\ifmmode\ddot{o}\else\"{o}\fi}sner}},\ }\bibfield  {title} {\bibinfo {title} {{Enhancing Plasmonic Superconductivity in Layered Materials via Dynamical Coulomb Engineering}},\ }\href {https://doi.org/10.48550/arXiv.2508.06195} {\bibfield  {journal} {\bibinfo  {journal} {arXiv:2508.06195}\ } (\bibinfo {year} {2025})}\BibitemShut {NoStop}%
\bibitem [{\citenamefont {Kobayashi}\ \emph {et~al.}(2022)\citenamefont {Kobayashi}, \citenamefont {Bhattacharya}, \citenamefont {Timm},\ and\ \citenamefont {Brydon}}]{Kobayashi2022Apr}%
  \BibitemOpen
  \bibfield  {author} {\bibinfo {author} {\bibfnamefont {S.}~\bibnamefont {Kobayashi}}, \bibinfo {author} {\bibfnamefont {A.}~\bibnamefont {Bhattacharya}}, \bibinfo {author} {\bibfnamefont {C.}~\bibnamefont {Timm}},\ and\ \bibinfo {author} {\bibfnamefont {P.~M.~R.}\ \bibnamefont {Brydon}},\ }\bibfield  {title} {\bibinfo {title} {{Bogoliubov Fermi surfaces from pairing of emergent $j=\frac{3}{2}$ fermions on the pyrochlore lattice}},\ }\href {https://doi.org/10.1103/PhysRevB.105.134507} {\bibfield  {journal} {\bibinfo  {journal} {Phys. Rev. B}\ }\textbf {\bibinfo {volume} {105}},\ \bibinfo {pages} {134507} (\bibinfo {year} {2022})}\BibitemShut {NoStop}%
\bibitem [{\citenamefont {Moriya}\ and\ \citenamefont {Ueda}(2003)}]{Moriya2003spinfluct}%
  \BibitemOpen
  \bibfield  {author} {\bibinfo {author} {\bibfnamefont {T.}~\bibnamefont {Moriya}}\ and\ \bibinfo {author} {\bibfnamefont {K.}~\bibnamefont {Ueda}},\ }\bibfield  {title} {\bibinfo {title} {{Antiferromagnetic spin fluctuation and superconductivity}},\ }\href {https://doi.org/10.1088/0034-4885/66/8/202} {\bibfield  {journal} {\bibinfo  {journal} {Rep. Prog. Phys.}\ }\textbf {\bibinfo {volume} {66}},\ \bibinfo {pages} {1299} (\bibinfo {year} {2003})}\BibitemShut {NoStop}%
\bibitem [{\citenamefont {Kuroki}\ \emph {et~al.}(2004)\citenamefont {Kuroki}, \citenamefont {Tanaka},\ and\ \citenamefont {Arita}}]{Kuroki2004Aug}%
  \BibitemOpen
  \bibfield  {author} {\bibinfo {author} {\bibfnamefont {K.}~\bibnamefont {Kuroki}}, \bibinfo {author} {\bibfnamefont {Y.}~\bibnamefont {Tanaka}},\ and\ \bibinfo {author} {\bibfnamefont {R.}~\bibnamefont {Arita}},\ }\bibfield  {title} {\bibinfo {title} {{Possible Spin-Triplet$f$-Wave Pairing Due to Disconnected Fermi Surfaces In ${\mathrm{N}\mathrm{a}}_{x}{\mathrm{C}\mathrm{o}\mathrm{O}}_{2}\ifmmode\cdot\else\textperiodcentered\fi{}y{\mathrm{H}}_{2}\mathrm{O}$}},\ }\href {https://doi.org/10.1103/PhysRevLett.93.077001} {\bibfield  {journal} {\bibinfo  {journal} {Phys. Rev. Lett.}\ }\textbf {\bibinfo {volume} {93}},\ \bibinfo {pages} {077001} (\bibinfo {year} {2004})}\BibitemShut {NoStop}%
\bibitem [{\citenamefont {Kuroki}\ \emph {et~al.}(2005)\citenamefont {Kuroki}, \citenamefont {Tanaka},\ and\ \citenamefont {Arita}}]{Kuroki2005Jan}%
  \BibitemOpen
  \bibfield  {author} {\bibinfo {author} {\bibfnamefont {K.}~\bibnamefont {Kuroki}}, \bibinfo {author} {\bibfnamefont {Y.}~\bibnamefont {Tanaka}},\ and\ \bibinfo {author} {\bibfnamefont {R.}~\bibnamefont {Arita}},\ }\bibfield  {title} {\bibinfo {title} {{Competition between singlet and triplet pairings in ${\mathrm{Na}}_{x}{\mathrm{CoO}}_{2}\ifmmode\cdot\else\textperiodcentered\fi{}y\phantom{\rule{0.3em}{0ex}}{\mathrm{H}}_{2}\mathrm{O}$}},\ }\href {https://doi.org/10.1103/PhysRevB.71.024506} {\bibfield  {journal} {\bibinfo  {journal} {Phys. Rev. B}\ }\textbf {\bibinfo {volume} {71}},\ \bibinfo {pages} {024506} (\bibinfo {year} {2005})}\BibitemShut {NoStop}%
\bibitem [{\citenamefont {R{\o}mer}\ \emph {et~al.}(2021)\citenamefont {R{\o}mer}, \citenamefont {Hirschfeld},\ and\ \citenamefont {Andersen}}]{Romer2021SROspinfluctuations}%
  \BibitemOpen
  \bibfield  {author} {\bibinfo {author} {\bibfnamefont {A.~T.}\ \bibnamefont {R{\o}mer}}, \bibinfo {author} {\bibfnamefont {P.~J.}\ \bibnamefont {Hirschfeld}},\ and\ \bibinfo {author} {\bibfnamefont {B.~M.}\ \bibnamefont {Andersen}},\ }\bibfield  {title} {\bibinfo {title} {{Superconducting state of ${\mathrm{Sr}}_{2}{\mathrm{RuO}}_{4}$ in the presence of longer-range Coulomb interactions}},\ }\href {https://doi.org/10.1103/PhysRevB.104.064507} {\bibfield  {journal} {\bibinfo  {journal} {Phys. Rev. B}\ }\textbf {\bibinfo {volume} {104}},\ \bibinfo {pages} {064507} (\bibinfo {year} {2021})}\BibitemShut {NoStop}%
\bibitem [{\citenamefont {Schnell}\ \emph {et~al.}(2006)\citenamefont {Schnell}, \citenamefont {Mazin},\ and\ \citenamefont {Liu}}]{Schnell2006dwavephonon}%
  \BibitemOpen
  \bibfield  {author} {\bibinfo {author} {\bibfnamefont {I.}~\bibnamefont {Schnell}}, \bibinfo {author} {\bibfnamefont {I.~I.}\ \bibnamefont {Mazin}},\ and\ \bibinfo {author} {\bibfnamefont {A.~Y.}\ \bibnamefont {Liu}},\ }\bibfield  {title} {\bibinfo {title} {{Unconventional superconducting pairing symmetry induced by phonons}},\ }\href {https://doi.org/10.1103/PhysRevB.74.184503} {\bibfield  {journal} {\bibinfo  {journal} {Phys. Rev. B}\ }\textbf {\bibinfo {volume} {74}},\ \bibinfo {pages} {184503} (\bibinfo {year} {2006})}\BibitemShut {NoStop}%
\bibitem [{\citenamefont {Scheurer}(2016)}]{Scheurer2016NCSTSC}%
  \BibitemOpen
  \bibfield  {author} {\bibinfo {author} {\bibfnamefont {M.~S.}\ \bibnamefont {Scheurer}},\ }\bibfield  {title} {\bibinfo {title} {{Mechanism, time-reversal symmetry, and topology of superconductivity in noncentrosymmetric systems}},\ }\href {https://doi.org/10.1103/PhysRevB.93.174509} {\bibfield  {journal} {\bibinfo  {journal} {Phys. Rev. B}\ }\textbf {\bibinfo {volume} {93}},\ \bibinfo {pages} {174509} (\bibinfo {year} {2016})}\BibitemShut {NoStop}%
\bibitem [{Sup()}]{Suppl}%
  \BibitemOpen
  \href@noop {} {}\bibinfo {note} {See Supplemental Material on page \pageref{sec:Suppl} for further details of the model, the case where surface state bandwidth is larger than phonon bandwidth, solutions of the gap equation without FS average, and the gaps in the spin basis, which includes Refs.~\cite{Bardeen1955Pines, Schrieffer1966Wolff, VinasBostrom2024May, AaseMaeland2023Dec, Protter2021funkintFE, Brekke2023Dec, Maeland2024Thesis, AdaptQuad, Maeland2022Aug, Thingstad2021, DiSante2023CoulombRange, Linder2010dwaveTI, Hutchinson2020Nov, Benestad}.}\BibitemShut {Stop}%
\bibitem [{\citenamefont {Georges}\ \emph {et~al.}(2013)\citenamefont {Georges}, \citenamefont {Medici},\ and\ \citenamefont {Mravlje}}]{Georges2013Hund}%
  \BibitemOpen
  \bibfield  {author} {\bibinfo {author} {\bibfnamefont {A.}~\bibnamefont {Georges}}, \bibinfo {author} {\bibfnamefont {L.~d.}\ \bibnamefont {Medici}},\ and\ \bibinfo {author} {\bibfnamefont {J.}~\bibnamefont {Mravlje}},\ }\bibfield  {title} {\bibinfo {title} {{Strong Correlations from Hund{'}s Coupling}},\ }\href {https://doi.org/10.1146/annurev-conmatphys-020911-125045} {\bibfield  {journal} {\bibinfo  {journal} {Annu. Rev. Condens. Matter Phys.}\ }\textbf {\bibinfo {volume} {4}},\ \bibinfo {pages} {137} (\bibinfo {year} {2013})}\BibitemShut {NoStop}%
\bibitem [{\citenamefont {Budich}\ \emph {et~al.}(2013)\citenamefont {Budich}, \citenamefont {Trauzettel},\ and\ \citenamefont {Sangiovanni}}]{Budich2013Hund}%
  \BibitemOpen
  \bibfield  {author} {\bibinfo {author} {\bibfnamefont {J.~C.}\ \bibnamefont {Budich}}, \bibinfo {author} {\bibfnamefont {B.}~\bibnamefont {Trauzettel}},\ and\ \bibinfo {author} {\bibfnamefont {G.}~\bibnamefont {Sangiovanni}},\ }\bibfield  {title} {\bibinfo {title} {{Fluctuation-driven topological Hund insulators}},\ }\href {https://doi.org/10.1103/PhysRevB.87.235104} {\bibfield  {journal} {\bibinfo  {journal} {Phys. Rev. B}\ }\textbf {\bibinfo {volume} {87}},\ \bibinfo {pages} {235104} (\bibinfo {year} {2013})}\BibitemShut {NoStop}%
\bibitem [{\citenamefont {Amaricci}\ \emph {et~al.}(2015)\citenamefont {Amaricci}, \citenamefont {Budich}, \citenamefont {Capone}, \citenamefont {Trauzettel},\ and\ \citenamefont {Sangiovanni}}]{Amaricci2015Hund}%
  \BibitemOpen
  \bibfield  {author} {\bibinfo {author} {\bibfnamefont {A.}~\bibnamefont {Amaricci}}, \bibinfo {author} {\bibfnamefont {J.~C.}\ \bibnamefont {Budich}}, \bibinfo {author} {\bibfnamefont {M.}~\bibnamefont {Capone}}, \bibinfo {author} {\bibfnamefont {B.}~\bibnamefont {Trauzettel}},\ and\ \bibinfo {author} {\bibfnamefont {G.}~\bibnamefont {Sangiovanni}},\ }\bibfield  {title} {\bibinfo {title} {{First-Order Character and Observable Signatures of Topological Quantum Phase Transitions}},\ }\href {https://doi.org/10.1103/PhysRevLett.114.185701} {\bibfield  {journal} {\bibinfo  {journal} {Phys. Rev. Lett.}\ }\textbf {\bibinfo {volume} {114}},\ \bibinfo {pages} {185701} (\bibinfo {year} {2015})}\BibitemShut {NoStop}%
\bibitem [{\citenamefont {Bruus}\ and\ \citenamefont {Flensberg}(2004)}]{BruusFlensberg}%
  \BibitemOpen
  \bibfield  {author} {\bibinfo {author} {\bibfnamefont {H.}~\bibnamefont {Bruus}}\ and\ \bibinfo {author} {\bibfnamefont {K.}~\bibnamefont {Flensberg}},\ }\href@noop {} {\emph {\bibinfo {title} {Many-Body Quantum Theory in Condensed Matter Physics: An Introduction}}}\ (\bibinfo  {publisher} {Oxford University Press},\ \bibinfo {address} {Oxford},\ \bibinfo {year} {2004})\BibitemShut {NoStop}%
\bibitem [{\citenamefont {Kl{\o}getvedt}(2023)}]{Klogetvedt2023}%
  \BibitemOpen
  \bibfield  {author} {\bibinfo {author} {\bibfnamefont {J.~N.}\ \bibnamefont {Kl{\o}getvedt}},\ }\href@noop {} {\bibinfo {title} {{Topological Magnon-Phonon Hybrid Excitations and Hall Effects in Two-Dimensional Ferromagnets}}},\ \bibinfo {howpublished} {Master thesis, Norwegian University of Science and Technology, \href{https://hdl.handle.net/11250/3097131}{https://hdl.handle.net/11250/3097131}} (\bibinfo {year} {2023})\BibitemShut {NoStop}%
\bibitem [{\citenamefont {Sylju{\aa}sen}(2024)}]{Syljuasen2024}%
  \BibitemOpen
  \bibfield  {author} {\bibinfo {author} {\bibfnamefont {E.}~\bibnamefont {Sylju{\aa}sen}},\ }\href@noop {} {\bibinfo {title} {{Transverse quantum transport in multiband Bose-systems}}},\ \bibinfo {howpublished} {Master thesis, Norwegian University of Science and Technology, \href{https://hdl.handle.net/11250/3155941}{https://hdl.handle.net/11250/3155941}} (\bibinfo {year} {2024})\BibitemShut {NoStop}%
\bibitem [{\citenamefont {Thingstad}\ \emph {et~al.}(2020)\citenamefont {Thingstad}, \citenamefont {Kamra}, \citenamefont {Wells},\ and\ \citenamefont {Sudb{\o}}}]{Thingstad2020Jun}%
  \BibitemOpen
  \bibfield  {author} {\bibinfo {author} {\bibfnamefont {E.}~\bibnamefont {Thingstad}}, \bibinfo {author} {\bibfnamefont {A.}~\bibnamefont {Kamra}}, \bibinfo {author} {\bibfnamefont {J.~W.}\ \bibnamefont {Wells}},\ and\ \bibinfo {author} {\bibfnamefont {A.}~\bibnamefont {Sudb{\o}}},\ }\bibfield  {title} {\bibinfo {title} {{Phonon-mediated superconductivity in doped monolayer materials}},\ }\href {https://doi.org/10.1103/PhysRevB.101.214513} {\bibfield  {journal} {\bibinfo  {journal} {Phys. Rev. B}\ }\textbf {\bibinfo {volume} {101}},\ \bibinfo {pages} {214513} (\bibinfo {year} {2020})}\BibitemShut {NoStop}%
\bibitem [{\citenamefont {Leraand}\ \emph {et~al.}(2025)\citenamefont {Leraand}, \citenamefont {M{\ae}land},\ and\ \citenamefont {Sudb{\o}}}]{Leraand2025Feb}%
  \BibitemOpen
  \bibfield  {author} {\bibinfo {author} {\bibfnamefont {K.}~\bibnamefont {Leraand}}, \bibinfo {author} {\bibfnamefont {K.}~\bibnamefont {M{\ae}land}},\ and\ \bibinfo {author} {\bibfnamefont {A.}~\bibnamefont {Sudb{\o}}},\ }\bibfield  {title} {\bibinfo {title} {{Phonon-mediated spin-polarized superconductivity in altermagnets}},\ }\href {https://doi.org/10.1103/g4dl-1ff2} {\bibfield  {journal} {\bibinfo  {journal} {Phys. Rev. B}\ }\textbf {\bibinfo {volume} {112}},\ \bibinfo {pages} {104510} (\bibinfo {year} {2025})}\BibitemShut {NoStop}%
\bibitem [{\citenamefont {Leraand}\ \emph {et~al.}(2026)\citenamefont {Leraand}, \citenamefont {M{\ae}land},\ and\ \citenamefont {Sudb{\o}}}]{Leraand2026Mar}%
  \BibitemOpen
  \bibfield  {author} {\bibinfo {author} {\bibfnamefont {K.}~\bibnamefont {Leraand}}, \bibinfo {author} {\bibfnamefont {K.}~\bibnamefont {M{\ae}land}},\ and\ \bibinfo {author} {\bibfnamefont {A.}~\bibnamefont {Sudb{\o}}},\ }\bibfield  {title} {\bibinfo {title} {{Spin-dependent quasiparticle lifetimes in altermagnets}},\ }\href {https://doi.org/10.1103/r3vm-m1m3} {\bibfield  {journal} {\bibinfo  {journal} {Phys. Rev. B}\ }\textbf {\bibinfo {volume} {113}},\ \bibinfo {pages} {115148} (\bibinfo {year} {2026})}\BibitemShut {NoStop}%
\bibitem [{\citenamefont {Lucas}(1968)}]{Lucas1968PhononOBC}%
  \BibitemOpen
  \bibfield  {author} {\bibinfo {author} {\bibfnamefont {A.~A.}\ \bibnamefont {Lucas}},\ }\bibfield  {title} {\bibinfo {title} {{Phonon Modes of an Ionic Crystal Slab}},\ }\href {https://doi.org/10.1063/1.1669588} {\bibfield  {journal} {\bibinfo  {journal} {J. Chem. Phys.}\ }\textbf {\bibinfo {volume} {48}},\ \bibinfo {pages} {3156} (\bibinfo {year} {1968})}\BibitemShut {NoStop}%
\bibitem [{\citenamefont {Benedek}\ \emph {et~al.}(2010)\citenamefont {Benedek}, \citenamefont {Bernasconi}, \citenamefont {Chis}, \citenamefont {Chulkov}, \citenamefont {Echenique}, \citenamefont {Hellsing},\ and\ \citenamefont {Toennies}}]{Benedek2010PhononOBC}%
  \BibitemOpen
  \bibfield  {author} {\bibinfo {author} {\bibfnamefont {G.}~\bibnamefont {Benedek}}, \bibinfo {author} {\bibfnamefont {M.}~\bibnamefont {Bernasconi}}, \bibinfo {author} {\bibfnamefont {V.}~\bibnamefont {Chis}}, \bibinfo {author} {\bibfnamefont {E.}~\bibnamefont {Chulkov}}, \bibinfo {author} {\bibfnamefont {P.~M.}\ \bibnamefont {Echenique}}, \bibinfo {author} {\bibfnamefont {B.}~\bibnamefont {Hellsing}},\ and\ \bibinfo {author} {\bibfnamefont {J.~P.}\ \bibnamefont {Toennies}},\ }\bibfield  {title} {\bibinfo {title} {{Theory of surface phonons at metal surfaces: recent advances}},\ }\href {https://doi.org/10.1088/0953-8984/22/8/084020} {\bibfield  {journal} {\bibinfo  {journal} {J. Phys.: Condens. Matter}\ }\textbf {\bibinfo {volume} {22}},\ \bibinfo {pages} {084020} (\bibinfo {year} {2010})}\BibitemShut {NoStop}%
\bibitem [{\citenamefont {Bardeen}\ \emph {et~al.}(1957)\citenamefont {Bardeen}, \citenamefont {Cooper},\ and\ \citenamefont {Schrieffer}}]{BCS}%
  \BibitemOpen
  \bibfield  {author} {\bibinfo {author} {\bibfnamefont {J.}~\bibnamefont {Bardeen}}, \bibinfo {author} {\bibfnamefont {L.~N.}\ \bibnamefont {Cooper}},\ and\ \bibinfo {author} {\bibfnamefont {J.~R.}\ \bibnamefont {Schrieffer}},\ }\bibfield  {title} {\bibinfo {title} {Theory of {Superconductivity}},\ }\href {https://doi.org/10.1103/PhysRev.108.1175} {\bibfield  {journal} {\bibinfo  {journal} {Phys. Rev.}\ }\textbf {\bibinfo {volume} {108}},\ \bibinfo {pages} {1175} (\bibinfo {year} {1957})}\BibitemShut {NoStop}%
\bibitem [{\citenamefont {Sigrist}\ and\ \citenamefont {Ueda}(1991)}]{Sigrist}%
  \BibitemOpen
  \bibfield  {author} {\bibinfo {author} {\bibfnamefont {M.}~\bibnamefont {Sigrist}}\ and\ \bibinfo {author} {\bibfnamefont {K.}~\bibnamefont {Ueda}},\ }\bibfield  {title} {\bibinfo {title} {Phenomenological theory of unconventional superconductivity},\ }\href {https://doi.org/10.1103/RevModPhys.63.239} {\bibfield  {journal} {\bibinfo  {journal} {Rev. Mod. Phys.}\ }\textbf {\bibinfo {volume} {63}},\ \bibinfo {pages} {239} (\bibinfo {year} {1991})}\BibitemShut {NoStop}%
\bibitem [{\citenamefont {Fossheim}\ and\ \citenamefont {Sudb{\o}}(2004)}]{SFsuperconductivity}%
  \BibitemOpen
  \bibfield  {author} {\bibinfo {author} {\bibfnamefont {K.}~\bibnamefont {Fossheim}}\ and\ \bibinfo {author} {\bibfnamefont {A.}~\bibnamefont {Sudb{\o}}},\ }\href@noop {} {\emph {\bibinfo {title} {Superconductivity: Physics and Applications}}}\ (\bibinfo  {publisher} {Wiley, Chichester, UK},\ \bibinfo {year} {2004})\BibitemShut {NoStop}%
\bibitem [{\citenamefont {Fu}\ and\ \citenamefont {Kane}(2008)}]{Fu2008Kane}%
  \BibitemOpen
  \bibfield  {author} {\bibinfo {author} {\bibfnamefont {L.}~\bibnamefont {Fu}}\ and\ \bibinfo {author} {\bibfnamefont {C.~L.}\ \bibnamefont {Kane}},\ }\bibfield  {title} {\bibinfo {title} {{Superconducting Proximity Effect and Majorana Fermions at the Surface of a Topological Insulator}},\ }\href {https://doi.org/10.1103/PhysRevLett.100.096407} {\bibfield  {journal} {\bibinfo  {journal} {Phys. Rev. Lett.}\ }\textbf {\bibinfo {volume} {100}},\ \bibinfo {pages} {096407} (\bibinfo {year} {2008})}\BibitemShut {NoStop}%
\bibitem [{\citenamefont {Fukui}\ and\ \citenamefont {Fujiwara}(2010)}]{Fukui2010FuKaneTopoStability}%
  \BibitemOpen
  \bibfield  {author} {\bibinfo {author} {\bibfnamefont {T.}~\bibnamefont {Fukui}}\ and\ \bibinfo {author} {\bibfnamefont {T.}~\bibnamefont {Fujiwara}},\ }\bibfield  {title} {\bibinfo {title} {{Topological Stability of Majorana Zero Modes in Superconductor{\textendash}Topological Insulator Systems}},\ }\href {https://doi.org/10.1143/JPSJ.79.033701} {\bibfield  {journal} {\bibinfo  {journal} {J. Phys. Soc. Jpn.}\ }\textbf {\bibinfo {volume} {79}},\ \bibinfo {pages} {033701} (\bibinfo {year} {2010})}\BibitemShut {NoStop}%
\bibitem [{\citenamefont {Venditti}\ \emph {et~al.}(2026)\citenamefont {Venditti}, \citenamefont {Berthod},\ and\ \citenamefont {Rademaker}}]{Venditti2026d+id}%
  \BibitemOpen
  \bibfield  {author} {\bibinfo {author} {\bibfnamefont {G.}~\bibnamefont {Venditti}}, \bibinfo {author} {\bibfnamefont {C.}~\bibnamefont {Berthod}},\ and\ \bibinfo {author} {\bibfnamefont {L.}~\bibnamefont {Rademaker}},\ }\bibfield  {title} {\bibinfo {title} {{Angular momentum of vortex-core Majorana zero modes}},\ }\href {https://doi.org/10.1103/fg3f-3stw} {\bibfield  {journal} {\bibinfo  {journal} {Phys. Rev. B}\ }\textbf {\bibinfo {volume} {113}},\ \bibinfo {pages} {014502} (\bibinfo {year} {2026})}\BibitemShut {NoStop}%
\bibitem [{\citenamefont {Andersen}\ \emph {et~al.}(2024)\citenamefont {Andersen}, \citenamefont {Kreisel},\ and\ \citenamefont {Hirschfeld}}]{Andersen2024TRSB}%
  \BibitemOpen
  \bibfield  {author} {\bibinfo {author} {\bibfnamefont {B.~M.}\ \bibnamefont {Andersen}}, \bibinfo {author} {\bibfnamefont {A.}~\bibnamefont {Kreisel}},\ and\ \bibinfo {author} {\bibfnamefont {P.~J.}\ \bibnamefont {Hirschfeld}},\ }\bibfield  {title} {\bibinfo {title} {{Spontaneous time-reversal symmetry breaking by disorder in superconductors}},\ }\href {https://doi.org/10.3389/fphy.2024.1353425} {\bibfield  {journal} {\bibinfo  {journal} {Front. Phys.}\ }\textbf {\bibinfo {volume} {12}},\ \bibinfo {pages} {1353425} (\bibinfo {year} {2024})}\BibitemShut {NoStop}%
\bibitem [{\citenamefont {Lapp}\ and\ \citenamefont {Timm}(2022)}]{Lapp2022nodalMajorana}%
  \BibitemOpen
  \bibfield  {author} {\bibinfo {author} {\bibfnamefont {C.~J.}\ \bibnamefont {Lapp}}\ and\ \bibinfo {author} {\bibfnamefont {C.}~\bibnamefont {Timm}},\ }\bibfield  {title} {\bibinfo {title} {{Majorana flat bands at structured surfaces of nodal noncentrosymmetric superconductors}},\ }\href {https://doi.org/10.1103/PhysRevB.105.184501} {\bibfield  {journal} {\bibinfo  {journal} {Phys. Rev. B}\ }\textbf {\bibinfo {volume} {105}},\ \bibinfo {pages} {184501} (\bibinfo {year} {2022})}\BibitemShut {NoStop}%
\bibitem [{\citenamefont {Lapp}\ \emph {et~al.}(2025)\citenamefont {Lapp}, \citenamefont {Link},\ and\ \citenamefont {Timm}}]{Lapp2025nodalMajorana}%
  \BibitemOpen
  \bibfield  {author} {\bibinfo {author} {\bibfnamefont {C.~J.}\ \bibnamefont {Lapp}}, \bibinfo {author} {\bibfnamefont {J.~M.}\ \bibnamefont {Link}},\ and\ \bibinfo {author} {\bibfnamefont {C.}~\bibnamefont {Timm}},\ }\bibfield  {title} {\bibinfo {title} {{Manipulation of Majorana wave packets at surfaces of nodal noncentrosymmetric superconductors}},\ }\href {https://doi.org/10.1103/mtc1-r12c} {\bibfield  {journal} {\bibinfo  {journal} {Phys. Rev. B}\ }\textbf {\bibinfo {volume} {112}},\ \bibinfo {pages} {094519} (\bibinfo {year} {2025})}\BibitemShut {NoStop}%
\bibitem [{\citenamefont {Steinke}\ \emph {et~al.}(2020)\citenamefont {Steinke}, \citenamefont {Wehling},\ and\ \citenamefont {R{\ifmmode\ddot{o}\else\"{o}\fi}sner}}]{Steinke2020CoulombEngineering}%
  \BibitemOpen
  \bibfield  {author} {\bibinfo {author} {\bibfnamefont {C.}~\bibnamefont {Steinke}}, \bibinfo {author} {\bibfnamefont {T.~O.}\ \bibnamefont {Wehling}},\ and\ \bibinfo {author} {\bibfnamefont {M.}~\bibnamefont {R{\ifmmode\ddot{o}\else\"{o}\fi}sner}},\ }\bibfield  {title} {\bibinfo {title} {{Coulomb-engineered heterojunctions and dynamical screening in transition metal dichalcogenide monolayers}},\ }\href {https://doi.org/10.1103/PhysRevB.102.115111} {\bibfield  {journal} {\bibinfo  {journal} {Phys. Rev. B}\ }\textbf {\bibinfo {volume} {102}},\ \bibinfo {pages} {115111} (\bibinfo {year} {2020})}\BibitemShut {NoStop}%
\bibitem [{\citenamefont {van Loon}\ \emph {et~al.}(2023)\citenamefont {van Loon}, \citenamefont {Sch{\ifmmode\ddot{u}\else\"{u}\fi}ler}, \citenamefont {Springer}, \citenamefont {Sangiovanni}, \citenamefont {Tomczak},\ and\ \citenamefont {Wehling}}]{vanLoon2023CoulombEngineering}%
  \BibitemOpen
  \bibfield  {author} {\bibinfo {author} {\bibfnamefont {E.~G. C.~P.}\ \bibnamefont {van Loon}}, \bibinfo {author} {\bibfnamefont {M.}~\bibnamefont {Sch{\ifmmode\ddot{u}\else\"{u}\fi}ler}}, \bibinfo {author} {\bibfnamefont {D.}~\bibnamefont {Springer}}, \bibinfo {author} {\bibfnamefont {G.}~\bibnamefont {Sangiovanni}}, \bibinfo {author} {\bibfnamefont {J.~M.}\ \bibnamefont {Tomczak}},\ and\ \bibinfo {author} {\bibfnamefont {T.~O.}\ \bibnamefont {Wehling}},\ }\bibfield  {title} {\bibinfo {title} {{Coulomb engineering of two-dimensional Mott materials}},\ }\href {https://doi.org/10.1038/s41699-023-00408-x} {\bibfield  {journal} {\bibinfo  {journal} {npj 2D Mater. Appl.}\ }\textbf {\bibinfo {volume} {7}},\ \bibinfo {pages} {47} (\bibinfo {year} {2023})}\BibitemShut {NoStop}%
\bibitem [{\citenamefont {Cappelluti}\ \emph {et~al.}(2023)\citenamefont {Cappelluti}, \citenamefont {Grimaldi},\ and\ \citenamefont {Pietronero}}]{Cappelluti2023nonadiabatic}%
  \BibitemOpen
  \bibfield  {author} {\bibinfo {author} {\bibfnamefont {E.}~\bibnamefont {Cappelluti}}, \bibinfo {author} {\bibfnamefont {C.}~\bibnamefont {Grimaldi}},\ and\ \bibinfo {author} {\bibfnamefont {L.}~\bibnamefont {Pietronero}},\ }\bibfield  {title} {\bibinfo {title} {{Electron{\textendash}phonon driven unconventional superconductivity: The role of small Fermi energies and of nonadiabatic processes}},\ }\href {https://doi.org/10.1016/j.physc.2023.1354343} {\bibfield  {journal} {\bibinfo  {journal} {Phys. C}\ }\textbf {\bibinfo {volume} {613}},\ \bibinfo {pages} {1354343} (\bibinfo {year} {2023})}\BibitemShut {NoStop}%
\bibitem [{\citenamefont {Migdal}(1958)}]{migdal1958interaction}%
  \BibitemOpen
  \bibfield  {author} {\bibinfo {author} {\bibfnamefont {A.~B.}\ \bibnamefont {Migdal}},\ }\bibfield  {title} {\bibinfo {title} {Interaction between electrons and lattice vibrations in a normal metal},\ }\href@noop {} {\bibfield  {journal} {\bibinfo  {journal} {Zh. Eksp. Teor. Fiz.}\ }\textbf {\bibinfo {volume} {34}},\ \bibinfo {pages} {1438} (\bibinfo {year} {1958})},\ \bibinfo {note} {[Sov. Phys. JETP \textbf{34}, 996 (1958)]}\BibitemShut {NoStop}%
\bibitem [{\citenamefont {Roy}\ \emph {et~al.}(2014)\citenamefont {Roy}, \citenamefont {Sau},\ and\ \citenamefont {Das~Sarma}}]{Roy2014Migdal2D}%
  \BibitemOpen
  \bibfield  {author} {\bibinfo {author} {\bibfnamefont {B.}~\bibnamefont {Roy}}, \bibinfo {author} {\bibfnamefont {J.~D.}\ \bibnamefont {Sau}},\ and\ \bibinfo {author} {\bibfnamefont {S.}~\bibnamefont {Das~Sarma}},\ }\bibfield  {title} {\bibinfo {title} {{Migdal's theorem and electron-phonon vertex corrections in Dirac materials}},\ }\href {https://doi.org/10.1103/PhysRevB.89.165119} {\bibfield  {journal} {\bibinfo  {journal} {Phys. Rev. B}\ }\textbf {\bibinfo {volume} {89}},\ \bibinfo {pages} {165119} (\bibinfo {year} {2014})}\BibitemShut {NoStop}%
\bibitem [{\citenamefont {Marsiglio}(2020)}]{Marsiglio2020ElishbergRev}%
  \BibitemOpen
  \bibfield  {author} {\bibinfo {author} {\bibfnamefont {F.}~\bibnamefont {Marsiglio}},\ }\bibfield  {title} {\bibinfo {title} {{Eliashberg theory: A short review}},\ }\href {https://doi.org/10.1016/j.aop.2020.168102} {\bibfield  {journal} {\bibinfo  {journal} {Ann. Phys.}\ }\textbf {\bibinfo {volume} {417}},\ \bibinfo {pages} {168102} (\bibinfo {year} {2020})}\BibitemShut {NoStop}%
\bibitem [{\citenamefont {Chubukov}\ \emph {et~al.}(2020)\citenamefont {Chubukov}, \citenamefont {Abanov}, \citenamefont {Esterlis},\ and\ \citenamefont {Kivelson}}]{Chubukov2020Eliashberg}%
  \BibitemOpen
  \bibfield  {author} {\bibinfo {author} {\bibfnamefont {A.~V.}\ \bibnamefont {Chubukov}}, \bibinfo {author} {\bibfnamefont {A.}~\bibnamefont {Abanov}}, \bibinfo {author} {\bibfnamefont {I.}~\bibnamefont {Esterlis}},\ and\ \bibinfo {author} {\bibfnamefont {S.~A.}\ \bibnamefont {Kivelson}},\ }\bibfield  {title} {\bibinfo {title} {{Eliashberg theory of phonon-mediated superconductivity {\ifmmode---\else\textemdash\fi} When it is valid and how it breaks down}},\ }\href {https://doi.org/10.1016/j.aop.2020.168190} {\bibfield  {journal} {\bibinfo  {journal} {Ann. Phys.}\ }\textbf {\bibinfo {volume} {417}},\ \bibinfo {pages} {168190} (\bibinfo {year} {2020})}\BibitemShut {NoStop}%
\bibitem [{\citenamefont {Schrodi}\ \emph {et~al.}(2020)\citenamefont {Schrodi}, \citenamefont {Oppeneer},\ and\ \citenamefont {Aperis}}]{Schrodi2020beyondMigdal}%
  \BibitemOpen
  \bibfield  {author} {\bibinfo {author} {\bibfnamefont {F.}~\bibnamefont {Schrodi}}, \bibinfo {author} {\bibfnamefont {P.~M.}\ \bibnamefont {Oppeneer}},\ and\ \bibinfo {author} {\bibfnamefont {A.}~\bibnamefont {Aperis}},\ }\bibfield  {title} {\bibinfo {title} {{Full-bandwidth Eliashberg theory of superconductivity beyond Migdal's approximation}},\ }\href {https://doi.org/10.1103/PhysRevB.102.024503} {\bibfield  {journal} {\bibinfo  {journal} {Phys. Rev. B}\ }\textbf {\bibinfo {volume} {102}},\ \bibinfo {pages} {024503} (\bibinfo {year} {2020})}\BibitemShut {NoStop}%
\bibitem [{\citenamefont {Miserev}\ \emph {et~al.}(2025)\citenamefont {Miserev}, \citenamefont {Hutchinson}, \citenamefont {Schoeller}, \citenamefont {Klinovaja},\ and\ \citenamefont {Loss}}]{Miserev2025range}%
  \BibitemOpen
  \bibfield  {author} {\bibinfo {author} {\bibfnamefont {D.}~\bibnamefont {Miserev}}, \bibinfo {author} {\bibfnamefont {J.}~\bibnamefont {Hutchinson}}, \bibinfo {author} {\bibfnamefont {H.}~\bibnamefont {Schoeller}}, \bibinfo {author} {\bibfnamefont {J.}~\bibnamefont {Klinovaja}},\ and\ \bibinfo {author} {\bibfnamefont {D.}~\bibnamefont {Loss}},\ }\bibfield  {title} {\bibinfo {title} {{High-Temperature Superconductivity from Finite-Range Attractive Interaction}},\ }\href {https://doi.org/10.1103/glch-3385} {\bibfield  {journal} {\bibinfo  {journal} {Phys. Rev. Lett.}\ }\textbf {\bibinfo {volume} {135}},\ \bibinfo {pages} {186502} (\bibinfo {year} {2025})}\BibitemShut {NoStop}%
\bibitem [{\citenamefont {Dsouza}\ \emph {et~al.}(2026)\citenamefont {Dsouza}, \citenamefont {Parthenios}, \citenamefont {Andersen},\ and\ \citenamefont {Christensen}}]{Dsouza2026KL}%
  \BibitemOpen
  \bibfield  {author} {\bibinfo {author} {\bibfnamefont {R.}~\bibnamefont {Dsouza}}, \bibinfo {author} {\bibfnamefont {N.}~\bibnamefont {Parthenios}}, \bibinfo {author} {\bibfnamefont {B.~M.}\ \bibnamefont {Andersen}},\ and\ \bibinfo {author} {\bibfnamefont {M.~H.}\ \bibnamefont {Christensen}},\ }\bibfield  {title} {\bibinfo {title} {{Kohn-Luttinger Superconductivity of Weyl Fermi Arcs in PtBi$_2$}},\ }\href {https://doi.org/10.48550/arXiv.2605.31501} {\bibfield  {journal} {\bibinfo  {journal} {arXiv:2605.31501}\ } (\bibinfo {year} {2026})}\BibitemShut {NoStop}%
\bibitem [{\citenamefont {Buccheri}\ \emph {et~al.}(2026)\citenamefont {Buccheri}, \citenamefont {de~Martino},\ and\ \citenamefont {Brink}}]{Buccheri2026ph}%
  \BibitemOpen
  \bibfield  {author} {\bibinfo {author} {\bibfnamefont {F.}~\bibnamefont {Buccheri}}, \bibinfo {author} {\bibfnamefont {A.}~\bibnamefont {de~Martino}},\ and\ \bibinfo {author} {\bibfnamefont {J.~v.~d.}\ \bibnamefont {Brink}},\ }\bibfield  {title} {\bibinfo {title} {{Phonon-driven nodal surface superconductivity of Fermi arcs}},\ }\href {https://doi.org/10.48550/arXiv.2606.02371} {\bibfield  {journal} {\bibinfo  {journal} {arXiv:2606.02371}\ } (\bibinfo {year} {2026})}\BibitemShut {NoStop}%
\bibitem [{\citenamefont {Bardeen}\ and\ \citenamefont {Pines}(1955)}]{Bardeen1955Pines}%
  \BibitemOpen
  \bibfield  {author} {\bibinfo {author} {\bibfnamefont {J.}~\bibnamefont {Bardeen}}\ and\ \bibinfo {author} {\bibfnamefont {D.}~\bibnamefont {Pines}},\ }\bibfield  {title} {\bibinfo {title} {{Electron-Phonon Interaction in Metals}},\ }\href {https://doi.org/10.1103/PhysRev.99.1140} {\bibfield  {journal} {\bibinfo  {journal} {Phys. Rev.}\ }\textbf {\bibinfo {volume} {99}},\ \bibinfo {pages} {1140} (\bibinfo {year} {1955})}\BibitemShut {NoStop}%
\bibitem [{\citenamefont {Schrieffer}\ and\ \citenamefont {Wolff}(1966)}]{Schrieffer1966Wolff}%
  \BibitemOpen
  \bibfield  {author} {\bibinfo {author} {\bibfnamefont {J.~R.}\ \bibnamefont {Schrieffer}}\ and\ \bibinfo {author} {\bibfnamefont {P.~A.}\ \bibnamefont {Wolff}},\ }\bibfield  {title} {\bibinfo {title} {{Relation between the Anderson and Kondo Hamiltonians}},\ }\href {https://doi.org/10.1103/PhysRev.149.491} {\bibfield  {journal} {\bibinfo  {journal} {Phys. Rev.}\ }\textbf {\bibinfo {volume} {149}},\ \bibinfo {pages} {491} (\bibinfo {year} {1966})}\BibitemShut {NoStop}%
\bibitem [{\citenamefont {Vi{\ifmmode\tilde{n}\else\~{n}\fi}as~Bostr{\ifmmode\ddot{o}\else\"{o}\fi}m}\ and\ \citenamefont {Vi{\ifmmode\tilde{n}\else\~{n}\fi}as~Bostr{\ifmmode\ddot{o}\else\"{o}\fi}m}(2024)}]{VinasBostrom2024May}%
  \BibitemOpen
  \bibfield  {author} {\bibinfo {author} {\bibfnamefont {F.}~\bibnamefont {Vi{\ifmmode\tilde{n}\else\~{n}\fi}as~Bostr{\ifmmode\ddot{o}\else\"{o}\fi}m}}\ and\ \bibinfo {author} {\bibfnamefont {E.}~\bibnamefont {Vi{\ifmmode\tilde{n}\else\~{n}\fi}as~Bostr{\ifmmode\ddot{o}\else\"{o}\fi}m}},\ }\bibfield  {title} {\bibinfo {title} {{Magnon-mediated topological superconductivity in a quantum wire}},\ }\href {https://doi.org/10.1103/PhysRevResearch.6.L022042} {\bibfield  {journal} {\bibinfo  {journal} {Phys. Rev. Res.}\ }\textbf {\bibinfo {volume} {6}},\ \bibinfo {pages} {L022042} (\bibinfo {year} {2024})}\BibitemShut {NoStop}%
\bibitem [{\citenamefont {Aase}\ \emph {et~al.}(2023)\citenamefont {Aase}, \citenamefont {M{\ae}land},\ and\ \citenamefont {Sudb{\o}}}]{AaseMaeland2023Dec}%
  \BibitemOpen
  \bibfield  {author} {\bibinfo {author} {\bibfnamefont {N.~H.}\ \bibnamefont {Aase}}, \bibinfo {author} {\bibfnamefont {K.}~\bibnamefont {M{\ae}land}},\ and\ \bibinfo {author} {\bibfnamefont {A.}~\bibnamefont {Sudb{\o}}},\ }\bibfield  {title} {\bibinfo {title} {{Multiband strong-coupling superconductors with spontaneously broken time-reversal symmetry}},\ }\href {https://doi.org/10.1103/PhysRevB.108.214508} {\bibfield  {journal} {\bibinfo  {journal} {Phys. Rev. B}\ }\textbf {\bibinfo {volume} {108}},\ \bibinfo {pages} {214508} (\bibinfo {year} {2023})}\BibitemShut {NoStop}%
\bibitem [{\citenamefont {Protter}\ \emph {et~al.}(2021)\citenamefont {Protter}, \citenamefont {Boyack},\ and\ \citenamefont {Marsiglio}}]{Protter2021funkintFE}%
  \BibitemOpen
  \bibfield  {author} {\bibinfo {author} {\bibfnamefont {M.}~\bibnamefont {Protter}}, \bibinfo {author} {\bibfnamefont {R.}~\bibnamefont {Boyack}},\ and\ \bibinfo {author} {\bibfnamefont {F.}~\bibnamefont {Marsiglio}},\ }\bibfield  {title} {\bibinfo {title} {{Functional-integral approach to Gaussian fluctuations in Eliashberg theory}},\ }\href {https://doi.org/10.1103/PhysRevB.104.014513} {\bibfield  {journal} {\bibinfo  {journal} {Phys. Rev. B}\ }\textbf {\bibinfo {volume} {104}},\ \bibinfo {pages} {014513} (\bibinfo {year} {2021})}\BibitemShut {NoStop}%
\bibitem [{\citenamefont {Brekke}\ \emph {et~al.}(2023)\citenamefont {Brekke}, \citenamefont {Brataas},\ and\ \citenamefont {Sudb{\o}}}]{Brekke2023Dec}%
  \BibitemOpen
  \bibfield  {author} {\bibinfo {author} {\bibfnamefont {B.}~\bibnamefont {Brekke}}, \bibinfo {author} {\bibfnamefont {A.}~\bibnamefont {Brataas}},\ and\ \bibinfo {author} {\bibfnamefont {A.}~\bibnamefont {Sudb{\o}}},\ }\bibfield  {title} {\bibinfo {title} {{Two-dimensional altermagnets: Superconductivity in a minimal microscopic model}},\ }\href {https://doi.org/10.1103/PhysRevB.108.224421} {\bibfield  {journal} {\bibinfo  {journal} {Phys. Rev. B}\ }\textbf {\bibinfo {volume} {108}},\ \bibinfo {pages} {224421} (\bibinfo {year} {2023})}\BibitemShut {NoStop}%
\bibitem [{\citenamefont {M{\ae}land}(2024)}]{Maeland2024Thesis}%
  \BibitemOpen
  \bibfield  {author} {\bibinfo {author} {\bibfnamefont {K.}~\bibnamefont {M{\ae}land}},\ }\emph {\bibinfo {title} {{Many-body effects and topology in magnets and superconductors}}},\ \href {https://hdl.handle.net/11250/3146878} {Ph.D. thesis},\ \bibinfo  {school} {NTNU}, \bibinfo {address} {Norway} (\bibinfo {year} {2024})\BibitemShut {NoStop}%
\bibitem [{\citenamefont {Mousavi}\ \emph {et~al.}(2012)\citenamefont {Mousavi}, \citenamefont {Pask},\ and\ \citenamefont {Sukumar}}]{AdaptQuad}%
  \BibitemOpen
  \bibfield  {author} {\bibinfo {author} {\bibfnamefont {S.~E.}\ \bibnamefont {Mousavi}}, \bibinfo {author} {\bibfnamefont {J.~E.}\ \bibnamefont {Pask}},\ and\ \bibinfo {author} {\bibfnamefont {N.}~\bibnamefont {Sukumar}},\ }\bibfield  {title} {\bibinfo {title} {{Efficient adaptive integration of functions with sharp gradients and cusps in n-dimensional parallelepipeds}},\ }\href {https://doi.org/10.1002/nme.4267} {\bibfield  {journal} {\bibinfo  {journal} {Int. J. Numer. Methods Eng.}\ }\textbf {\bibinfo {volume} {91}},\ \bibinfo {pages} {343} (\bibinfo {year} {2012})}\BibitemShut {NoStop}%
\bibitem [{\citenamefont {M{\ae}land}\ and\ \citenamefont {Sudb{\o}}(2022)}]{Maeland2022Aug}%
  \BibitemOpen
  \bibfield  {author} {\bibinfo {author} {\bibfnamefont {K.}~\bibnamefont {M{\ae}land}}\ and\ \bibinfo {author} {\bibfnamefont {A.}~\bibnamefont {Sudb{\o}}},\ }\bibfield  {title} {\bibinfo {title} {{Quantum topological phase transitions in skyrmion crystals}},\ }\href {https://doi.org/10.1103/PhysRevResearch.4.L032025} {\bibfield  {journal} {\bibinfo  {journal} {Phys. Rev. Res.}\ }\textbf {\bibinfo {volume} {4}},\ \bibinfo {pages} {L032025} (\bibinfo {year} {2022})}\BibitemShut {NoStop}%
\bibitem [{\citenamefont {Thingstad}(2021)}]{Thingstad2021}%
  \BibitemOpen
  \bibfield  {author} {\bibinfo {author} {\bibfnamefont {E.}~\bibnamefont {Thingstad}},\ }\emph {\bibinfo {title} {{Collective effects in low-dimensional systems with coupled quasiparticles}}},\ \href {https://hdl.handle.net/11250/2823203} {Ph.D. thesis},\ \bibinfo  {school} {NTNU}, \bibinfo {address} {Norway} (\bibinfo {year} {2021})\BibitemShut {NoStop}%
\bibitem [{\citenamefont {Di~Sante}\ \emph {et~al.}(2023)\citenamefont {Di~Sante}, \citenamefont {Kim}, \citenamefont {Hanke}, \citenamefont {Wehling}, \citenamefont {Franchini}, \citenamefont {Thomale},\ and\ \citenamefont {Sangiovanni}}]{DiSante2023CoulombRange}%
  \BibitemOpen
  \bibfield  {author} {\bibinfo {author} {\bibfnamefont {D.}~\bibnamefont {Di~Sante}}, \bibinfo {author} {\bibfnamefont {B.}~\bibnamefont {Kim}}, \bibinfo {author} {\bibfnamefont {W.}~\bibnamefont {Hanke}}, \bibinfo {author} {\bibfnamefont {T.}~\bibnamefont {Wehling}}, \bibinfo {author} {\bibfnamefont {C.}~\bibnamefont {Franchini}}, \bibinfo {author} {\bibfnamefont {R.}~\bibnamefont {Thomale}},\ and\ \bibinfo {author} {\bibfnamefont {G.}~\bibnamefont {Sangiovanni}},\ }\bibfield  {title} {\bibinfo {title} {{Electronic correlations and universal long-range scaling in kagome metals}},\ }\href {https://doi.org/10.1103/PhysRevResearch.5.L012008} {\bibfield  {journal} {\bibinfo  {journal} {Phys. Rev. Res.}\ }\textbf {\bibinfo {volume} {5}},\ \bibinfo {pages} {L012008} (\bibinfo {year} {2023})}\BibitemShut {NoStop}%
\bibitem [{\citenamefont {Linder}\ \emph {et~al.}(2010)\citenamefont {Linder}, \citenamefont {Tanaka}, \citenamefont {Yokoyama}, \citenamefont {Sudb{\o}},\ and\ \citenamefont {Nagaosa}}]{Linder2010dwaveTI}%
  \BibitemOpen
  \bibfield  {author} {\bibinfo {author} {\bibfnamefont {J.}~\bibnamefont {Linder}}, \bibinfo {author} {\bibfnamefont {Y.}~\bibnamefont {Tanaka}}, \bibinfo {author} {\bibfnamefont {T.}~\bibnamefont {Yokoyama}}, \bibinfo {author} {\bibfnamefont {A.}~\bibnamefont {Sudb{\o}}},\ and\ \bibinfo {author} {\bibfnamefont {N.}~\bibnamefont {Nagaosa}},\ }\bibfield  {title} {\bibinfo {title} {{Unconventional Superconductivity on a Topological Insulator}},\ }\href {https://doi.org/10.1103/PhysRevLett.104.067001} {\bibfield  {journal} {\bibinfo  {journal} {Phys. Rev. Lett.}\ }\textbf {\bibinfo {volume} {104}},\ \bibinfo {pages} {067001} (\bibinfo {year} {2010})}\BibitemShut {NoStop}%
\bibitem [{\citenamefont {Hutchinson}\ and\ \citenamefont {Marsiglio}(2020)}]{Hutchinson2020Nov}%
  \BibitemOpen
  \bibfield  {author} {\bibinfo {author} {\bibfnamefont {J.}~\bibnamefont {Hutchinson}}\ and\ \bibinfo {author} {\bibfnamefont {F.}~\bibnamefont {Marsiglio}},\ }\bibfield  {title} {\bibinfo {title} {{Mixed temperature-dependent order parameters in the extended Hubbard model}},\ }\href {https://doi.org/10.1088/1361-648X/abc801} {\bibfield  {journal} {\bibinfo  {journal} {J. Phys.: Condens. Matter}\ }\textbf {\bibinfo {volume} {33}},\ \bibinfo {pages} {065603} (\bibinfo {year} {2020})}\BibitemShut {NoStop}%
\bibitem [{\citenamefont {Benestad}(2022)}]{Benestad}%
  \BibitemOpen
  \bibfield  {author} {\bibinfo {author} {\bibfnamefont {J.}~\bibnamefont {Benestad}},\ }\href@noop {} {\bibinfo {title} {Electron-magnon coupling and magnon-induced superconductivity in hybrid structures of metals and magnets with non-collinear ground states}},\ \bibinfo {howpublished} {Master thesis, Norwegian University of Science and Technology, \href{https://hdl.handle.net/11250/3015244}{https://hdl.handle.net/11250/3015244}} (\bibinfo {year} {2022})\BibitemShut {NoStop}%
\bibitem [{\citenamefont {Waje}\ \emph {et~al.}(2025)\citenamefont {Waje}, \citenamefont {Jakubczyk}, \citenamefont {van~den Brink},\ and\ \citenamefont {Timm}}]{Waje2025PtBi2GL}%
  \BibitemOpen
  \bibfield  {author} {\bibinfo {author} {\bibfnamefont {H.}~\bibnamefont {Waje}}, \bibinfo {author} {\bibfnamefont {F.}~\bibnamefont {Jakubczyk}}, \bibinfo {author} {\bibfnamefont {J.}~\bibnamefont {van~den Brink}},\ and\ \bibinfo {author} {\bibfnamefont {C.}~\bibnamefont {Timm}},\ }\bibfield  {title} {\bibinfo {title} {{Ginzburg-Landau theory for unconventional surface superconductivity in ${\mathrm{PtBi}}_{2}$}},\ }\href {https://doi.org/10.1103/kkqg-ntcz} {\bibfield  {journal} {\bibinfo  {journal} {Phys. Rev. B}\ }\textbf {\bibinfo {volume} {112}},\ \bibinfo {pages} {144519} (\bibinfo {year} {2025})}\BibitemShut {NoStop}%
\bibitem [{\citenamefont {Erlandsen}\ \emph {et~al.}(2019)\citenamefont {Erlandsen}, \citenamefont {Kamra}, \citenamefont {Brataas},\ and\ \citenamefont {Sudb{\o}}}]{EirikAFMNM}%
  \BibitemOpen
  \bibfield  {author} {\bibinfo {author} {\bibfnamefont {E.}~\bibnamefont {Erlandsen}}, \bibinfo {author} {\bibfnamefont {A.}~\bibnamefont {Kamra}}, \bibinfo {author} {\bibfnamefont {A.}~\bibnamefont {Brataas}},\ and\ \bibinfo {author} {\bibfnamefont {A.}~\bibnamefont {Sudb{\o}}},\ }\bibfield  {title} {\bibinfo {title} {{Enhancement of superconductivity mediated by antiferromagnetic squeezed magnons}},\ }\href {https://doi.org/10.1103/PhysRevB.100.100503} {\bibfield  {journal} {\bibinfo  {journal} {Phys. Rev. B}\ }\textbf {\bibinfo {volume} {100}},\ \bibinfo {pages} {100503} (\bibinfo {year} {2019})}\BibitemShut {NoStop}%
\bibitem [{\citenamefont {Thingstad}\ \emph {et~al.}(2021)\citenamefont {Thingstad}, \citenamefont {Erlandsen},\ and\ \citenamefont {Sudb{\o}}}]{Thingstad2021AFMNMAFM}%
  \BibitemOpen
  \bibfield  {author} {\bibinfo {author} {\bibfnamefont {E.}~\bibnamefont {Thingstad}}, \bibinfo {author} {\bibfnamefont {E.}~\bibnamefont {Erlandsen}},\ and\ \bibinfo {author} {\bibfnamefont {A.}~\bibnamefont {Sudb{\o}}},\ }\bibfield  {title} {\bibinfo {title} {{Eliashberg study of superconductivity induced by interfacial coupling to antiferromagnets}},\ }\href {https://doi.org/10.1103/PhysRevB.104.014508} {\bibfield  {journal} {\bibinfo  {journal} {Phys. Rev. B}\ }\textbf {\bibinfo {volume} {104}},\ \bibinfo {pages} {014508} (\bibinfo {year} {2021})}\BibitemShut {NoStop}%
\bibitem [{\citenamefont {M{\ae}land}\ and\ \citenamefont {Sudb{\o}}(2023{\natexlab{a}})}]{Maeland2023AprPRL}%
  \BibitemOpen
  \bibfield  {author} {\bibinfo {author} {\bibfnamefont {K.}~\bibnamefont {M{\ae}land}}\ and\ \bibinfo {author} {\bibfnamefont {A.}~\bibnamefont {Sudb{\o}}},\ }\bibfield  {title} {\bibinfo {title} {{Topological Superconductivity Mediated by Skyrmionic Magnons}},\ }\href {https://doi.org/10.1103/PhysRevLett.130.156002} {\bibfield  {journal} {\bibinfo  {journal} {Phys. Rev. Lett.}\ }\textbf {\bibinfo {volume} {130}},\ \bibinfo {pages} {156002} (\bibinfo {year} {2023}{\natexlab{a}})}\BibitemShut {NoStop}%
\bibitem [{\citenamefont {M{\ae}land}\ \emph {et~al.}(2023)\citenamefont {M{\ae}land}, \citenamefont {Abnar}, \citenamefont {Benestad},\ and\ \citenamefont {Sudb{\o}}}]{Maeland2023DecTSC}%
  \BibitemOpen
  \bibfield  {author} {\bibinfo {author} {\bibfnamefont {K.}~\bibnamefont {M{\ae}land}}, \bibinfo {author} {\bibfnamefont {S.}~\bibnamefont {Abnar}}, \bibinfo {author} {\bibfnamefont {J.}~\bibnamefont {Benestad}},\ and\ \bibinfo {author} {\bibfnamefont {A.}~\bibnamefont {Sudb{\o}}},\ }\bibfield  {title} {\bibinfo {title} {{Topological superconductivity mediated by magnons of helical magnetic states}},\ }\href {https://doi.org/10.1103/PhysRevB.108.224515} {\bibfield  {journal} {\bibinfo  {journal} {Phys. Rev. B}\ }\textbf {\bibinfo {volume} {108}},\ \bibinfo {pages} {224515} (\bibinfo {year} {2023})}\BibitemShut {NoStop}%
\bibitem [{\citenamefont {Sun}\ \emph {et~al.}(2023)\citenamefont {Sun}, \citenamefont {M{\ae}land},\ and\ \citenamefont {Sudb{\o}}}]{SunMaeland2023Aug}%
  \BibitemOpen
  \bibfield  {author} {\bibinfo {author} {\bibfnamefont {C.}~\bibnamefont {Sun}}, \bibinfo {author} {\bibfnamefont {K.}~\bibnamefont {M{\ae}land}},\ and\ \bibinfo {author} {\bibfnamefont {A.}~\bibnamefont {Sudb{\o}}},\ }\bibfield  {title} {\bibinfo {title} {{Stability of superconducting gap symmetries arising from antiferromagnetic magnons}},\ }\href {https://doi.org/10.1103/PhysRevB.108.054520} {\bibfield  {journal} {\bibinfo  {journal} {Phys. Rev. B}\ }\textbf {\bibinfo {volume} {108}},\ \bibinfo {pages} {054520} (\bibinfo {year} {2023})}\BibitemShut {NoStop}%
\bibitem [{\citenamefont {Sun}\ \emph {et~al.}(2024)\citenamefont {Sun}, \citenamefont {M{\ae}land}, \citenamefont {Thingstad},\ and\ \citenamefont {Sudb{\o}}}]{SunMaeland2024Mar}%
  \BibitemOpen
  \bibfield  {author} {\bibinfo {author} {\bibfnamefont {C.}~\bibnamefont {Sun}}, \bibinfo {author} {\bibfnamefont {K.}~\bibnamefont {M{\ae}land}}, \bibinfo {author} {\bibfnamefont {E.}~\bibnamefont {Thingstad}},\ and\ \bibinfo {author} {\bibfnamefont {A.}~\bibnamefont {Sudb{\o}}},\ }\bibfield  {title} {\bibinfo {title} {{Strong-coupling approach to temperature dependence of competing orders of superconductivity: Possible time-reversal symmetry breaking and nontrivial topology}},\ }\href {https://doi.org/10.1103/PhysRevB.109.174520} {\bibfield  {journal} {\bibinfo  {journal} {Phys. Rev. B}\ }\textbf {\bibinfo {volume} {109}},\ \bibinfo {pages} {174520} (\bibinfo {year} {2024})}\BibitemShut {NoStop}%
\bibitem [{\citenamefont {M{\ae}land}\ \emph {et~al.}(2024)\citenamefont {M{\ae}land}, \citenamefont {Brekke},\ and\ \citenamefont {Sudb{\o}}}]{Maeland2024Feb}%
  \BibitemOpen
  \bibfield  {author} {\bibinfo {author} {\bibfnamefont {K.}~\bibnamefont {M{\ae}land}}, \bibinfo {author} {\bibfnamefont {B.}~\bibnamefont {Brekke}},\ and\ \bibinfo {author} {\bibfnamefont {A.}~\bibnamefont {Sudb{\o}}},\ }\bibfield  {title} {\bibinfo {title} {{Many-body effects on superconductivity mediated by double-magnon processes in altermagnets}},\ }\href {https://doi.org/10.1103/PhysRevB.109.134515} {\bibfield  {journal} {\bibinfo  {journal} {Phys. Rev. B}\ }\textbf {\bibinfo {volume} {109}},\ \bibinfo {pages} {134515} (\bibinfo {year} {2024})}\BibitemShut {NoStop}%
\bibitem [{\citenamefont {Pfleiderer}(2009)}]{Pfleiderer2009HeavyFermion}%
  \BibitemOpen
  \bibfield  {author} {\bibinfo {author} {\bibfnamefont {C.}~\bibnamefont {Pfleiderer}},\ }\bibfield  {title} {\bibinfo {title} {{Superconducting phases of $f$-electron compounds}},\ }\href {https://doi.org/10.1103/RevModPhys.81.1551} {\bibfield  {journal} {\bibinfo  {journal} {Rev. Mod. Phys.}\ }\textbf {\bibinfo {volume} {81}},\ \bibinfo {pages} {1551} (\bibinfo {year} {2009})}\BibitemShut {NoStop}%
\bibitem [{\citenamefont {Nomani}\ and\ \citenamefont {Hosur}(2023)}]{Nomani2023FermiArcSCDOS}%
  \BibitemOpen
  \bibfield  {author} {\bibinfo {author} {\bibfnamefont {A.}~\bibnamefont {Nomani}}\ and\ \bibinfo {author} {\bibfnamefont {P.}~\bibnamefont {Hosur}},\ }\bibfield  {title} {\bibinfo {title} {{Intrinsic surface superconducting instability in type-I Weyl semimetals}},\ }\href {https://doi.org/10.1103/PhysRevB.108.165144} {\bibfield  {journal} {\bibinfo  {journal} {Phys. Rev. B}\ }\textbf {\bibinfo {volume} {108}},\ \bibinfo {pages} {165144} (\bibinfo {year} {2023})}\BibitemShut {NoStop}%
\bibitem [{\citenamefont {Bai}\ \emph {et~al.}(2025)\citenamefont {Bai}, \citenamefont {LiMing},\ and\ \citenamefont {Zhou}}]{Bai2025WSMSC}%
  \BibitemOpen
  \bibfield  {author} {\bibinfo {author} {\bibfnamefont {X.}~\bibnamefont {Bai}}, \bibinfo {author} {\bibfnamefont {W.}~\bibnamefont {LiMing}},\ and\ \bibinfo {author} {\bibfnamefont {T.}~\bibnamefont {Zhou}},\ }\bibfield  {title} {\bibinfo {title} {{Superconductivity in Weyl semimetals with time reversal symmetry}},\ }\href {https://doi.org/10.1088/1367-2630/ada574} {\bibfield  {journal} {\bibinfo  {journal} {New J. Phys.}\ }\textbf {\bibinfo {volume} {27}},\ \bibinfo {pages} {013003} (\bibinfo {year} {2025})}\BibitemShut {NoStop}%
\bibitem [{\citenamefont {Huang}\ \emph {et~al.}(2026)\citenamefont {Huang}, \citenamefont {Wang},\ and\ \citenamefont {Zhou}}]{Huang2025WeylIISC}%
  \BibitemOpen
  \bibfield  {author} {\bibinfo {author} {\bibfnamefont {J.}~\bibnamefont {Huang}}, \bibinfo {author} {\bibfnamefont {Z.~D.}\ \bibnamefont {Wang}},\ and\ \bibinfo {author} {\bibfnamefont {T.}~\bibnamefont {Zhou}},\ }\bibfield  {title} {\bibinfo {title} {{Higher-order topological superconductivity in type-II time-reversal symmetric Weyl semimetals with a hybrid pairing}},\ }\href {https://doi.org/10.1103/xy6g-q2f5} {\bibfield  {journal} {\bibinfo  {journal} {Phys. Rev. B}\ }\textbf {\bibinfo {volume} {113}},\ \bibinfo {pages} {054523} (\bibinfo {year} {2026})}\BibitemShut {NoStop}%
\bibitem [{\citenamefont {M{\ae}land}\ and\ \citenamefont {Sudb{\o}}(2023{\natexlab{b}})}]{Maeland2023DecCC}%
  \BibitemOpen
  \bibfield  {author} {\bibinfo {author} {\bibfnamefont {K.}~\bibnamefont {M{\ae}land}}\ and\ \bibinfo {author} {\bibfnamefont {A.}~\bibnamefont {Sudb{\o}}},\ }\bibfield  {title} {\bibinfo {title} {{Exceeding the Chandrasekhar-Clogston limit in flat-band superconductors: A multiband strong-coupling approach}},\ }\href {https://doi.org/10.1103/PhysRevB.108.214511} {\bibfield  {journal} {\bibinfo  {journal} {Phys. Rev. B}\ }\textbf {\bibinfo {volume} {108}},\ \bibinfo {pages} {214511} (\bibinfo {year} {2023}{\natexlab{b}})}\BibitemShut {NoStop}%
\end{thebibliography}%

%--------------End matter
\let\section\oldsection %reset section definition
\onecolumngrid
\section*{End Matter}
\twocolumngrid
\let\oldsection\section
\renewcommand{\section}[1]{\hspace*{1em}\textit{#1}---}
\section{Electron Hamiltonian}%
The part of the electron Hamiltonian describing hopping is
$
    H_{\text{hop}} = -\sum_{i\Bar{\boldsymbol{\delta}} \ell \ell' \sigma} t_{\ell\ell'}(\Bar{\boldsymbol{\delta}}) c_{i+\delta,\ell\sigma}^\dagger c_{i\ell'\sigma}.
$
We only include NN hopping $t_{\ell\ell'}(\Bar{\boldsymbol{\delta}})$, whose strength depends on orbital indices and the direction of the hopping. 
The eight NN vectors are
\begin{align*}
    &\Bar{\boldsymbol{\delta}}_1 = (\sqrt{3}/2, 1/2,0), \Bar{\boldsymbol{\delta}}_2 = (0,1,0), \Bar{\boldsymbol{\delta}}_3 = (-\sqrt{3}/2, 1/2,0), \\
    & \Bar{\boldsymbol{\delta}}_4 = (-\sqrt{3}/2, -1/2,0),  \Bar{\boldsymbol{\delta}}_5 = (0,-1,0),\\
    & \Bar{\boldsymbol{\delta}}_6 = (\sqrt{3}/2, -1/2,0), \Bar{\boldsymbol{\delta}}_7 = (0,0,1), \Bar{\boldsymbol{\delta}}_8 = (0,0,-1).
\end{align*}
For the hopping strengths, we set $t_{AA}(\Bar{\boldsymbol{\delta}}_{i=1,\dots,6}) = t/2, t_{AB}(\Bar{\boldsymbol{\delta}}_{i=1,3,5}) = t_o/2,$ $ t_{AB}(\Bar{\boldsymbol{\delta}}_{i=2,4,6}) = -t_o/2$, $t_{AA}(\Bar{\boldsymbol{\delta}}_{i = 7,8}) = -\beta/2,$ $t_{AB}(\Bar{\boldsymbol{\delta}}_7) = -\beta_o/2$, $t_{AB}(\Bar{\boldsymbol{\delta}}_8) = \beta_o/2$,  $t_{BB}(\Bar{\boldsymbol{\delta}}_i) = -t_{AA}(\Bar{\boldsymbol{\delta}}_i)$, and $t_{BA}(\Bar{\boldsymbol{\delta}}_{i}) = -t_{AB}(\Bar{\boldsymbol{\delta}}_{i})$. 
Note that $p$- and $d$-wave orbitals dominate in PtBi$_2$ \cite{Vocaturo2024PtBi2Effective}. While the orbitals are unspecified in the effective model, the sign of hopping terms indicate that one orbital has even and the other has odd parity.
 
We work in a slab geometry with periodic boundary conditions in the $x$ and $y$ directions and open boundary conditions in the $z$ direction. The partial FT is $c_{i\ell\sigma} = (1/\sqrt{N_L})\sum_{\boldsymbol{k}} c_{\boldsymbol{k} z_i \ell\sigma}e^{\mathrm{i}\boldsymbol{k}\cdot \boldsymbol{r}_i}$, where $\boldsymbol{r}_i = (x_i, y_i)$ is the in-plane position of lattice site $i$.
The SOC term is
\begin{align}
    H_{\text{SOC}} &= \sum_{\boldsymbol{k}z_i\ell} \bigg[ s_{\boldsymbol{k}} \Big( c_{\boldsymbol{k} z_i \ell \uparrow}^\dagger c_{\boldsymbol{k} z_i \bar{\ell}\downarrow} - \frac{1}{2}\sum_{\delta = \pm1}  c_{\boldsymbol{k} z_i \ell\uparrow}^\dagger c_{\boldsymbol{k}, z_i+\delta, \bar{\ell}\downarrow} \Big) \nonumber \\
    &+\text{H.c.}\bigg],
\end{align}
where $s_{\boldsymbol{k}} = \alpha [ \sin k_y + \cos(\sqrt{3}k_x/2)\sin(k_y/2) +\mathrm{i} \sqrt{3} \sin(\sqrt{3}k_x/2)\cos(k_y/2)]$ and H.c.\ indicates the Hermitian conjugate of the preceding term. The orbital index $\bar{\ell} = B(A)$ when $\ell = A(B)$.
The inversion symmetry breaking term in the Hamiltonian is $H_{\gamma} = \sum_{\boldsymbol{k} z_i \ell \sigma} \gamma c_{\boldsymbol{k} z_i \ell\sigma}^\dagger c_{\boldsymbol{k} z_i  \bar{\ell}\sigma}$.

\begin{figure}
    \centering
    \includegraphics[width=\linewidth]{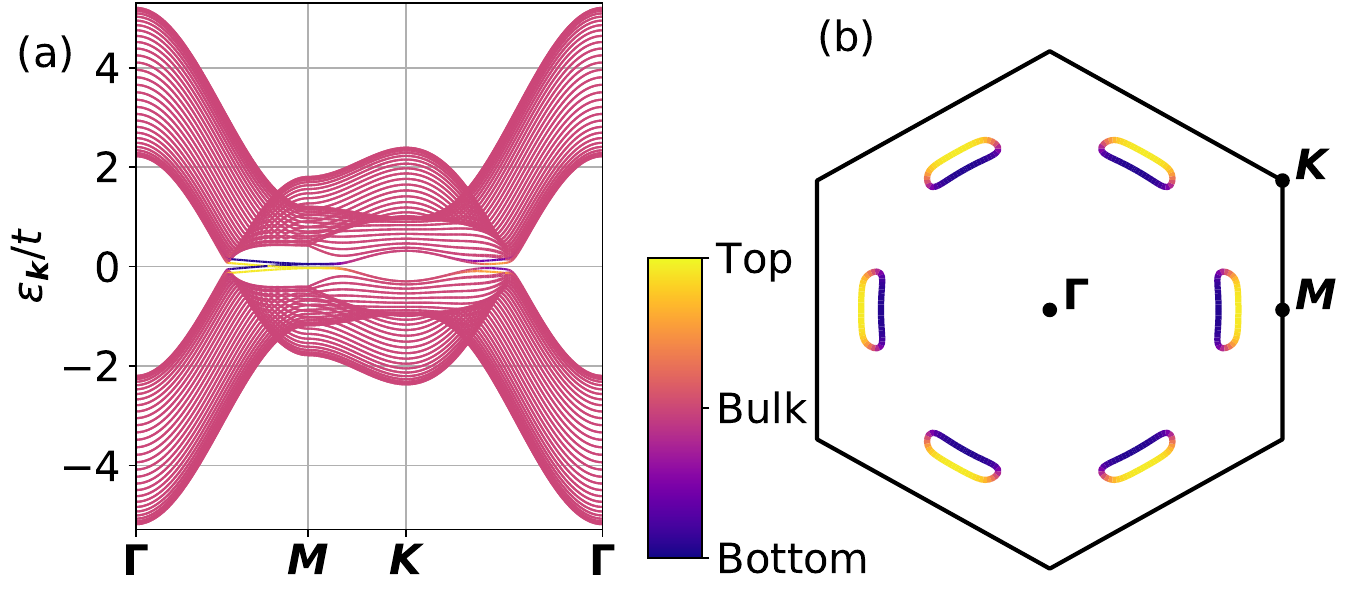}
    \caption{
    (a) Electron bands in slab geometry shown along a path between high symmetry points in the 1BZ. 
    The bands are colored by the location of their eigenstate $W_{\boldsymbol{k}n}$ along the $z$ direction, as indicated in the color bar. Panel (b) shows the full FS, also with colors denoting $W_{\boldsymbol{k}n}$. The parameters are the same as Fig.~\ref{fig:bands}. }
    \label{fig:bandsFS}
\end{figure}

Figure \ref{fig:bandsFS} shows the full band structure in the slab geometry along with the full FS. With nonzero $\mu$ we see separate Fermi arcs for top and bottom surfaces, six on each surface between the $\boldsymbol{\Gamma}$ and $\boldsymbol{M}$ points, which corresponds to measurements in PtBi$_2$. Additionally there are bulklike parts of the FS, which are the projection of the bulk FS. Parameter choices tune the location and length of the Fermi arc. While the model does not allow for matching the shape of the Fermi arc exactly to the ARPES measurements in PtBi$_2$ \cite{Kuibarov2024FermiArcSCNat, Changdar2025iwave}, we have confirmed that our main results are the same for many different parameter choices and correspondingly many different Fermi arc shapes and locations in the 1BZ. We can identify the five different gap symmetries in Fig.~\ref{fig:PD} with the symmetry analyses in Refs.~\cite{Changdar2025iwave, Waje2025PtBi2GL}. They belong to three different irreps of the P31m space group \cite{Suppl}. The $i$-wave state is the lowest order gap function in the $A_2$ irrep. This gap function is further special for the threefold rotation symmetric system as its nodes occur on high-symmetry lines $\boldsymbol{\Gamma M}$. Thus, the specific result that $i$-wave pairing ends up dominating the parameter space likely relies also on the symmetry of the model and material. In other Weyl semimetals with other symmetries, nodal gaps may appear for the same reasons discussed herein, but they are not necessarily $i$-wave.

\section{Coulomb Hamiltonian}%
Let us define the number operator $n_{i\ell\sigma} = c_{i\ell\sigma}^\dagger c_{i\ell\sigma}$ and the total number operator
$N_i = \sum_{\ell\sigma} c_{i\ell\sigma}^\dagger c_{i\ell\sigma}.$
For a two-orbital system, the on-site Hubbard repulsion follows Hund's rules \cite{Georges2013Hund, Budich2013Hund, Amaricci2015Hund}, 
\begin{align}
    H_{U} &=  (U-2J) \sum_{i\ell} n_{i\ell\uparrow}n_{i\bar{\ell}\downarrow} + (U-3J) \sum_{i\sigma} n_{iA\sigma}n_{iB\sigma}  \nonumber \\
    &+U\sum_{i\ell} n_{i\ell\uparrow}n_{i\ell\downarrow}+J\sum_{i\ell\ell'} c_{i\ell\uparrow}^\dagger c_{i\ell'\downarrow}^\dagger c_{i\bar{\ell}'\downarrow}c_{i\bar{\ell}\uparrow} . \label{eq:Hon-site}
\end{align}
The parameters $U\geq 0$ and $0 \leq J < U/3$ model the strength of the on-site repulsion. We set $J=0.2U$ throughout, but find that the main qualitative results do not care about the value of $J$.
We model the NN repulsion as
\begin{align}
    H_{V} &=  V\sum_{\langle i,j \rangle } N_i N_j = V\sum_{\langle i,j \rangle \ell \ell' \sigma\sigma'} c_{i\ell\sigma}^\dagger c_{j\ell'\sigma'}^\dagger c_{j\ell'\sigma'} c_{i\ell\sigma}.
\end{align}

\section{Phonon model}%
We employ a force constant approach to model the phonon modes in the hexagonal crystal used for the electron model. The full derivation is shown in Ref.~\cite{Maeland2025Jun}, here we repeat the main points.
We consider small displacements $\Bar{\boldsymbol{u}}_{i}(t)$ of ions away from the equilibrium positions $\Bar{\boldsymbol{R}}_{i}$, such that the instantaneous position is $\Bar{\boldsymbol{r}}_{i}(t) = \Bar{\boldsymbol{R}}_{i} +\Bar{\boldsymbol{u}}_{i}(t)$. From a Taylor expansion of the potential energy term $E_p$, the relevant part is the term
\begin{align}
    E_p &= \frac{1}{2}\sum_{i\mu, j \nu} \Phi_{\mu\nu}(\Bar{\boldsymbol{R}}_j-\Bar{\boldsymbol{R}}_i) u_{i\mu}  u_{j\nu}.
\end{align}
If ion $j$ moves in the $\nu$ direction, then $\Phi_{\mu\nu}(\Bar{\boldsymbol{R}}_j-\Bar{\boldsymbol{R}}_i)$ is the force constant on ion $i$ in the $\mu$ direction. By using the symmetries of the P31m space group, we can limit the number of free force constants. We also limit the description to NN and next-NN distances $\Bar{\boldsymbol{R}}_j-\Bar{\boldsymbol{R}}_i$. The force constants are summarized in Table I of Ref.~\cite{Maeland2025Jun}. For the NN force constants $\gamma_{i\in\{1,3,4,5,6,7\}}$ we choose $\gamma_7 = \gamma_6$. We choose all next-NN force constant parameters equal, i.e., $\rho_i = \rho = -\gamma_1/10\sqrt{2}$. These choices are made for simplicity, leaving us with five free parameters $\gamma_1, \gamma_3, \gamma_4, \gamma_5, $ and $\gamma_6$. 

Armed with the force coefficients, we can write down the dynamical matrix, which in the slab geometry is defined as
\begin{equation}
    D_{\mu\nu}^{z_i z_j}(\boldsymbol{q}) = \sum_{x_j y_j} \frac{1}{M}\Phi_{\mu\nu}(\Bar{\boldsymbol{R}}_j-\Bar{\boldsymbol{R}}_i)e^{\mathrm{i}\boldsymbol{q}\cdot (\boldsymbol{R}_{j}-\boldsymbol{R}_{i}) }.
\end{equation}
The ion mass $M$ is measured in units of inverse energy, $M = Ma^2/\hbar^2$. To choose a value, we use $a = 1$~{\AA} and the average atomic mass in PtBi$_2$. The $3L$ eigenvalues of the dynamical matrix are $\omega_{\boldsymbol{q}m}^2$ and their corresponding normalized eigenvectors are $\hat{e}_{\boldsymbol{q}m}$. 

After quantizing lattice displacements through
\begin{equation}
\label{eq:uphononOBC}
    u_{i\mu} = \sum_{\boldsymbol{q} m} \frac{1}{\sqrt{2N_L M \omega_{\boldsymbol{q}m}}} e_{\boldsymbol{q}m}^{z_i \mu} (a_{-\boldsymbol{q},m}^\dagger+a_{\boldsymbol{q}m})e^{\mathrm{i}\boldsymbol{q}\cdot \boldsymbol{R}_{i}},
\end{equation}
the phonon Hamiltonian in the slab geometry is
$
    H_{\text{ph}} = \sum_{\boldsymbol{q} m} \omega_{\boldsymbol{q}m} a_{\boldsymbol{q}m}^\dagger a_{\boldsymbol{q}m},
$
where $\omega_{\boldsymbol{q}m}$ is the phonon energy in mode $m$.

Since the effective model has a single atom in the unit cell we get three acoustic phonon modes in the bulk. PtBi$_2$ has 9 atoms in the basis, so it has an additional 24 optical modes.  
We suggest that the 3 acoustic modes provide an effective model of the pairing due to all modes in the material. Thus, we do not match our acoustic modes exactly to the acoustic modes in PtBi$_2$, but rather take the maximum energy of the acoustic modes to correspond to the maximum energy of phonons in PtBi$_2$. The higher number of phonon modes in the material could indicate an overall stronger EPC, which we then capture in the effective model by increasing $\chi$. 

\section{Electron-phonon coupling derivation}%
By Taylor expanding the hopping terms around small ionic displacements we get an EPC
\begin{align}
\label{eq:EPC}
    H_{\text{EPC}} &= -\sum_{i \Bar{\boldsymbol{\delta}} \ell \ell' \sigma} (\Bar{\boldsymbol{u}}_{i+\delta}-\Bar{\boldsymbol{u}}_i) \cdot \nabla_{\Bar{\boldsymbol{\delta}}}t_{\ell\ell'}(\Bar{\boldsymbol{\delta}}) c_{i+\delta,\ell\sigma}^\dagger c_{i\ell' \sigma}.
\end{align}
We model the derivative of the hopping term as $\nabla_{\Bar{\boldsymbol{\delta}}}t_{\ell\ell'}(\Bar{\boldsymbol{\delta}}) = -\chi \Bar{\boldsymbol{\delta}} t_{\ell\ell'}(\Bar{\boldsymbol{\delta}})$ \cite{Thingstad2020Jun}. If the orbitals are modeled as Gaussians with standard deviations $s$, we have $\chi \sim 1/s^2$ \cite{Leraand2025Feb}. 
To get from Eq.~\eqref{eq:EPC} to Eq.~\eqref{eq:gEPC}, we quantize the lattice displacements and do a partial FT.

As explained in the main text, EPC originating with out-of-plane hopping does not behave according to the usual EPC form known from the jellium model, and is instead maximal for small momentum transfer. In other words, it results in a mostly long-range electron-electron interaction similar to magnon-mediated electron-electron interactions \cite{EirikAFMNM, Thingstad2021AFMNMAFM, Maeland2023AprPRL, Maeland2023DecTSC, SunMaeland2023Aug, SunMaeland2024Mar, Maeland2024Feb}. That is unexpected for phonons and is related to the slab geometry with electronic surface states. The sum over the $3L$ phonon modes in the slab geometry loosely corresponds to a sum over all $q_z$ in the bulk. Hence, phonon-mediated interactions appear as long ranged in plane but would correspond to usual short-range pairing from phonons if we were in the bulk (with only bulk states). Thus, the existence of electronic surface states in the slab geometry is essential for the enhancement of EPC for small in-plane momentum transfers. 

While there are parameter sets where the phonon-mediated attraction shows a local minimum in the center of the Fermi arc, we find that the most common case is strongest attraction in the center of the arc. The attraction then decreases when moving out along the arc as the states become more bulklike. We expect the out-of-plane type EPC to provide the strongest contribution. Even in the center of the Fermi arc, the layer closest to the surface has a nonzero occupation. Hence, it appears having concentrated occupation on almost only two layers enhances the strength of the EPC. The phonon-mediated electron-electron interaction involves a sum over many complex terms. Thus, we conjecture that making only two layers provide the main contribution is best at avoiding cancellations between terms in the sum.  

\section{High-angular momentum pairing from phonons}%
In the context of heavy-fermion superconductors \cite{Pfleiderer2009HeavyFermion}, electron-phonon coupling is often ruled out as a mechanism because the electron bandwidth is comparable to the phonon bandwidth. Then, the quasiclassical picture of electrons avoiding each other in time, not space, to avoid Coulomb repulsion and attract via ion movement breaks down. Unlike heavy-fermion systems, strong correlations are not expected in PtBi$_2$ \cite{Changdar2025iwave}, so phonons are worth exploring in detail. We take the sum of Coulomb repulsion and phonon-mediated attraction directly. We then find a nontrivial solution of the gap equation with a nonnegligible $T_c$ due to the anisotropic phonon-mediated interaction. We interpret the result as electrons avoiding each other in space by choosing a pairing with higher angular momentum. That way, they avoid the Coulomb repulsion which is strongest on site and decreases with spatial separation. At the same time, the phonon-mediated interaction contains a part that is sufficiently long ranged to support the $i$-wave pairing. 

\section{Surface superconductivity}%
In this Letter, we focus on identifying a mechanism for nodal topological superconductiivty on PtBi$_2$ surfaces rather than addressing the question why surface superconductivity dominates over bulk superconductivity \cite{Nomani2023FermiArcSCDOS, Bai2025WSMSC, Huang2025WeylIISC, Trama2024TRSWeylSM_SC, Maeland2025Jun}. It is likely not purely a DOS argument, since PtBi$_2$ also has topologically trivial bands crossing the Fermi level in the bulk \cite{Vocaturo2024PtBi2Effective}. 
However, the surface bands 
contain flatter regions close
to the Fermi level in PtBi$_2$ \cite{Kuibarov2024FermiArcSCNat, Changdar2025iwave, Kuibarov2025ARPES, Kuibarov2025SepHighTc}, 
which should boost the surface superconductivity \cite{Maeland2023DecCC}.
Most likely, the dominant surface superconductivity is due to a combination of larger surface DOS and stronger coupling on the surface compared with the bulk. EPC is generally believed to be stronger on surfaces \cite{Benedek2010PhononOBC}.

\newpage
\let\section\oldsection %reset section definition
\setcounter{secnumdepth}{3} %enable section numbering, turned off by default in prl
\setcounter{section}{0} 
%change figure numbering to S1, S2 etc.
\renewcommand{\thefigure}{S\arabic{figure}}
\renewcommand{\theHfigure}{S\arabic{figure}}
\setcounter{figure}{0} 
%change table numbering to S1, S2 etc.
\renewcommand{\thetable}{S\Roman{table}}
\renewcommand{\theHtable}{S\Roman{table}}
\setcounter{table}{0} 
%change equation numbering to S1 etc.
\renewcommand{\theequation}{S\arabic{equation}}
\renewcommand{\theHequation}{S\arabic{equation}}
\setcounter{equation}{0}  
%-----Change to numerical section numbering------
%-----Note: not following revtex, Phys Rev format, but ok for Supplementary
%-----Including S gives section S1, S2 etc.
% Usual (decimal) numbering
\renewcommand{\thesection}{S\arabic{section}}
\renewcommand{\thesubsection}{\thesection.\arabic{subsection}}
\renewcommand{\thesubsubsection}{\thesubsection.\arabic{subsubsection}}
% Fix references to sections when changing section numbering
\makeatletter
\renewcommand{\p@subsection}{}
\renewcommand{\p@subsubsection}{}
\makeatother

\onecolumngrid

\phantomsection
\begin{large}
\begin{center}
    \textbf{Supplemental Material for ``Mechanism for Nodal Topological Superconductivity on PtBi$_2$ Surface''} \label{sec:Suppl}
\end{center}
\end{large}

\section{Introduction}
Here we give more details of the electronic model in Secs.~\ref{sec:Weylmodel} and \ref{sec:Coulomb}. We discuss the electron-electron interaction in Sec.~\ref{sec:eeint}, show solutions of the gap equation in the full first Brillouin zone in Sec.~\ref{sec:SCfull}, and study a Fermi surface (FS) average gap equation for the case where the electron bandwidth exceeds the maximum phonon energy in Sec.~\ref{sec:FSavg}. In Sec.~\ref{sec:bandwidthEqual}, we further explore the case of equal surface state electron bandwidth and maximum phonon energy. We also characterize the five different gap symmetries in more detail, including their symmetry classification and behavior in the original spin basis in Sec.~\ref{sec:gaps}.

\section{Electron model} \label{sec:Weylmodel}
The effective model is introduced in Ref.~\cite{Vocaturo2024PtBi2Effective} and the specific case of a slab geometry is explained in detail in Ref.~\cite{Maeland2025Jun}. For the sake of being self-contained we repeat the main points here. 
The bulk version is a four-band model of a Weyl semimetal with 12 Weyl nodes that split in pairs of opposite chirality away from the $\boldsymbol{\Gamma M}$ line. Fig.~11 in Ref.~\cite{Vocaturo2024PtBi2Effective} shows Weyl nodes and bulk bands for the same model.

We have a hexagonal crystal with a single-atomic basis and two orbitals per site. The lattice vectors are $\Bar{\boldsymbol{a}}_1 = (0,1,0)$, $\Bar{\boldsymbol{a}}_2 = (\sqrt{3}/2, -1/2, 0)$, and $\Bar{\boldsymbol{a}}_3 = (0,0,1)$, giving high-symmetry points $\boldsymbol{\Gamma} = (0,0)$, $\boldsymbol{M} = (2\pi/\sqrt{3}, 0)$, and $\boldsymbol{K} = (2\pi/\sqrt{3}, 2\pi/3)$ in 2D momentum space for the slab geometry.

Working in the slab, let us define
\begin{equation}
    f_{\boldsymbol{k}} = \mu_o -t\bqty{\cos (k_y) +2\cos(\frac{\sqrt{3}k_x}{2})\cos(\frac{k_y}{2})},
\end{equation}
\begin{equation}
    g_{\boldsymbol{k}} = t_o \bqty{\sin(k_y)-2 \cos(\frac{\sqrt{3}k_x}{2})\sin(\frac{k_y}{2})},
\end{equation}
and the vector $\boldsymbol{c}_{\boldsymbol{k}} = (c_{\boldsymbol{k},L,A\uparrow}, c_{\boldsymbol{k},L,A\downarrow}, c_{\boldsymbol{k},L,B\uparrow}, c_{\boldsymbol{k},L,B\downarrow}, $ $c_{\boldsymbol{k},L-1,A\uparrow}, c_{\boldsymbol{k},L-1,A\downarrow}, c_{\boldsymbol{k},L-1,B\uparrow}, c_{\boldsymbol{k},L-1,B\downarrow}, \dots, $ $ c_{\boldsymbol{k},1,A\uparrow}, c_{\boldsymbol{k},1,A\downarrow}, c_{\boldsymbol{k},1,B\uparrow}, c_{\boldsymbol{k},1,B\downarrow})^T$. Then, the normal state electron Hamiltonian is
\begin{equation}
    H_{\text{el}} = \sum_{\boldsymbol{k}} \boldsymbol{c}_{\boldsymbol{k}}^\dagger H(\boldsymbol{k}) \boldsymbol{c}_{\boldsymbol{k}}.
\end{equation}
The Hamiltonian matrix is block tridiagonal
\begin{equation}
    H(\boldsymbol{k}) = \begin{pmatrix}
        H_D(\boldsymbol{k}) & H_U(\boldsymbol{k}) & 0 & \cdots &   0 \\
        H_L(\boldsymbol{k}) & H_D(\boldsymbol{k}) & H_U(\boldsymbol{k}) & \ddots & \vdots \\
        0 & H_L(\boldsymbol{k}) & \ddots & \ddots & 0  \\
        \vdots & \ddots & \ddots & \ddots & H_U(\boldsymbol{k}) \\
        0 & \cdots & 0 & H_L(\boldsymbol{k}) & H_D(\boldsymbol{k})  
    \end{pmatrix},
\end{equation}
with diagonal block $H_D(\boldsymbol{k})$ and upper diagonal block $H_U(\boldsymbol{k})$ given by
\begin{equation}
    H_D(\boldsymbol{k}) = \begin{pmatrix}
        -\mu + f_{\boldsymbol{k}} & 0 & \gamma-\mathrm{i}g_{\boldsymbol{k}} & s_{\boldsymbol{k}}  \\
        0 & -\mu + f_{\boldsymbol{k}} & s_{\boldsymbol{k}}^* & \gamma-\mathrm{i}g_{\boldsymbol{k}}  \\
        \gamma+\mathrm{i}g_{\boldsymbol{k}} & s_{\boldsymbol{k}} & -\mu-f_{\boldsymbol{k}} & 0  \\
        s_{\boldsymbol{k}}^* & \gamma +\mathrm{i}g_{\boldsymbol{k}} & 0 & -\mu - f_{\boldsymbol{k}} 
    \end{pmatrix}, \quad H_U(\boldsymbol{k}) = \frac{1}{2}\begin{pmatrix}
         \beta & 0 & \beta_o & -s_{\boldsymbol{k}}  \\
         0 &\beta & -s_{\boldsymbol{k}}^* & \beta_o  \\
         -\beta_o & -s_{\boldsymbol{k}} & -\beta & 0  \\
         -s_{\boldsymbol{k}}^* & -\beta_o & 0 & -\beta 
    \end{pmatrix}.
\end{equation}
The lower diagonal block $H_L(\boldsymbol{k})$ is the hermitian conjugate of the upper diagonal block.

\section{Coulomb interaction} \label{sec:Coulomb}

We now perform a partial FT and a transform to the band basis of $H_C = H_U + H_V$ defined in the main text. Using $c_{\boldsymbol{k} z_i \ell\sigma} = \sum_n v_{\boldsymbol{k} n z_i \ell\sigma}d_{\boldsymbol{k}n}$, we can write
\begin{equation}
\label{eq:HCgeneral}
    H_C = \sum_{\boldsymbol{kk}'\boldsymbol{q}n_1 n_2 n_3 n_4} V_{\boldsymbol{k}+\boldsymbol{q}, \boldsymbol{k}'-\boldsymbol{q}, \boldsymbol{k}', \boldsymbol{k}}^{n_1 n_2 n_3 n_4} d_{\boldsymbol{k}+\boldsymbol{q},n_1}^\dagger d_{\boldsymbol{k}'-\boldsymbol{q},n_2}^\dagger d_{\boldsymbol{k}'n_3}d_{\boldsymbol{k}n_4},
\end{equation}
with
\begin{align}
    V_{\boldsymbol{k}+\boldsymbol{q}, \boldsymbol{k}'-\boldsymbol{q}, \boldsymbol{k}', \boldsymbol{k}}^{n_1 n_2 n_3 n_4} &= \frac{U}{N_L}\sum_{z_i\ell} v_{\boldsymbol{k}+\boldsymbol{q},n_1 z_i \ell \uparrow}^* v_{\boldsymbol{k}'-\boldsymbol{q},n_2 z_i \ell \downarrow}^* v_{\boldsymbol{k}'n_3z_i \ell \downarrow} v_{\boldsymbol{k}n_4 z_i\ell\uparrow} \nonumber \\
    &+ \frac{U-2J}{N_L} \sum_{z_i\ell} v_{\boldsymbol{k}+\boldsymbol{q},n_1 z_i \ell \uparrow}^* v_{\boldsymbol{k}'-\boldsymbol{q},n_2 z_i \bar{\ell} \downarrow}^* v_{\boldsymbol{k}' n_3 z_i \bar{\ell} \downarrow} v_{\boldsymbol{k}n_4 z_i\ell\uparrow}  \nonumber \\
    &+ \frac{U-3J}{N_L} \sum_{z_i\sigma} v_{\boldsymbol{k}+\boldsymbol{q},n_1 z_i A\sigma}^* v_{\boldsymbol{k}'-\boldsymbol{q},n_2 z_i B\sigma}^* v_{\boldsymbol{k}' n_3z_i B\sigma} v_{\boldsymbol{k} n_4 z_iA\sigma} \nonumber \\
    &+\frac{J}{N_L}\sum_{z_i\ell\ell'} v_{\boldsymbol{k}+\boldsymbol{q},n_1 z_i \ell \uparrow}^* v_{\boldsymbol{k}'-\boldsymbol{q},n_2 z_i \ell' \downarrow}^* v_{\boldsymbol{k}' n_3 z_i \bar{\ell}' \downarrow} v_{\boldsymbol{k} n_4 z_i\bar{\ell}\uparrow} \nonumber \\
    &+\frac{V}{N_L}\sum_{z_i, \ell \ell' \sigma\sigma'} \gamma(\boldsymbol{q})  v_{\boldsymbol{k}+\boldsymbol{q},n_1 z_i\ell\sigma}^* v_{\boldsymbol{k}'-\boldsymbol{q},n_2 z_i,\ell'\sigma'}^* v_{\boldsymbol{k}', n_3 z_i, \ell'\sigma'} v_{\boldsymbol{k}, n_4 z_i \ell\sigma} \nonumber \\
    &+ \frac{V}{N_L}\sum_{ z_i, \delta_z=\pm1, \ell \ell' \sigma\sigma'} v_{\boldsymbol{k}+\boldsymbol{q},n_1 z_i\ell\sigma}^* v_{\boldsymbol{k}'-\boldsymbol{q},n_2, z_i+\delta_z,\ell'\sigma'}^* v_{\boldsymbol{k}',n_3, z_i +\delta_z, \ell'\sigma'} v_{\boldsymbol{k}, n_4 z_i \ell\sigma} ,
\end{align}
where $\gamma(\boldsymbol{q}) = \sum_{\boldsymbol{\delta}} e^{\mathrm{i}\boldsymbol{q}\cdot \boldsymbol{\delta}} = 2\cos q_y + 4 \cos(\frac{\sqrt{3}q_x}{2})\cos(\frac{q_y}{2})$. The extra momentum dependence through $\gamma(\boldsymbol{q})$ for the in-plane parts of the NN repulsion makes the Coulomb repulsion more pronounced at small momentum transfers. Hence the repulsion is even stronger in the center of the Fermi arcs when compared to pure onsite repulsion. That is the momentum space understanding of why NN repulsion results in the nodal $i$-wave pairing. Note that the Coulomb interaction depends on all momentum indices in the band basis due to the transformation coefficients $v_{\boldsymbol{k} n z_i \ell\sigma}$.

\section{Electron-electron interaction} \label{sec:eeint}
A Schrieffer-Wolff (SW) transformation \cite{Bardeen1955Pines, Schrieffer1966Wolff} results in an effective electron-electron interaction mediated by phonons of the form
\begin{equation}
    V_{\boldsymbol{k}'\boldsymbol{k}}^{\text{ph, SW}} = \sum_m \frac{g_{\boldsymbol{k}'\boldsymbol{k}}^m g_{-\boldsymbol{k}',-\boldsymbol{k}}^m \omega_{\boldsymbol{k}'-\boldsymbol{k},m}}{(\epsilon_{\boldsymbol{k}}-\epsilon_{\boldsymbol{k}'})^2-\omega_{\boldsymbol{k}'-\boldsymbol{k},m}^2}.
\end{equation}
If $\epsilon_{\boldsymbol{k}}\neq \epsilon_{\boldsymbol{k}'}$ this interaction has many singularities which are not necessarily physical. The reason is that the Schrieffer-Wolff transformation treats the phonon-mediated interaction as instantaneous, while in fact it is retarded. Hence, a frequency description can be more appropriate. The electron-electron interaction can also be derived via functional integral methods by integrating out the phonons \cite{VinasBostrom2024May, AaseMaeland2023Dec, Protter2021funkintFE, Maeland2023DecCC}, or via Green's function techniques \cite{Marsiglio2020ElishbergRev, Maeland2023DecCC}. Both approaches show that the frequency resolved phonon-mediated interaction is
\begin{equation}
    V_{\boldsymbol{k}'\boldsymbol{k}}^{\text{ph}}(i\omega_\nu) = \frac{1}{2}\sum_m g_{\boldsymbol{k}'\boldsymbol{k}}^m g_{-\boldsymbol{k}',-\boldsymbol{k}}^m D_m(\boldsymbol{k}'-\boldsymbol{k}, i\omega_\nu) = \sum_m \frac{g_{\boldsymbol{k}'\boldsymbol{k}}^m g_{-\boldsymbol{k}',-\boldsymbol{k}}^m \omega_{\boldsymbol{k}'-\boldsymbol{k},m}}{(i\omega_\nu)^2-\omega_{\boldsymbol{k}'-\boldsymbol{k},m}^2},
\end{equation}
where $i\omega_\nu$ is a bosonic Matsubara frequency and $D_m(\boldsymbol{q}, i\omega_\nu)$ is the Green's function for phonon mode $m$.
An analytic continuation to real frequencies gives
\begin{equation}
    V_{\boldsymbol{k}'\boldsymbol{k}}^{\text{ph}}(\omega) = \sum_m \frac{g_{\boldsymbol{k}'\boldsymbol{k}}^m g_{-\boldsymbol{k}',-\boldsymbol{k}}^m \omega_{\boldsymbol{k}'-\boldsymbol{k},m}}{\omega^2-\omega_{\boldsymbol{k}'-\boldsymbol{k},m}^2}.
\end{equation}
Hence, the singularities occur in frequency space and are not necessarily on-shell. As an approximation to the frequency dependence we introduce a box potential, 
\begin{equation}
    V_{\boldsymbol{k}'\boldsymbol{k}}^{\text{ph}} = \begin{cases}
        V_{\boldsymbol{k}'\boldsymbol{k}}^{\text{ph}}(\omega = 0) = -\sum_m \frac{g_{\boldsymbol{k}'\boldsymbol{k}}^m g_{-\boldsymbol{k}',-\boldsymbol{k}}^m }{\omega_{\boldsymbol{k}'-\boldsymbol{k},m}}, \qquad \text{if } |\epsilon_{\boldsymbol{k}}|, |\epsilon_{\boldsymbol{k}'}| < \omega_D \\
        0, \qquad \text{otherwise}
    \end{cases}.
\end{equation}
The energy $\omega_D$ is the maximum phonon energy in the material. The box potential for the frequency is similar to the one used in BCS theory \cite{BCS, SFsuperconductivity}. Unlike the original BCS theory, we keep the momentum dependence of the phonon-mediated interaction. We expect this approximation gives good predictions of the momentum dependence of the gap along the FS, while the box potential for the frequency dependence yields an approximate result for the momentum dependence perpendicular to the FS. The momentum dependence parallel to the FS is our main focus. 

For acoustic phonons, $V_{\boldsymbol{k}'\boldsymbol{k}}^{\text{ph}}$ has a zero in the denominator at $\boldsymbol{k}=\boldsymbol{k}'$. The numerator also goes to zero at $\boldsymbol{k}=\boldsymbol{k}'$ for acoustic phonons \cite{Maeland2025Jun}, so we find the value of $V_{\boldsymbol{k}\boldsymbol{k}}^{\text{ph}}$ by limits. Comparing to the SW transformation result, we interpret the $\omega=0$ limit as $\epsilon_{\boldsymbol{k}}=\epsilon_{\boldsymbol{k}'}$. Therefore, we take the limit along constant energy contours,
\begin{equation}
    V_{\boldsymbol{k}\boldsymbol{k}}^{\text{ph}} = \frac{V_{\boldsymbol{k}+l_\parallel, \boldsymbol{k}}^{\text{ph}} + V_{\boldsymbol{k}-l_\parallel, \boldsymbol{k}}^{\text{ph}}}{2},
\end{equation}
where $l_\parallel$ is a short momentum displacement along a constant energy contour.

The contribution from the Coulomb interactions to the BCS-type electron-electron interaction term $H_{\text{BCS}}$, namely $V_{\boldsymbol{k}'\boldsymbol{k}}^{C}$, results from choosing the band index with a FS, setting $\boldsymbol{k}' = -\boldsymbol{k}$ and then redefining $\boldsymbol{k}+\boldsymbol{q} \to \boldsymbol{k}'$ in Eq.~\eqref{eq:HCgeneral}.

As discussed in Refs.~\cite{Scheurer2016NCSTSC, Maeland2025Jun}, the transformation into the band basis leads to a gauge dependence in the electron-electron interaction with no observable consequences. We set a global gauge of the eigenvectors $v_{\boldsymbol{k}, z_i \ell\sigma}$ by choosing the element related to orbital $\ell = B$, spin $\sigma = \downarrow$, and bottom layer $z_i = 1$ to be real and positive at all $\boldsymbol{k}$.

\begin{figure}
    \centering
    \includegraphics[width=\linewidth]{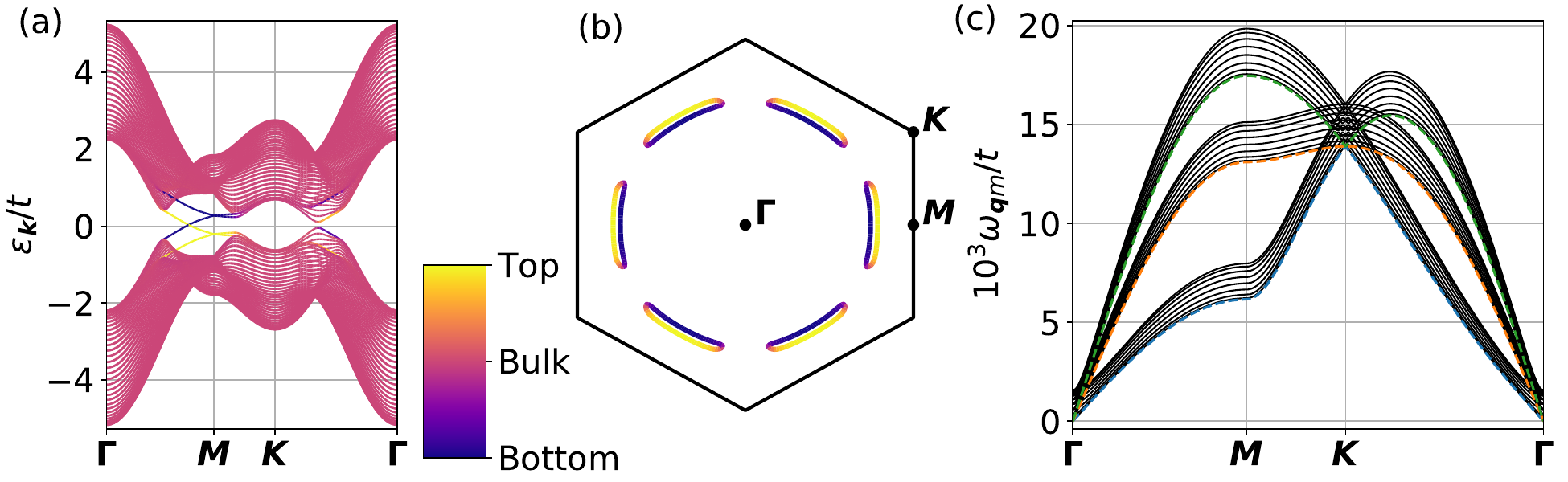}
    \caption{(a) Electron bands in slab geometry shown along a path between high symmetry points in the 1BZ. The bands are colored by the location of their eigenstate $W_{\boldsymbol{k}n}$ along the $z$-direction, as indicated in the colorbar. Panel (b) shows the full FS, also with colors denoting $W_{\boldsymbol{k}n}$. (c) The phonon spectrum along a path between high symmetry points in the 1BZ. The parameters are $t_o/t = 0.24$, $\beta/t = -1.5$, $\beta_o = -0.4$, $\mu/t = -0.03, \mu_o/t = -0.7, \alpha/t = -0.21$, $\gamma/t = -0.28$, $\gamma_1 = -(0.005t)^2$, $\gamma_3 = 0.448\gamma_1$, $\gamma_4 = \gamma_1$, $\gamma_5 = 0.45\gamma_1$, $\gamma_6 = 1.3 \gamma_1$, (a,b) $L=30$, and (c) $L=8$. }
    \label{fig:bands9}
\end{figure}

\begin{figure}
    \centering
    \includegraphics[width=\linewidth]{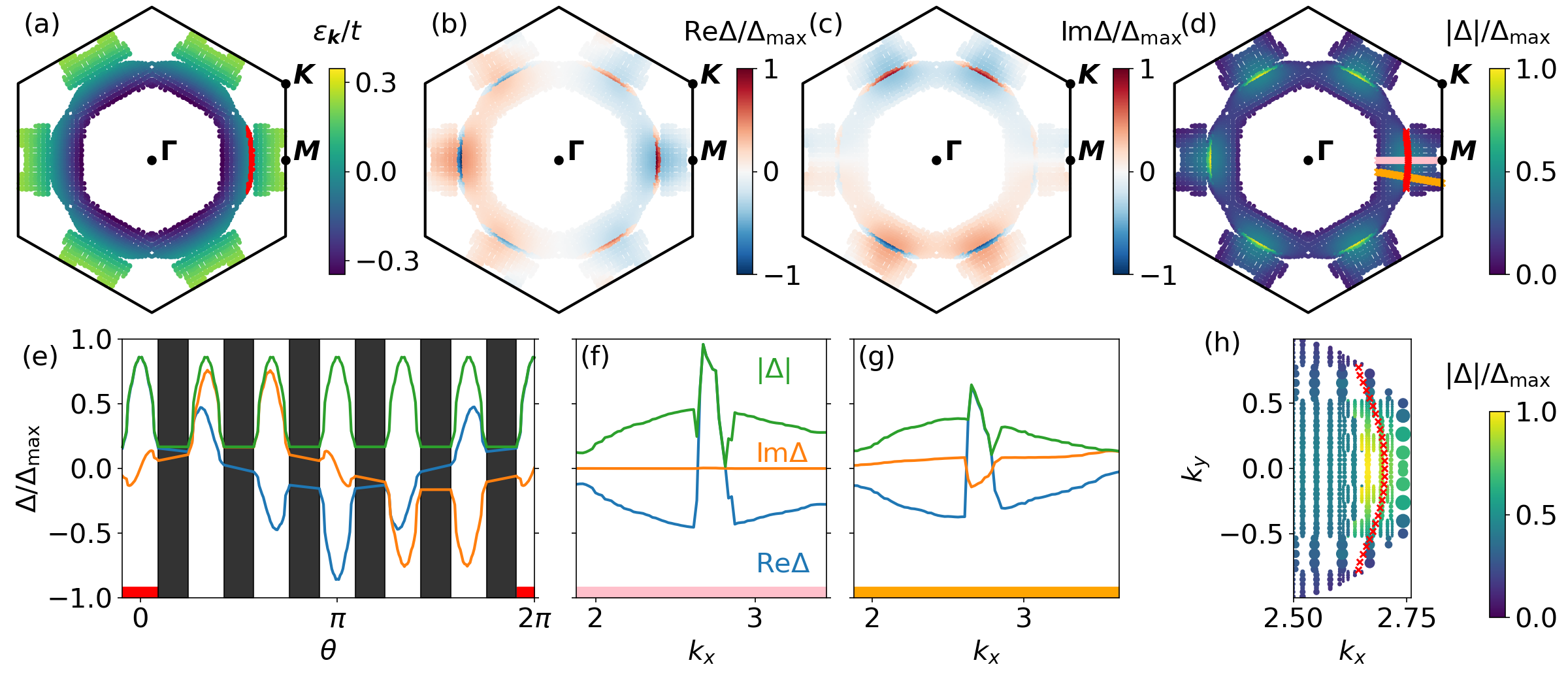}
    \caption{(a) The surface band $\epsilon_{\boldsymbol{k}}$ in the 1BZ with high-symmetry points marked. The values are shown at the 8016 points used in the adaptive quadrature and the size of each marker is scaled by the weight of the point. We found the adaptive quadrature by integrating $\chi_{\boldsymbol{k}}$ at $T=t/500$ with a tolerance of $0.02$. In the white regions there is no surface state. The red points between $\boldsymbol{\Gamma}$ and $\boldsymbol{M}$ show where $|\epsilon_{\boldsymbol{k}}| < \omega_D$ within the lines $k_y = \pm k_x/\sqrt{3}$. (b) Real part, (c) imaginary part, and (d) absolute value of the gap, all scaled by the largest absolute value $\Delta_{\text{max}}$. (e) Gap on the FS shown as a function of the angle $\theta$ that $\boldsymbol{k}_F$ makes with the $k_x$-axis. Note that a bit more than one complete revolution is shown. The red regions show where points correspond to the red crosses in (d). The black regions show directions where there is no FS and so the gap is simply linearly interpolated here and has no meaning. (f) [(g)] shows the gap along the pink [orange] crosses shown in (d). In (e), (f), and (g) the real part of the gap is shown in blue, the imaginary part in orange, and the absolute value in green, as indicated in (f). Panel (h) is a zoomed in version of (d) close to one Fermi arc, denoted by red crosses. The parameters are $Mt = 4.87\times 10^{4}$, $\chi = 9$, $L=30$, $U = 0.8t$, $J=0.2U$, $V = 0.3t$, and otherwise the same as Fig.~\ref{fig:bands9}. If $t = 1$~eV, we get $T_c \approx 0.81$~K. }
    \label{fig:fullgapPRB}
\end{figure}

\section{Full momentum gap equation} \label{sec:SCfull}
The linearized gap equation for a non-degenerate FS is \cite{SFsuperconductivity, Maeland2024Feb, Brekke2023Dec, Leraand2025Feb, Maeland2024Thesis}
\begin{equation}
\label{eq:lingapeq}
    \Delta_{\boldsymbol{k}} = -\sum_{\boldsymbol{k}'} \bar{V}_{\boldsymbol{k}\boldsymbol{k}'} \Delta_{\boldsymbol{k}'} \frac{1}{2|\epsilon_{\boldsymbol{k}'}|}\tanh\frac{\beta_{c}|\epsilon_{\boldsymbol{k}'}|}{2}.
\end{equation}
Here, $\beta_{c} = 1/k_B T_c$ is the inverse critical temperature.
The main contributions to the momentum sum comes from points close to the FS where $\chi_{\boldsymbol{k}'} = \tanh(\beta_{c}|\epsilon_{\boldsymbol{k}'}|/2)/(2|\epsilon_{\boldsymbol{k}'}|)$ is peaked. Hence, we turn the sum into an integral \cite{Thingstad2020Jun},
\begin{equation}
    \Delta_{\boldsymbol{k}} = - \frac{N_L}{A_{\text{BZ}}} \int d\boldsymbol{k}' \bar{V}_{\boldsymbol{k}\boldsymbol{k}'} \Delta_{\boldsymbol{k}'} \chi_{\boldsymbol{k}'},
\end{equation}
where $A_{\text{BZ}}$ is the area of the first Brillouin zone (1BZ), and then approximate the integral by an adaptive quadrature. That gives a sum over unevenly spaced points $\boldsymbol{k}'$ multiplied by weights $w_{\boldsymbol{k}'}$,
\begin{equation}
    \Delta_{\boldsymbol{k}} = - \frac{N_L}{A_{\text{BZ}}} \sum_{\boldsymbol{k}'} w_{\boldsymbol{k}'} \bar{V}_{\boldsymbol{k}\boldsymbol{k}'} \Delta_{\boldsymbol{k}'} \chi_{\boldsymbol{k}'}.
\end{equation}
We find the points and weights by integrating $\chi_{\boldsymbol{k}'}$ at a low temperature, giving a denser grid of points close to the FS. For the adaptive quadrature we follow the method in Ref.~\cite{AdaptQuad}, with some additional details of our implementation explained in Ref.~\cite{Maeland2022Aug}. We start with a square box with the range $k_x \in [0,2\pi/\sqrt{3}), k_y \in [-2\pi/3, 2\pi/3)$. From that we keep the points within the lines $k_y = \pm k_x/\sqrt{3}$. Finally, we rotate these points successively by $\pi/3$ until we have filled the 1BZ.

The gap equation is an eigenvalue problem where we can view the sum over $\boldsymbol{k}'$ as a matrix with indices $\boldsymbol{k}, \boldsymbol{k}'$ times $\Delta_{\boldsymbol{k}'}$ as a vector. 
This matrix will be large due to the large number of discrete momentum space points, so it will be more efficient to solve the gap equation if the matrix is hermitian. Following Ref.~\cite{Thingstad2020Jun}, we can obtain a hermitian matrix by symmetrizing, using that $\bar{V}_{\boldsymbol{k}\boldsymbol{k}'} = \bar{V}_{\boldsymbol{k}'\boldsymbol{k}}^*$ due to hermiticity of $H_{\text{BCS}}$. We multiply both sides of the equation by $\sqrt{w_{\boldsymbol{k}}\chi_{\boldsymbol{k}}}$, and define $\Bar{\Delta}_{\boldsymbol{k}} = \Delta_{\boldsymbol{k}}\sqrt{w_{\boldsymbol{k}}\chi_{\boldsymbol{k}}}$ and the hermitian matrix
\begin{equation}
    M_{\boldsymbol{k}\boldsymbol{k}'} = -\frac{N_L}{A_{\text{BZ}}} \sqrt{w_{\boldsymbol{k}}\chi_{\boldsymbol{k}}} \bar{V}_{\boldsymbol{k}\boldsymbol{k}'} \sqrt{w_{\boldsymbol{k}'}\chi_{\boldsymbol{k}'}},
\end{equation}
giving
\begin{equation}
\label{eq:fullmomentumgap}
    \Bar{\Delta}_{\boldsymbol{k}} = \sum_{\boldsymbol{k}'} M_{\boldsymbol{k}\boldsymbol{k}'} \Bar{\Delta}_{\boldsymbol{k}'}.
\end{equation}
We then search by bisection method in temperature until the highest temperature where the maximum eigenvalue of $M_{\boldsymbol{k}\boldsymbol{k}'}$ is 1. That temperature is the estimate of the critical temeprature. The corresponding eigenvector $\Bar{\Delta}_{\boldsymbol{k}}$ is a solution of the symmetrized gap equation. Remember that $\Delta_{\boldsymbol{k}} = \Bar{\Delta}_{\boldsymbol{k}}/\sqrt{w_{\boldsymbol{k}}\chi_{\boldsymbol{k}}}$ is the gap function.

If at some $\boldsymbol{k}$ there is no surface state, $\chi_{\boldsymbol{k}}$ is undefined. We assume all states that are not surface states are completely decoupled from the gap close to the FS. We find that this holds for both phonon-mediated and Coulomb interactions to a very good approximation. The decoupling is due to the spatial separation of the electronic states, giving negligible overlap of their eigenstates. So, assuming surface superconductivity dominates, $\Delta_{\boldsymbol{k}} = 0$ at momenta where there is no surface state. Hence, those $\boldsymbol{k}$ where there is no surface state should be excluded from the sum in the gap equation, without changing any weights.

We first consider a case with large surface state bandwidth compared to the maximum phonon energy, as shown in Fig.~\ref{fig:bands9}. Specifically,
the surface state bandwidth is $W \approx 0.35t$ and $\omega_D \approx 0.020t$. Then, the phonon-mediated interaction is only active in a small momentum region close to the FS shown by red in Fig.~\ref{fig:fullgapPRB}(a). We find that the gapped $p_x+\mathrm{i}p_y$-wave pairing predicted from phonons alone, survives Coulomb repulsion when $\omega_D \ll W$. Note from the real and imaginary parts in Figs.~\ref{fig:fullgapPRB}(b) and \ref{fig:fullgapPRB}(c) how, to a simplest approximation, the gap outside the region $|\epsilon_{\boldsymbol{k}}| < \omega_D$ is the negative of the gap inside that region. 
The BCS estimate of $T_c$ is decreased by the Coulomb repulsion. With zero Coulomb repulsion and otherwise same parameters as Fig.~\ref{fig:fullgapPRB}, an FS averaged gap equation gives $\lambda \approx 0.311$ and $T_c \approx 10.8$~K. 
This amounts to only quantitative changes from Coulomb repulsion, as expected from Morel-Anderson physics \cite{Bogoliubov1958, Morel1962Anderson}.

For Figs.~\ref{fig:fullgapPRB}(e), \ref{fig:fullgapPRB}(f), and \ref{fig:fullgapPRB}(g) we find the gap from the results in Figs.~\ref{fig:fullgapPRB}(b), \ref{fig:fullgapPRB}(c), and \ref{fig:fullgapPRB}(d) along specified lines. These lines do not correspond exactly to the points in the adaptive quadrature. Hence, to make the curves smoother we take an average of several points that are close to the wanted point. The average is weighted by the inverse distance between each point on the desired line and the chosen points from the adaptive quadrature.

Solving the gap in the full 1BZ is time consuming, so in the next section we introduce an FS average that reproduces the main results of Fig.~\ref{fig:fullgapPRB}. We then use this to understand how we can tune parameters to obtain $i\times (p_x+\mathrm{i}p_y)$-wave pairing close to the FS.

\section{General Fermi surface average gap equation when combining phonons and Coulomb} \label{sec:FSavg}

Typically, the fermion bandwidth is larger than the maximum phonon energy. Then, a way to understand the Morel-Anderson pseudopotential for $s$-wave superconductors \cite{Morel1962Anderson} is to imagine a gap $\Delta_1$ close to the FS, and a gap $\Delta_2$ farther from the FS. Typically, $\Delta_2 \sim -\Delta_1$, thus taking advantage of the Coulomb repulsion \cite{Bogoliubov1958, Thingstad2021, Thingstad2020Jun}. That way, Coulomb repulsion has a reduced effect on $\Delta_1$ and $T_c$, expressed via the pseudopotential $\mu^* = N_F U/[1+N_F U \ln(W/\omega_D)]$, where $N_F$ is the density of states (DOS) on the FS, $W$ is the fermion bandwidth, and $U$ is the onsite Coulomb repulsion.

Here, we generalize this idea to a case where we keep the angular momentum dependence of interactions and gaps. We start from the linearized gap equation in Eq.~\eqref{eq:lingapeq}, but make it more general by replacing $N_L$ with $N$ and imagining arbitrary dimensions. We then split the momentum integral into a part parallel $k_\parallel$ and part perpendicular $k_\perp$ to a constant energy contour $C(k'_{\perp})$ of $\epsilon_{\boldsymbol{k}'} = \epsilon_{k'_{\perp}}$,
\begin{equation}
    \sum_{\boldsymbol{k}'} = \frac{N}{A_{\text{BZ}}}\int_{\text{1BZ}} d\boldsymbol{k}' = \frac{N}{A_{\text{BZ}}}\int dk'_{\perp} \int_{C(k'_{\perp})}d k'_{\parallel}.
\end{equation}
Here, $S(\epsilon_{k'_{\perp}}) = \int_{C(k'_{\perp})}d k'_{\parallel}$ is the length of the constant energy contour and $N$ is the total number of lattice sites \cite{Maeland2024Thesis}. 

We now introduce the following approximation to the perpendicular momentum dependence of the interaction,
\begin{equation}
    \bar{V}_{\boldsymbol{k}\boldsymbol{k}'} = \begin{cases}
        \bar{V}_{\boldsymbol{k}\boldsymbol{k}'}^{\text{attr}} = \bar{V}_{\boldsymbol{k}\boldsymbol{k}'}^{\text{ph, FS}} + \bar{V}_{\boldsymbol{k}\boldsymbol{k}'}^{C, \text{FS}}, \qquad |\epsilon_{\boldsymbol{k}}|,  |\epsilon_{\boldsymbol{k}'}| < \omega_D,\\
        \bar{V}_{\boldsymbol{k}\boldsymbol{k}'}^{\text{rep}} = \bar{V}_{\boldsymbol{k}\boldsymbol{k}'}^{C, \text{FS}}, \qquad (\omega_D <|\epsilon_{\boldsymbol{k}}| < W \text{ and } |\epsilon_{\boldsymbol{k}'}| < W) \text{ or } (\omega_D <|\epsilon_{\boldsymbol{k}'}| < W \text{ and } |\epsilon_{\boldsymbol{k}}| < W),\\
        0, \qquad \text{otherwise},
    \end{cases}
\end{equation}
where $ \bar{V}_{\boldsymbol{k}\boldsymbol{k}'}^{\text{ph, FS}}$ and $\bar{V}_{\boldsymbol{k}\boldsymbol{k}'}^{C, \text{FS}}$ are the phonon and Coulomb contributions to the interaction, with momenta restricted to the FS. Note that $\bar{V}_{\boldsymbol{k}\boldsymbol{k}'}^{C, \text{FS}}$ is used for the repulsive potential $\bar{V}_{\boldsymbol{k}\boldsymbol{k}'}^{\text{rep}}$ even though one or two if its momenta are not on the FS in the cases where it is used. This is a rough approximation, but usually a good one since Coulomb interactions are weakly dependent on momentum. Also, note that the electron bandwidth is technically $2W$, but the relevant parameter is how far away from the FS it maximally extends in either direction, which is $W$.

Similarly, the gap now takes two separate values close to the FS and farther away from it
\begin{equation}
    \Delta_{\boldsymbol{k}} = \begin{cases}
        \Delta_{k_\parallel}^{(1)}, \qquad |\epsilon_{\boldsymbol{k}}|< \omega_D, \\
        \Delta_{k_\parallel}^{(2)}, \qquad \omega_D < |\epsilon_{\boldsymbol{k}}|< W, \\
        0, \qquad \text{otherwise}.
    \end{cases}
\end{equation}
It is natural to guess that while $\Delta_{k_\parallel}^{(1)}$ and $\Delta_{k_\parallel}^{(2)}$ may both show sign changes in their angular dependence, we will typically have $\Delta_{k_\parallel}^{(2)} \sim -\Delta_{k_\parallel}^{(1)}$.

Given the above approximations, we perform the parallel integral on the FS,
\begin{align}
    \Delta_{k_\parallel}^{(1)} &= -\frac{N}{A_{\text{BZ}}}\int dk'_{\perp} \frac{S(\epsilon_{k'_{\perp}})}{2|\epsilon_{k'_\perp}|} \tanh\frac{\beta_{c}|\epsilon_{k'_\perp}|}{2} \frac{1}{S_{\text{FS}}} \int_{\text{FS}}d k'_{\parallel} \bar{V}_{k_{\parallel}k'_\parallel}^{\text{attr}} \Delta_{k'_\parallel}^{(1)} \nonumber \\
    & -\frac{N}{A_{\text{BZ}}}\int dk'_{\perp} \frac{S(\epsilon_{k'_{\perp}})}{2|\epsilon_{k'_\perp}|} \tanh\frac{\beta_{c}|\epsilon_{k'_\perp}|}{2} \frac{1}{S_{\text{FS}}} \int_{\text{FS}}d k'_{\parallel} \bar{V}_{k_{\parallel}k'_\parallel}^{\text{rep}} \Delta_{k'_\parallel}^{(2)},
\end{align}
\begin{align}
    \Delta_{k_\parallel}^{(2)} &= -\frac{N}{A_{\text{BZ}}}\int dk'_{\perp} \frac{S(\epsilon_{k'_{\perp}})}{2|\epsilon_{k'_\perp}|} \tanh\frac{\beta_{c}|\epsilon_{k'_\perp}|}{2} \frac{1}{S_{\text{FS}}} \int_{\text{FS}}d k'_{\parallel} \bar{V}_{k_{\parallel}k'_\parallel}^{\text{rep}} \Delta_{k'_\parallel}^{(1)} \nonumber \\
    & -\frac{N}{A_{\text{BZ}}}\int dk'_{\perp} \frac{S(\epsilon_{k'_{\perp}})}{2|\epsilon_{k'_\perp}|} \tanh\frac{\beta_{c}|\epsilon_{k'_\perp}|}{2} \frac{1}{S_{\text{FS}}} \int_{\text{FS}}d k'_{\parallel} \bar{V}_{k_{\parallel}k'_\parallel}^{\text{rep}} \Delta_{k'_\parallel}^{(2)},
\end{align}
where $S_{\text{FS}}$ is the length of the FS. Keep in mind that when $\Delta_{k'_\parallel}^{(2)}$ is used, $\omega_D < |\epsilon_{k'_{\perp}}| < W$.
While $\Delta_{k_\parallel}^{(2)}$ is the angular dependence of the gap away from the FS, we find it for momenta on the FS, due to the assumptions in the form of the electron-electron interaction.
We now introduce a change of variables to energy in the perpendicular integral, 
\begin{equation}
    \int d k'_\perp = \int d\epsilon \frac{dk'_\perp}{d\epsilon} \to \int d\epsilon \abs{\frac{\partial \epsilon}{\partial k'_{\perp}}}^{-1} = \int d\epsilon v_{k'_\parallel}^{-1}.
\end{equation}
Here, $v_{k'_\parallel} = |\partial \epsilon/\partial k'_{\perp}|$ is the slope of the band perpendicular to the FS, which depends on $k'_\parallel$ \cite{Maeland2024Thesis}.
Performing the change of variables $\epsilon_{k'_{\perp}} \to \epsilon$, the gap equation becomes
\begin{align}
    \Delta_{k_\parallel}^{(1)} &= -\frac{N}{A_{\text{BZ}}}\int_{\text{FS}}d k'_{\parallel} v_{k'_\parallel}^{-1} \bar{V}_{k_{\parallel}k'_\parallel}^{\text{attr}} \Delta_{k'_\parallel}^{(1)} \int_{0}^{\omega_D} d \epsilon \frac{\tanh(\beta_{c}\epsilon/2)}{\epsilon} \nonumber \\
    & -\frac{N}{A_{\text{BZ}}}\int_{\text{FS}}d k'_{\parallel} v_{k'_\parallel}^{-1} \bar{V}_{k_{\parallel}k'_\parallel}^{\text{rep}} \Delta_{k'_\parallel}^{(2)} \int_{\omega_D}^{W} d \epsilon \frac{\tanh(\beta_{c}\epsilon/2)}{\epsilon},
\end{align}
\begin{align}
    \Delta_{k_\parallel}^{(2)} &= -\frac{N}{A_{\text{BZ}}}\int_{\text{FS}}d k'_{\parallel} v_{k'_\parallel}^{-1} \bar{V}_{k_{\parallel}k'_\parallel}^{\text{rep}} \Delta_{k'_\parallel}^{(1)} \int_{0}^{\omega_D} d \epsilon \frac{\tanh(\beta_{c}\epsilon/2)}{\epsilon} \nonumber \\
    & -\frac{N}{A_{\text{BZ}}}\int_{\text{FS}}d k'_{\parallel} v_{k'_\parallel}^{-1} \bar{V}_{k_{\parallel}k'_\parallel}^{\text{rep}} \Delta_{k'_\parallel}^{(2)} \int_{\omega_D}^{W} d \epsilon \frac{\tanh(\beta_{c}\epsilon/2)}{\epsilon},
\end{align}
where the integration limits of $\epsilon$ come from the region where the associated gap $\Delta_{k'_\parallel}^{(1,2)}$ is set to be nonzero. We approximate $S(\epsilon_{k'_{\perp}}) = S_{\text{FS}}$ which is often a good approximation since the behavior close to the FS dominates \cite{Maeland2024Thesis}. Here, however, we note that it could be a weakness of the FS average since $\Delta_{k_\parallel}^{(2)}$ is the gap farther away from the FS. The approximation becomes better the more constant the DOS is as a function of energy. 

Within a weak-coupling approach we should expect $k_B T_c \ll \omega_D$, i.e., $\beta_c \omega_D \gg 1$. Then we can set $\tanh(\beta_{c}\epsilon/2) \approx 1$ in the integral
\begin{equation}
    \int_{\omega_D}^{W} d \epsilon \frac{\tanh(\beta_{c}\epsilon/2)}{\epsilon} \approx \int_{\omega_D}^{W} d \epsilon \frac{1}{\epsilon} = \ln(\frac{W}{\omega_D}),
\end{equation}
a term familiar from the Morel-Anderson pseudopotential. Meanwhile, within the weak-coupling assumption $\beta_c \omega_D \gg 1$ the integral
\begin{equation}
    \int_{0}^{\omega_D} d \epsilon \frac{\tanh(\beta_{c}\epsilon/2)}{\epsilon} \approx \ln(\frac{2}{\pi}e^{\gamma_E} \beta_c \omega_D),
\end{equation}
with $\gamma_E = 0.577\dots$ being the Euler-Mascheroni constant \cite{SFsuperconductivity}. Let us then write the gap equation as a matrix equation
\begin{equation}
    \begin{pmatrix}
        \Delta_{k_\parallel}^{(1)} \\
        \Delta_{k_\parallel}^{(2)}
    \end{pmatrix} = -\frac{NS_{\text{FS}}}{A_{\text{BZ}}N_{\text{samp}}} \sum_{k'_\parallel} v_{k'_\parallel}^{-1} \begin{pmatrix}
        \bar{V}_{k_{\parallel}k'_\parallel}^{\text{attr}}\ln(\frac{2}{\pi}e^{\gamma_E} \beta_c \omega_D) & \bar{V}_{k_{\parallel}k'_\parallel}^{\text{rep}}\ln(\frac{W}{\omega_D}) \\
        \bar{V}_{k_{\parallel}k'_\parallel}^{\text{rep}}\ln(\frac{2}{\pi}e^{\gamma_E} \beta_c \omega_D) & \bar{V}_{k_{\parallel}k'_\parallel}^{\text{rep}}\ln(\frac{W}{\omega_D})
    \end{pmatrix} \begin{pmatrix}
        \Delta_{k'_\parallel}^{(1)} \\
        \Delta_{k'_\parallel}^{(2)}
    \end{pmatrix}
\end{equation}
We approximate the $k'_\parallel$ integral with the trapezoidal rule, using $N_{\text{samp}}$ evenly spaced points on the FS. 
The gap equation is an eigenvalue problem. We solve it by the bisection method, searching for the value of $\beta_c$ where the greatest eigenvalue is one, giving the critical temperature $T_c$. Then, the corresponding eigenvector represents the vector that solves the eigenvalue problem and yields the momentum dependence of the two gap functions.

Let us comment on differences to more known approaches. References \cite{Bogoliubov1958, Morel1962Anderson} additionally neglect the angular dependence of the interaction assuming constant $s$-wave interactions and gaps. Then, the matrix gap equation can be rewritten as a linear homogeneous equation set, which only has nontrivial (nonzero gap) solutions when the determinant of the matrix is zero. That puts restrictions on the elements in the matrix and gives an analytic result $\Delta_2 = -\mu^* \Delta_1/(\lambda-\mu^*) $, where $\lambda$ is the dimensionless coupling from phonons alone. The gap equation then reduces to one for $\Delta_1$ only, where the only change is that $\lambda \to \lambda -\mu^*$ compared to BCS gap equations without Coulomb, see, e.g., Ref.~\cite{Thingstad2021} for details. Since we keep the angular momentum dependence, our equation set is not a linear homogeneous set and so $\Delta_{k_\parallel}^{(1)}$ and $\Delta_{k_\parallel}^{(2)}$ must be solved for together. We cannot define a pseudopotential, but the effect of the Coulomb interaction is covered by the coupled set of equations for two gaps. The existence of $\Delta_{k_\parallel}^{(2)}$ should boost $\Delta_{k_\parallel}^{(1)}$. That is the momentum resolved version of how the pseudopotential reduces the detrimental effects of Coulomb repulsion on $s$-wave gaps.

Adapting to the Weyl semimetal surface state, $N \to N_L$, i.e., total number of sites goes to total number of sites per layer. We rewrite the gap equation as
\begin{equation}
\label{eq:D12gapeq}
    \begin{pmatrix}
        \Delta_{k_\parallel}^{(1)} \\
        \Delta_{k_\parallel}^{(2)}
    \end{pmatrix} = - \sum_{k'_\parallel} N_{k'_\parallel}\begin{pmatrix}
        \bar{V}_{k_{\parallel}k'_\parallel}^{\text{attr}}\ln(\frac{2}{\pi}e^{\gamma_E} \beta_c \omega_D) & \bar{V}_{k_{\parallel}k'_\parallel}^{\text{rep}}\ln(\frac{W}{\omega_D}) \\
        \bar{V}_{k_{\parallel}k'_\parallel}^{\text{rep}}\ln(\frac{2}{\pi}e^{\gamma_E} \beta_c \omega_D) & \bar{V}_{k_{\parallel}k'_\parallel}^{\text{rep}}\ln(\frac{W}{\omega_D})
    \end{pmatrix} \begin{pmatrix}
        \Delta_{k'_\parallel}^{(1)} \\
        \Delta_{k'_\parallel}^{(2)}
    \end{pmatrix}.
\end{equation}
Here, $N_{k'_\parallel} = N_L S_{\text{FS}}v_{k'_\parallel}^{-1}/N_{\text{samp}}A_{\text{BZ}}$ is a momentum-dependent DOS factor.
The bulk-like states decouple from the surface states and experience only Coulomb repulsion. Therefore the gap for bulk states is zero. Hence, the electron bandwidth $W$ is now the range of energy of the surface states.

\begin{figure}
    \centering
    \includegraphics[width=\linewidth]{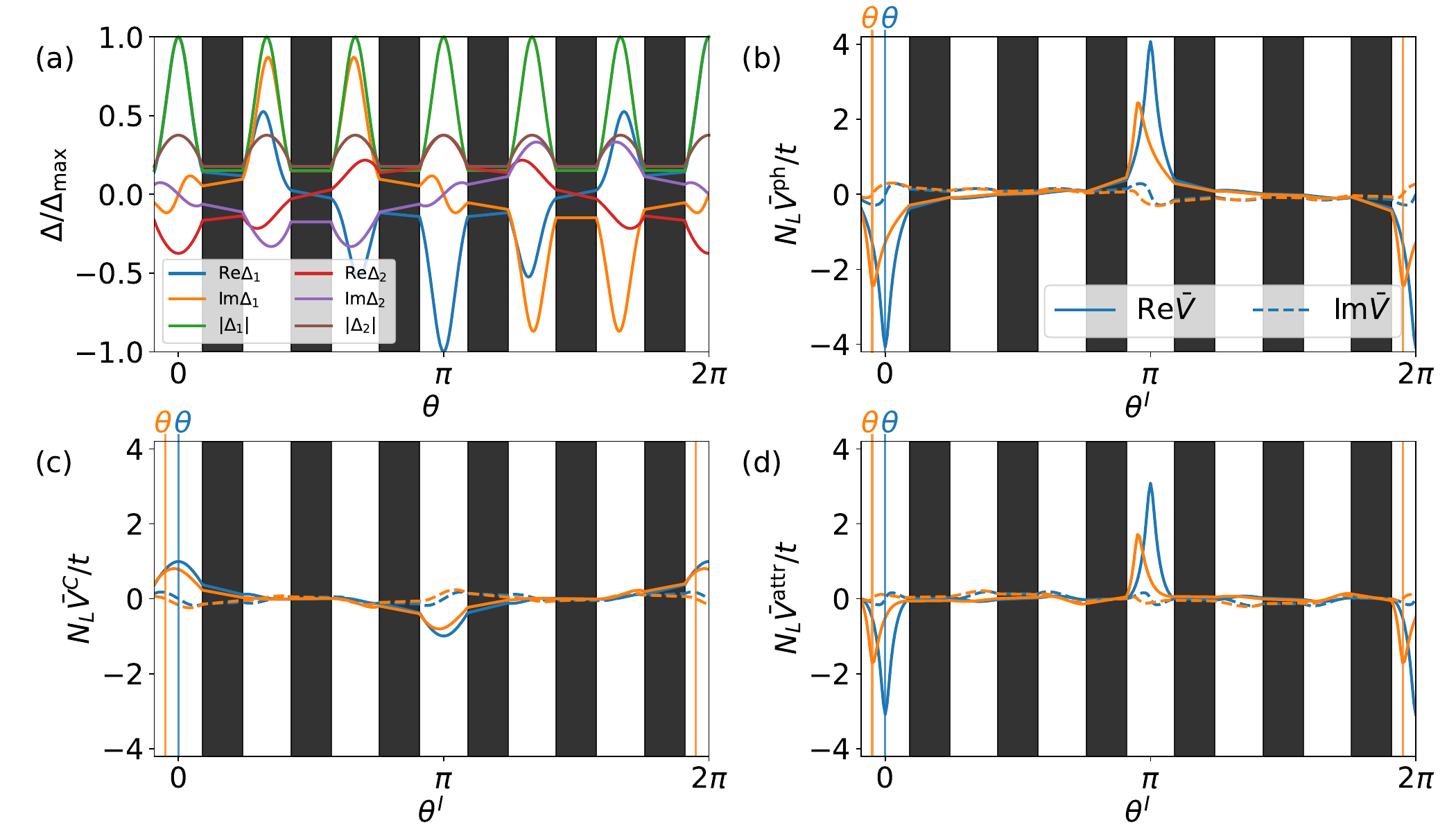}
    \caption{Solution of the coupled FS average gap equation in Eq.~\eqref{eq:D12gapeq}. (a) Real, imaginary and absolute values of $\Delta_{\boldsymbol{k}}^{(1)}$ and $\Delta_{\boldsymbol{k}}^{(2)}$ shown as functions of the angle $\theta$ that $\boldsymbol{k}$ makes with the $k_x$ axis. (b) Real part (solid lines) and imaginary part (dashed lines) of the phonon mediated interaction $\bar{V}_{\boldsymbol{k} \boldsymbol{k}'}^{\text{ph}}$ as a function of the angle $\theta'$ that $\boldsymbol{k}'$ makes with the $k_x$ axis. The color denotes at what angle $\theta$ that $\boldsymbol{k}$ is fixed. All momenta are on the FS. For the blue curves, $\boldsymbol{k}$ is in the center of the rightmost Fermi arc. For the orange curves, $\boldsymbol{k}$ is fixed at the midpoint between the center of the Fermi arc and its endpoint. (c) $\bar{V}_{\boldsymbol{k} \boldsymbol{k}'}^{\text{rep}} = \bar{V}_{\boldsymbol{k} \boldsymbol{k}'}^{C}$ and (d) $\bar{V}_{\boldsymbol{k} \boldsymbol{k}'}^{\text{attr}} = \bar{V}_{\boldsymbol{k} \boldsymbol{k}'}^{\text{ph}} + \bar{V}_{\boldsymbol{k} \boldsymbol{k}'}^{C}$ plotted in the same way. The parameters are $W = 0.35t$, $\omega_D = 0.020t$, $N_{\text{samp}} = 162$, and otherwise the same as in Fig.~\ref{fig:fullgapPRB}.}
    \label{fig:D12VPRB}
\end{figure}

In Fig.~\ref{fig:D12VPRB} we show a solution of the coupled FS average gap equation in Eq.~\eqref{eq:D12gapeq}.

We see from Fig.~\ref{fig:D12VPRB}(a) that $\Delta_{\boldsymbol{k}}^{(1)}$ corresponds well with the results in Fig.~\ref{fig:fullgapPRB}(e) from solving the gap equation in the full 1BZ. Hence, the FS averaged equation gives a good approximation to the results. We get a slight overestimate of the critical temperature, $T_c \approx 1.5$~K. Considering Fig.~\ref{fig:fullgapPRB}(h), it seems the magnitude of the gap has a radial momentum dependence within the region $|\epsilon_{\boldsymbol{k}}| < \omega_D$ while the FS average treats the radial dependence of the gap as constant here. 
As expected, a simplest approximation of the result is $\Delta_{\boldsymbol{k}}^{(2)} \sim -\Delta_{\boldsymbol{k}}^{(1)}$.

The key to achieve  $i\times (p_x +\mathrm{i}p_y)$ pairing for $\Delta_{\boldsymbol{k}}^{(1)}$ is to decouple it from $\Delta_{\boldsymbol{k}}^{(2)}$. Then, $\Delta_{\boldsymbol{k}}^{(2)}$ experiences only repulsion and is zero. Since surface states and bulk states nearly decouple, this situation can in fact be realized in Weyl semimetals if the surface state bandwidth is comparable to the phonon bandwidth.

\begin{figure}
    \centering
    \includegraphics[width=\linewidth]{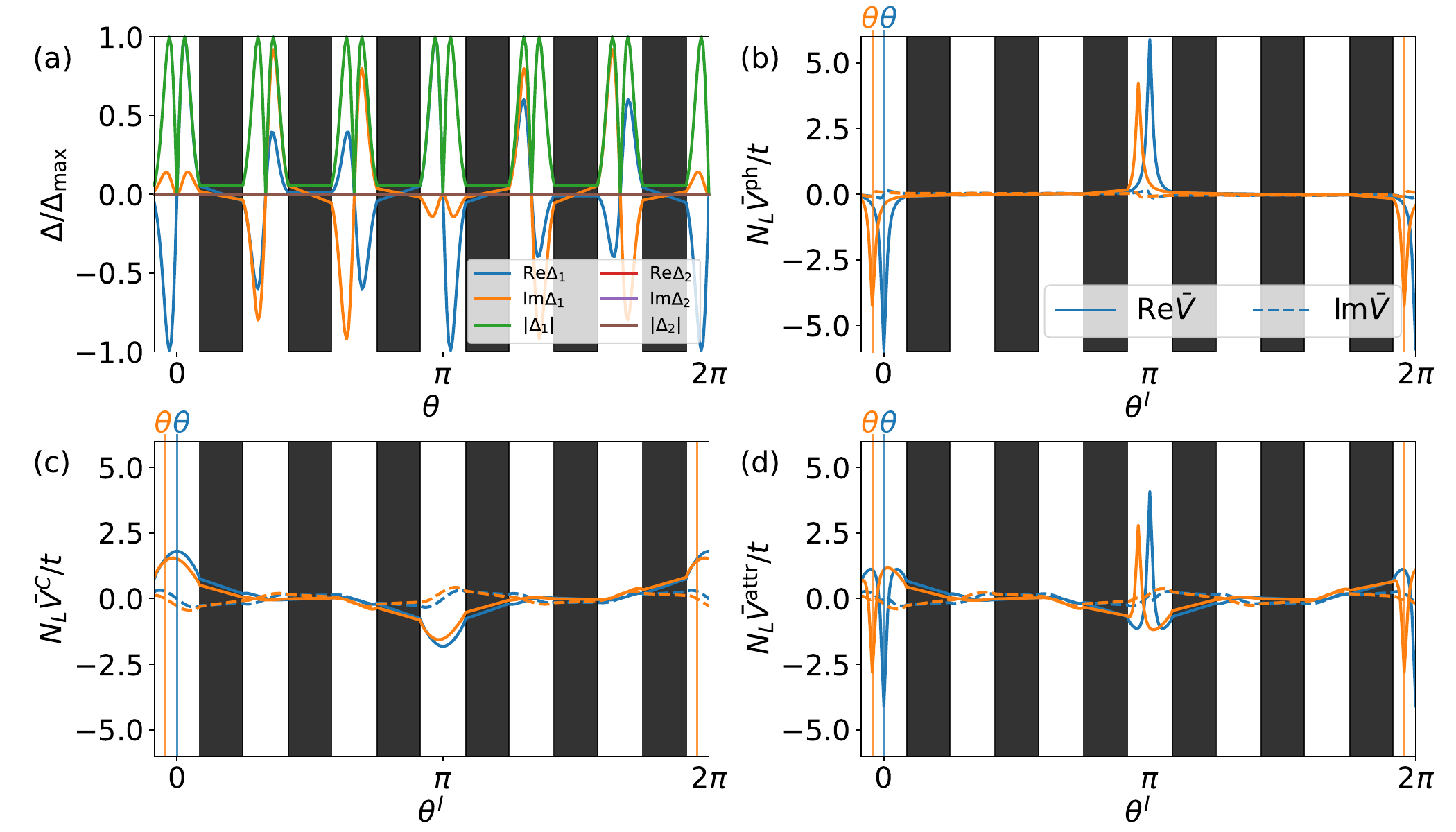}
    \caption{Solution of the coupled FS average gap equation in Eq.~\eqref{eq:D12gapeq}. Real, imaginary and absolute values of $\Delta_{\boldsymbol{k}}^{(1)}$ and $\Delta_{\boldsymbol{k}}^{(2)}$ shown as functions of the angle $\theta$ that $\boldsymbol{k}$ makes with the $k_x$ axis. $\Delta_{\boldsymbol{k}}^{(2)}$ is negligible and  $\Delta_{\boldsymbol{k}}^{(1)}$ is $i\times (p_x +\mathrm{i}p_y)$-wave. We count 10 sign changing nodes in the real and imaginary part of $\Delta_{\boldsymbol{k}}^{(1)}$ indicative of $h_y +\mathrm{i} h_x$-wave paring. At the same time, we see the non-sign-changing nodes in the imaginary part at $\theta = 0, \pi$. We also imagine that there are nodes in the gap in the center of all black regions where there is no FS. That gives in total 12 nodes of the absolute value, like an $i$-wave. (b), (c), and (d) Interactions plotted in the same fashion as Fig.~\ref{fig:D12VPRB}. $T_c \approx 6.7$~K if $t = 0.5$~eV. The parameters are $t_o/t = 0.24$, $\beta/t = -1.5$, $\beta_o = -0.4$, $\mu/t = -0.02, \mu_o/t = -0.7, \alpha/t = -0.06$, $\gamma/t = -0.08$, $\gamma_1 = -(0.01t)^2$, $\gamma_3 = 0.43\gamma_1$, $\gamma_4 = 1.5\gamma_1$, $\gamma_5 = 0.432\gamma_1$, $\gamma_6 = 0.6 \gamma_1$, $Mt = 24350$, $\chi = 20$, $U=t$, $J=0.2U$, $V_N = 0.6t$, $W = 0.10t$, $\omega_D = 0.040t$, $L = 30$, and $N_{\text{samp}} = 150$.}
    \label{fig:D12set2}
\end{figure}

\begin{figure}
    \centering
    \includegraphics[width=\linewidth]{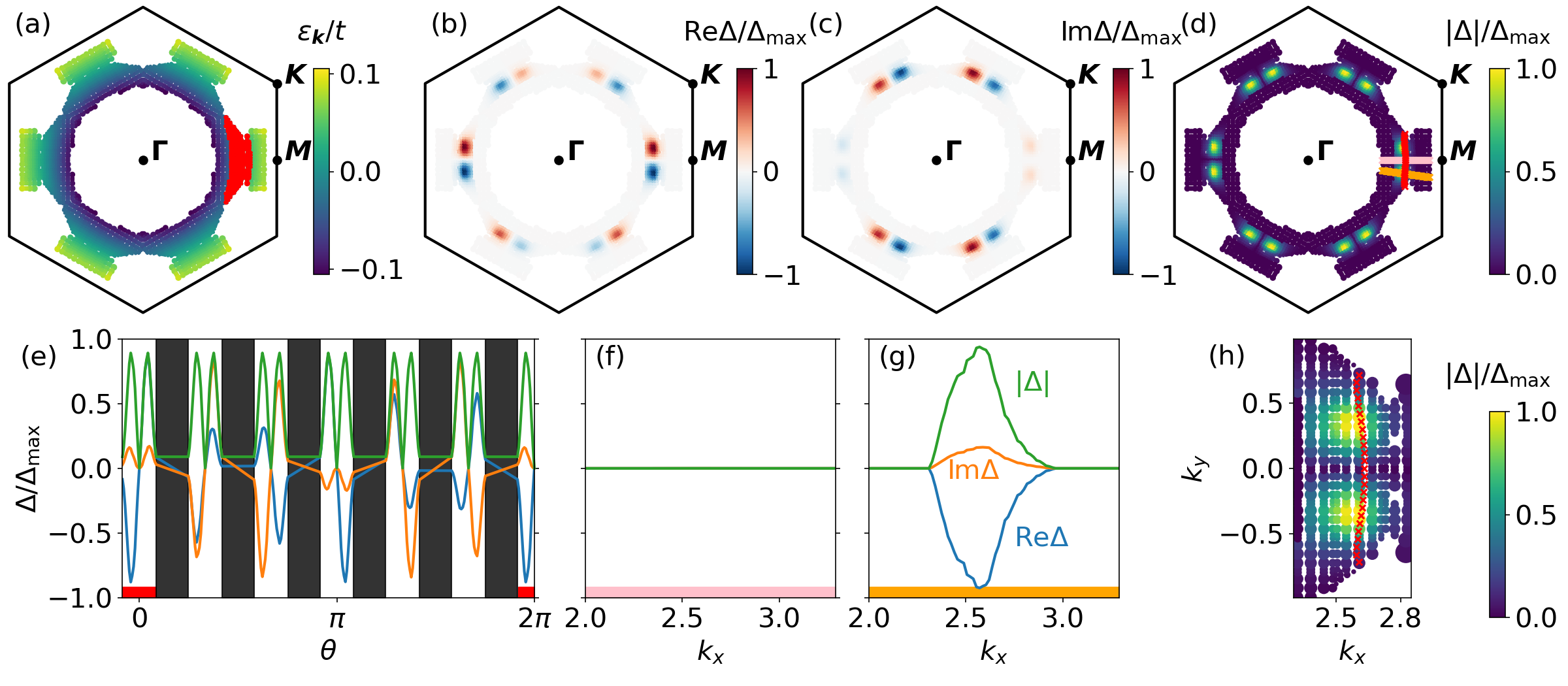}
    \caption{(a) The surface band $\epsilon_{\boldsymbol{k}}$ in the 1BZ. The values are shown at the 4200 points used in the adaptive quadrature and the size of each marker is scaled by the weight of the point. We found the adaptive quadrature by integrating $\chi_{\boldsymbol{k}}$ at $T=t/300$ with a tolerance of $0.02$. In the white regions there is no surface state. The red points between $\boldsymbol{\Gamma}$ and $\boldsymbol{M}$ show the points where $|\epsilon_{\boldsymbol{k}}| < \omega_D$ within the lines $k_y = \pm k_x/\sqrt{3}$. (b) Real part, (c) imaginary part, and (d) absolute value of the gap, all scaled by the largest absolute value $\Delta_{\text{max}}$. (e) Gap on the FS shown as a function of the angle $\theta$. The red regions show where points correspond to the red crosses in (d). The black regions show directions where there is no FS and so the gap is simply linearly interpolated here and has no meaning. (f) [(g)] shows the gap along the pink [orange] crosses shown in (d). In (e), (f), and (g) the real part of the gap is shown in blue, the imaginary part in orange, and the absolute value in green, as indicated in (g). Panel (h) is a zoomed in version of (d) close to one Fermi arc, denoted by red crosses. $T_c \approx 2.8$~K if $t = 0.5$~eV. The parameters are the same as in Fig.~\ref{fig:D12set2}.}
    \label{fig:fullgap2}
\end{figure}

In Fig.~\ref{fig:D12set2} we choose parameters such that $W = 0.105t$ and $\omega_D = 0.040t$. Then, we find that $\Delta_{\boldsymbol{k}}^{(2)}$ is negligible while $\Delta_{\boldsymbol{k}}^{(1)}$ shows $i\times (p_x +\mathrm{i}p_y)$-wave pairing. We interpret this as $\Delta_{\boldsymbol{k}}^{(1)}$ being $i\times (p_x +\mathrm{i}p_y)$ effectively leads to a decoupling between $\Delta_{\boldsymbol{k}}^{(1)}$ and $\Delta_{\boldsymbol{k}}^{(2)}$ because Coulomb interaction multiplied by the $i\times (p_x +\mathrm{i}p_y)$-wave gap is approximately zero [$\sum_{k'_\parallel} N_{k'_\parallel} \bar{V}_{k_\parallel k'_\parallel}^{C,\text{FS}}\Delta_{k'_\parallel } \approx 0$ when $\Delta_{k'_\parallel }$ is $i\times(p_x+\mathrm{i}p_y)$-wave]. A solution like Fig.~\ref{fig:D12VPRB}, where $\Delta_{\boldsymbol{k}}^{(2)}$ is not negligible and $\Delta_{\boldsymbol{k}}^{(1)}$ is fully gapped is now probably a subdominant solution with a lower $T_c$. This is due to a relatively larger area for $\Delta_{\boldsymbol{k}}^{(1)}$ compared to $\Delta_{\boldsymbol{k}}^{(2)}$ at the lower surface state bandwidth, and so rather than having Morel-Anderson-like physics the gap solution that gives best pairing for $\Delta_{\boldsymbol{k}}^{(1)}$ alone gives the highest $T_c$. 

We confirm the result by a solution of the full momentum gap equation in Fig.~\ref{fig:fullgap2}. Note that there, the gap is zero outside the region where $|\epsilon_{\boldsymbol{k}}| < \omega_D$, corresponding to $\Delta_{\boldsymbol{k}}^{(2)}  = 0$ from the FS averaged gap equation. We take this result to indicate that the surface state bandwidth need not be exactly equal to the maximum phonon energy to prefer $i$-wave pairing. $W$ can be larger than $\omega_D$, and an $i\times (p_x +\mathrm{i}p_y)$-wave gap close to the FS can still dominate.

We now move on to a case where $W \approx \omega_D$ such that we can focus exclusively on the pairing close to the FS for simplicity. However, with the result in Fig.~\ref{fig:D12set2} in mind, the following results should be qualitatively valid also when $W$ is at least twice as large as $\omega_D$. Based on the results discussed in the main text, we expect that a stronger Coulomb interaction will be needed to give nodal pairing the larger $W$ is relative to $\omega_D$. The reason is that stronger Coulomb will make $i\times (p_x +\mathrm{i}p_y)$-pairing more and more favorable compared to fully gapped pairing close to the FS ($\Delta_{\boldsymbol{k}}^{(1)}$).

\begin{figure}
    \centering
    \includegraphics[width=\linewidth]{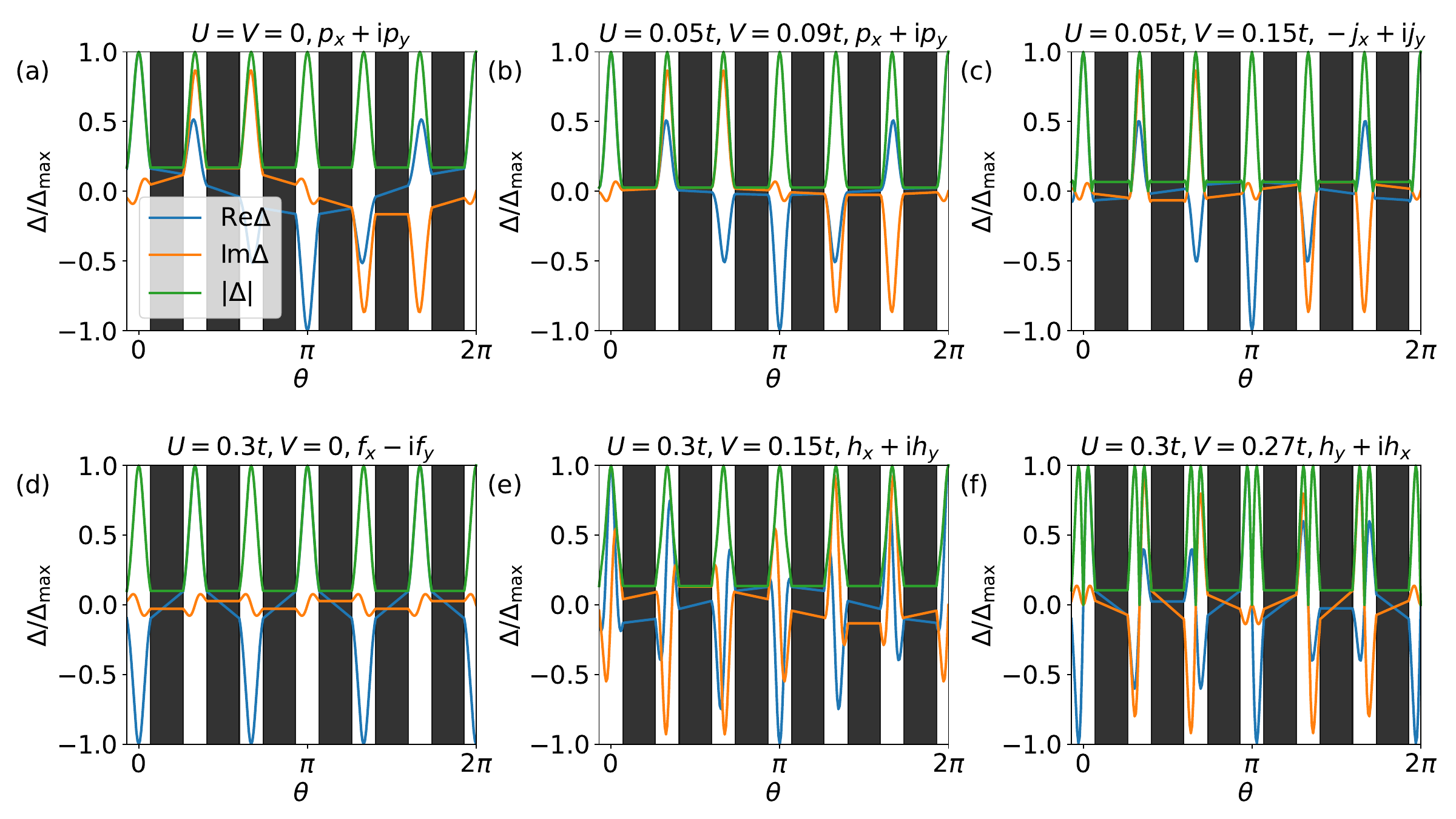}
    \caption{The gap functions $\Delta_{\boldsymbol{k}}$ with largest critical temperature at different values of $U$ and $V$ plotted as a function of the azimuthal angle of $\boldsymbol{k}$. These are solutions of the linearized gap equation with the same parameters as Fig.~3 in the main text. Specifically, the parameters are $t_o/t = 0.24$, $\beta/t = -1.5$, $\beta_o = -0.4$, $\mu/t = -0.01, \mu_o/t = -0.7, \alpha/t = -0.03$, $\gamma/t = -0.04$, $\gamma_1 = -(0.01t)^2$, $\gamma_3 = 0.43\gamma_1$, $\gamma_4 = 1.5\gamma_1$, $\gamma_5 = 0.432\gamma_1$, $\gamma_6 = 0.6 \gamma_1$, $\chi = 30$, $Mt = 24350$, $J=0.2U$, $L = 30$, and $N_{\text{samp}} = 126$. }
    \label{fig:gaps8}
\end{figure}

\section{Gap solutions when surface state bandwidth is comparable to phonon energy range} \label{sec:bandwidthEqual}

Taking for simplicity the case $W = \omega_D$, Coulomb interactions and phonon-mediated interactions have the same range. We can then perform an FS average by assuming the gap function to take the form 
\begin{equation}
    \Delta_{\boldsymbol{k}} = \begin{cases}
        \Delta_{k_\parallel}, \qquad |\epsilon_{\boldsymbol{k}}|< \omega_D, \\
        0, \qquad \text{otherwise}.
    \end{cases}
\end{equation}
Then, the FS averaged gap equation is
\begin{equation}
\label{eq:lingapeqFS}
    \lambda \Delta_{k_\parallel} =  -\sum_{k'_\parallel } N_{k'_\parallel} \bar{V}_{k_\parallel k'_\parallel}^{\text{FS}}\Delta_{k'_\parallel }, \qquad \text{with} \quad \bar{V}_{k_\parallel k'_\parallel}^{\text{FS}} = \bar{V}_{k_\parallel k'_\parallel}^{\text{ph, FS}} + \bar{V}_{k_\parallel k'_\parallel}^{C, \text{FS}}.
\end{equation}
For the case $W = \omega_D$, we could also manually set $\Delta_{\boldsymbol{k}}^{(2)} = 0$ in the coupled gap equation Eq.~\eqref{eq:D12gapeq}. The remaining equation for $\Delta_{\boldsymbol{k}}^{(1)}$ is then equivalent to Eq.~\eqref{eq:lingapeqFS}. 
We show results for this gap equation in Fig.~3 of the main text. Also, Fig.~\ref{fig:gaps8} shows the gap solutions for all five different gap symmetries plotted as a function of the angle on the FS. Figs.~\ref{fig:gaps8}(a), (b), and (c) show the gradual change from $p_x+ip_y$-wave to $j_x-ij_y$-wave by increasing $V$. As we discuss further in the next section, these states are closely related.

Having already stressed in the main text that the BCS prediction of $T_c$ is unreliable, we here provide the interested reader with some numbers. If we assume $t=0.5$~eV, we have $\omega_D \approx 20$~meV and $M = 204$~u, which is realistic for PtBi$_2$ \cite{Bashlakov2022PtBi2phononExp}. Then, with parameters following the stars in Fig.~3 of the main text, we obtain $T_c \approx 36.5$~K for $p$-, $T_c \approx 30.1$ for $f$-, $T_c \approx 11.3$~K for $h$-, and $T_c \approx 9.6$~K for $i$-wave pairing.

\begin{figure}
    \centering
    \includegraphics[width=0.7\linewidth]{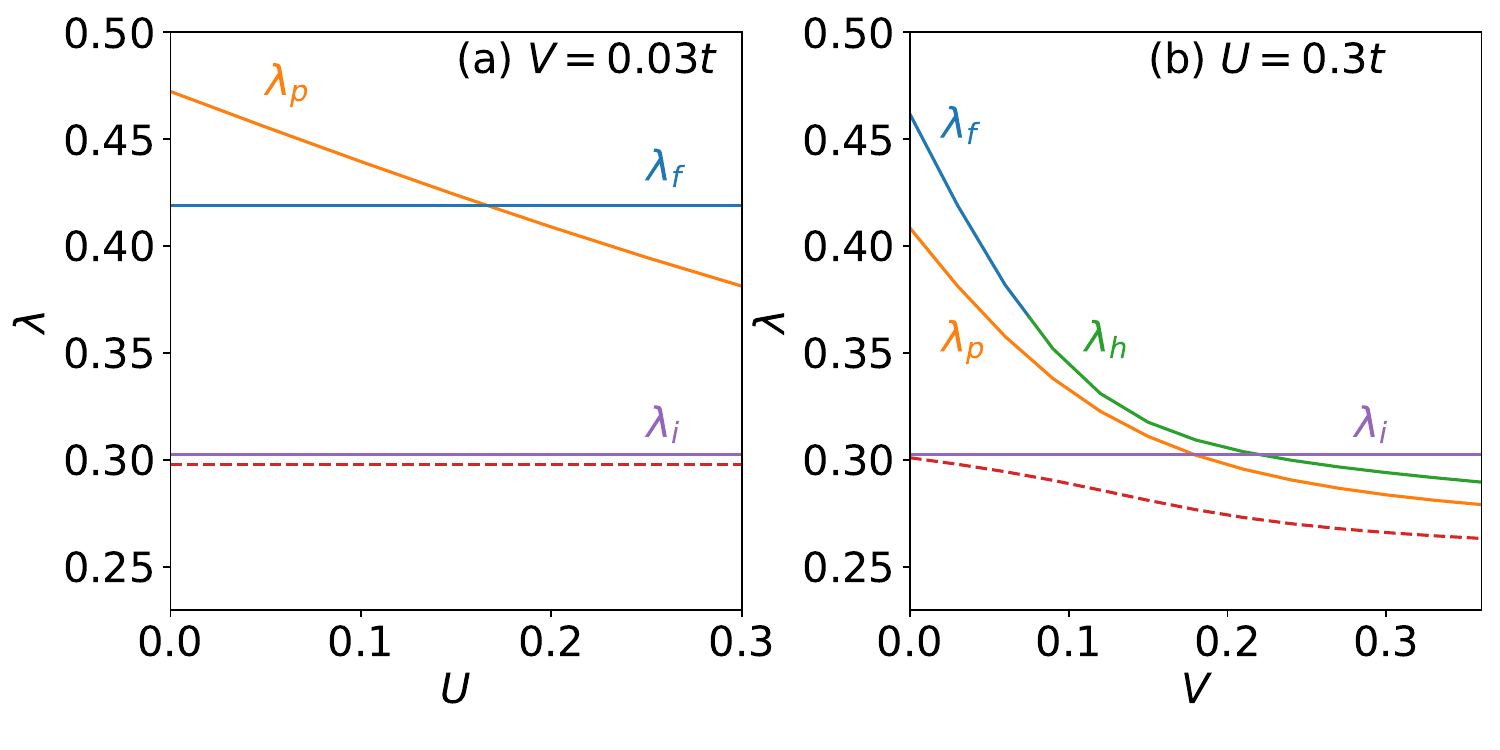}
    \caption{The four largest eigenvalues in the gap equation \eqref{eq:lingapeqFS}. (a) As a function of $U$ when $V = 0.03t$. The orange curve represents the $p_x+ip_y$-wave gap. Once $\lambda_p < \lambda_f$, we do not focus on any additional sign changes that may appear, and view the then subdominant solution to be an extension of the same gap symmetry. The purple curve is the $i$-wave state, while the red curve shows a doubly degenerate eigenvalue which is always subdominant. (b) As a function of $V$ when $U = 0.3t$. The orange curve is connected to the state which has a $p_x+ip_y$-wave gap when it is dominant. Here, it is subdominant and might have developed more nodes. The largest and doubly degenerate eigenvalue changes symmetry when it is dominating. Therefore we mark its symmetry change from fully gapped $f$- to $h$-wave by going from green to blue. The purple and red curves are the same as in panel (a). The parameters are the same as Fig.~\ref{fig:gaps8}}
    \label{fig:lamUV}
\end{figure}

Figure \ref{fig:lamUV} shows the four largest eigenvalues as a function of $U$ at a fixed $V$ and as a function of $V$ at a fixed $U$. Meanwhile Fig.~3 of the main text shows the largest eigenvalue at all considered $U$ and $V$. We note that the characterization of different lines in Fig.~\ref{fig:lamUV} in terms of the gap symmetries discussed in the main text is nontrivial. Once the fully gapped $p_x+ip_y$-wave state becomes subdominant, its corresponding gap function tends to develop more nodes indicating an increase in the angular momentum. This is similar to the nodal $j$-wave state shown in gray in Fig.~3 in the main text. Here, we still label it as $\lambda_p$ as it is continuously connected to the gap solution that is $p_x+ip_y$-wave at weak Coulomb repulsion. The same applies to the other lines in Fig.~\ref{fig:lamUV} once they become subdominant.
We include a red curve for the fourth largest eigenvalue in Fig.~\ref{fig:lamUV}. It is a doubly degenerate eigenvalue, as in the fully gapped $f$- and $h$-wave states. Since it is always subdominant, we do not characterize its gap function further except to note that it is likely a higher angular momentum solution within the same irreducible representation (irrep) as the fully gapped $f$- and $h$-wave states, see a reference to the symmetry classification below. Figure \ref{fig:lamUV} confirms the argument from Fig.~3 in the main text, namely, that upon increasing the strength or Coulomb repulsion, all states will eventually have a lower dimensionless coupling than the $i$-wave pairing ($\lambda_i$). This includes gap functions with both lower and higher angular momentum.

\begin{figure}
    \centering
    \includegraphics[width=\linewidth]{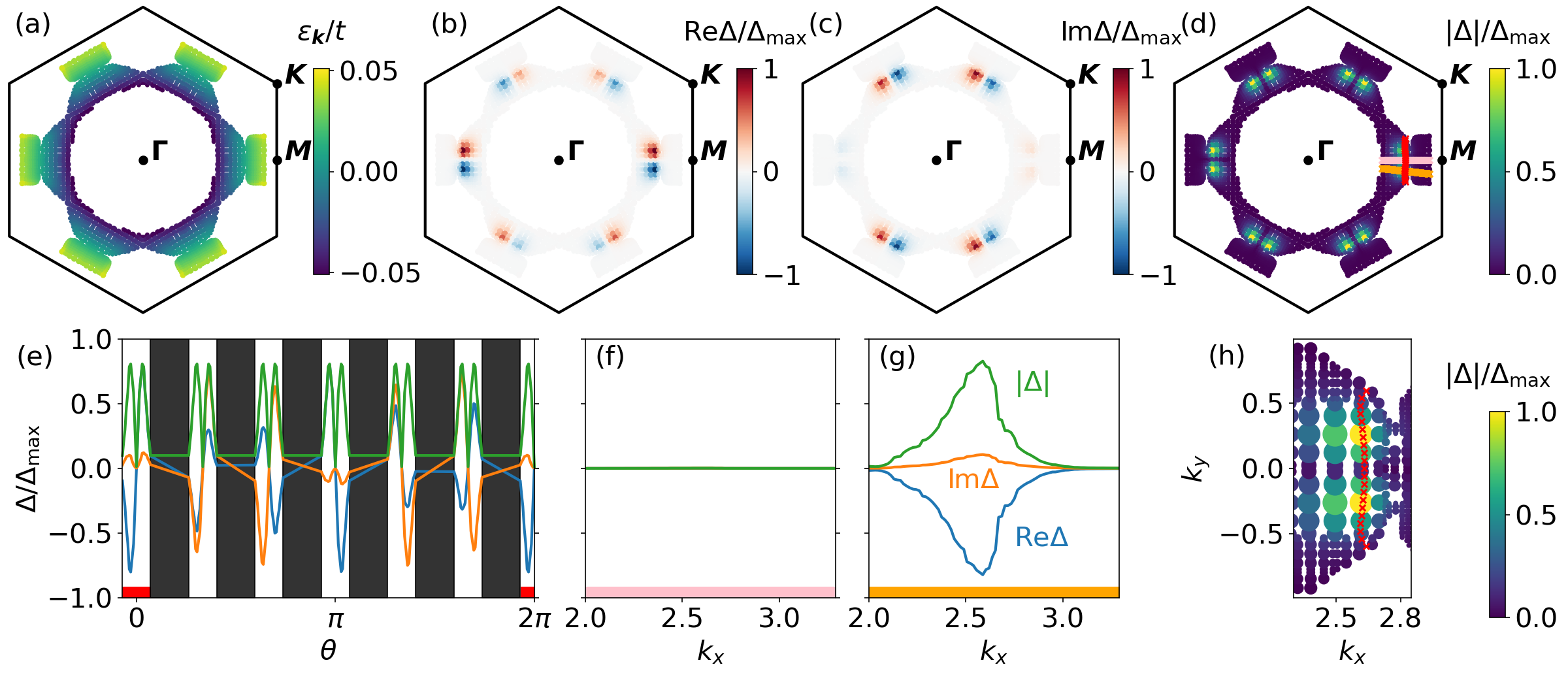}
    \caption{(a) The surface band $\epsilon_{\boldsymbol{k}}$ in the 1BZ with high-symmetry points marked. The values are shown at the 3984 points used in the adaptive quadrature and the size of each marker is scaled by the weight of the point. We found the adaptive quadrature by integrating $\chi_{\boldsymbol{k}}$ at $T=t/300$ with a tolerance of $0.02$. In the white regions there is no surface state. (b) Real part, (c) imaginary part, and (d) absolute value of the gap, all scaled by the largest absolute value $\Delta_{\text{max}}$. (e) Gap on the FS shown as a function of $\theta$. The red regions show where points correspond to the red crosses in (d). The black regions show directions where there is no FS and so the gap is simply linearly interpolated here and has no meaning. (f) [(g)] shows the gap along the pink [orange] crosses shown in (d). In (e), (f), and (g) the real part of the gap is shown in blue, the imaginary part in orange, and the absolute value in green, as indicated in (g). Panel (h) is a zoomed in version of (d) close to the Fermi arc, denoted by red crosses. $T_c \approx 4.0$~K if $t = 0.5$~eV. The parameters are $U = 0.6t$, $V = 0.3t$, and otherwise the same as Fig.~\ref{fig:gaps8}.}
    \label{fig:fullgap1}
\end{figure}

The case $W = \omega_D$ means that the validity of an FS average could be questioned since the region in momentum space where $|\epsilon_{\boldsymbol{k}}| < \omega_D$ is not small. Figure \ref{fig:fullgap1} shows the same qualitative results with a full momentum solution of the gap equation following Eq.~\eqref{eq:fullmomentumgap}. I.e., the gap is $i\times(p_x+\mathrm{i}p_y)$-wave at $U = 0.6t$ and $V = 0.3t$, just like the result from the FS averaged gap equation. The prediction of $T_c$ is reduced to $4.0$~K. That is because the gap decreases away from the FS in the radial direction [see Fig.~\ref{fig:fullgap1}(g)], while it is treated as a constant radially in the FS average. With the chosen parameters, $W \approx 0.055t$ and $\omega_D \approx 0.040t$. The region in momentum space where $|\epsilon_{\boldsymbol{k}}| > \omega_D$ is very small.

The case $W = \omega_D$ further seems to invalidate Migdal's theorem \cite{migdal1958interaction}, stating that vertex corrections are negligible if $W \gg \omega_D$. However, in 2D, Migdal's theorem can be reformulated as $v_F \gg u$, i.e., that the Fermi velocity is much greater than the phonon speed of sound \cite{Roy2014Migdal2D}. From Fig.~2 in the main text we estimate $v_F \approx 5u$ in our case. The key here is that while the surface state has a comparable bandwidth to the phonons, its extent in mometnum space is smaller than for the phonons, giving a larger slope. Also, we follow BCS theory, which is a weak coupling theory valid when $\lambda \ll 1$. Typically $\lambda < 0.5$ is a reasonable limit \cite{Chubukov2020Eliashberg}. Unlike Eliashberg theory, BCS theory neglects all renormalization of the normal state. The assumption is that the interactions are sufficiently weak that any small normal state renormalization will only offer small quantitative corrections. In such a situation, a reliance on Migdal's theorem is less crucial. Typically, Eliashberg theory does not reveal changes in the symmetry of the gap functions \cite{Marsiglio2020ElishbergRev, Chubukov2020Eliashberg, Thingstad2021AFMNMAFM, SunMaeland2023Aug, SunMaeland2024Mar, Brekke2023Dec, Maeland2024Feb, Maeland2023DecCC}. Furthermore, as explained in the End Matter, the pairing in the situation $W = \omega_D$ does not rely on the retarded nature of the phonon-mediated attraction. Thus, we expect only quantitative corrections from Eliashberg theory to our results.

In the effective model, we tune the strength of SOC to mimic the small bandwidth of the surface states in PtBi$_2$.
The reduced strength of SOC needed to get a small surface state bandwidth also reduces the strength of EPC on the surface. SOC is not weak in PtBi$_2$, so the low surface bandwidth there must be due to other factors not contained in the effective model. EPC is enhanced on surfaces due to changes in atomic distances close to the interface with vacuum \cite{Benedek2010PhononOBC}. This effect is not captured by our effective model.
Hence, if the bulk SC should be competitive, we view it as an artifact of weak SOC in the effective model. We could give $\chi$ a layer dependence to ensure stronger EPC on the surface. Also, we could increase the Coulomb repulsion in the bulk, which is expected due to a lower DOS \cite{DiSante2023CoulombRange}. Together these changes would ensure that surface superconductivity dominates.  
Rather than making the model more complicated, we focus exclusively on pairing on the bottom surface. We define bottom surface states by $W_{\boldsymbol{k}n} < 0.3$.

\section{Analysis of gap functions} \label{sec:gaps}

\begin{figure}
    \centering
    \includegraphics[width=0.9\linewidth]{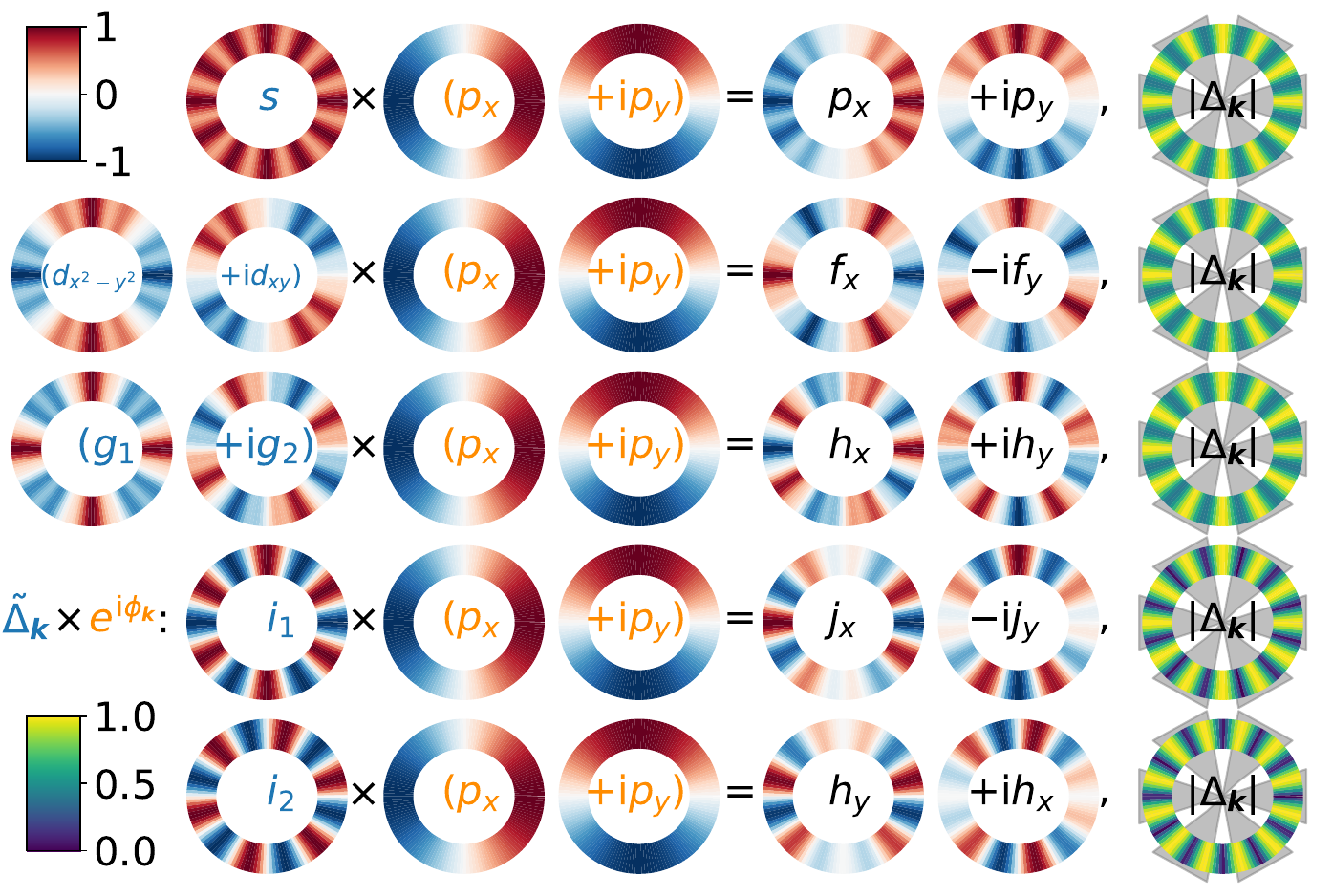}
    \caption{Illustration of gap symmetries on a circular FS around the $\boldsymbol{\Gamma}$ point. We imagine different even-parity pairings $\Tilde{\Delta}_{\boldsymbol{k}}$ multiplied by $p_x+\mathrm{i}p_y$-wave momentum dependence in $e^{\mathrm{i}\phi_{\boldsymbol{k}}}$ coming from spin-orbit coupling to yield odd-parity gaps $\Delta_{\boldsymbol{k}} = \Tilde{\Delta}_{\boldsymbol{k}}e^{\mathrm{i}\phi_{\boldsymbol{k}}}$ in the band basis. The angular extent of the Fermi arcs is indicated in gray for the absolute value of the gap. To model the anisotropic pairing, we multiply the even-parity gaps $\Tilde{\Delta}_{\boldsymbol{k}}$ in the top three rows by $0.7+0.3\cos(12\theta)$. This is to schematically show that the gap is maximal in the center of the Fermi arcs. The absolute value of the gap on the last row reproduces the experimentally observed gap profile in the Fermi arcs of PtBi$_2$ \cite{Changdar2025iwave}.}
    \label{fig:gapsymfull}
\end{figure}

Figure \ref{fig:gapsymfull} provides a way to interpret the five gap profiles identified in the main text. To deconvolve the $p_x+\mathrm{i}p_y$-wave momentum dependence from SOC, we imagine different even-parity gaps $\Tilde{\Delta}_{\boldsymbol{k}}$ multiplied by $p_x +\mathrm{i}p_y$-wave to give odd-parity gaps $\Delta_{\boldsymbol{k}} = \Tilde{\Delta}_{\boldsymbol{k}}e^{\mathrm{i}\phi_{\boldsymbol{k}}}$ in the band basis \cite{Scheurer2016NCSTSC}. 
The $p_x+\mathrm{i}p_y$-wave momentum dependence of $e^{\mathrm{i}\phi_{\boldsymbol{k}}}$ represents the chirality of the Fermi arc surface state \cite{Venditti2026d+id}. For this reason, all possible superconducting states are chiral, something Ref.~\cite{Huang2025PtBi2STM} provides experimental evidence of.
The first row shows the interpretation of the $p$-wave state as anisotropic $s$-wave pairing from phonons multiplied by chiral $p$-wave momentum dependence from SOC \cite{Maeland2025Jun}. The last row shows the $i$-wave state with an unusually nodal $h$-wave gap in the band basis as also shown in Fig.~1 of the main text. 
While $h$-waves have 10 nodes, here both the real and imaginary part have an additional two non-sign-changing nodes. Also, the nodes of the real and imaginary part line up so that the absolute value is nodal.

We have simplified the names of the $f$-, $h$- and $j$-waves to denote if they have sign-changing nodes at $k_x= 0$ ($f_x, h_x, j_x$) or $k_y=0$ ($f_y, h_y, j_y$). These are cubic, quintic and septic functions. Similarly, we simply refer to the two quartic $g$-wave functions as $g_1$ and $g_2$. 
Also, there are two possible $i$-wave functions, which we call $i_1$ and $i_2$. The state with nodes in the center of the Fermi arcs is $i_2$-wave for $\Tilde{\Delta}_{\boldsymbol{k}}$ and we use the separate names $i_1$ and $i_2$ henceforth.
For the schematic picture, we use simple functions of the angle $\theta$: $s: 1$, $p_{x}: \cos(\theta)$, $p_y: \sin(\theta)$, $d_{x^2-y^2}: -\cos(2\theta)$, $d_{x^2-y^2}: -\sin(2\theta)$, $g_1: \cos(4\theta)$, $g_2: \sin(4\theta)$, $i_1: -\cos(6\theta)$, and $i_2: \sin(6\theta)$. Similar simple functions for what we recognize as $f$-, $h$- and $j$-waves are $f_x: -\cos(3\theta)$, $f_y: \sin(3\theta)$, $h_x: \cos(5\theta)$, $h_y: \sin(5\theta)$, $j_x: -\cos(7\theta)$, and $j_y: \sin(7\theta)$.

The second and third rows in Fig.~\ref{fig:gapsymfull} provide an understanding of the fully gapped $f$- and $h$-wave states. These states are special, since here the even-parity pairing we consider is also complex. This is because $d_{x^2-y^2}$, $ d_{xy}$, $g_1$, or $g_2$ alone break the $C_{3z}$ symmetry of the material. The chosen complex linear combinations do not. This has consequences also in the gap equation, where we find two degenerate solutions to the gap equation in these cases. For the fully gapped $f$-wave case, we could interpret the two degenerate solutions as $d_{xy}\times (p_x+\mathrm{i}p_y)$ and $d_{x^2-y^2}\times (p_x+\mathrm{i}p_y)$. Any linear combination of these is also a solution. We choose the one where the absolute value of the gap obeys the $C_{3z}$ symmetry, i.e., the absolute value of the gap is the same on all the Fermi arcs. 
That gap represents a spontaneous breaking of time-reversal symmetry (TRS), which is an active research field \cite{Andersen2024TRSB, AaseMaeland2023Dec}. TRS is broken because only the chiral $p$-wave form of SOC is incorporated into the time-reversal operator in the band basis \cite{Scheurer2016NCSTSC}. As illustrated in Fig.~\ref{fig:gapsymfull}, even after deconvolving the chiral $p$-wave part, a complex gap $\Tilde{\Delta}_{\boldsymbol{k}}$ remains. 
The $(d_{xy}+id_{x^2-y^2}) \times (p_x+\mathrm{i}p_y)$-wave gap is an intrinsic realization of the state discussed in Ref.~\cite{Venditti2026d+id}.
If we instead choose, e.g., $d_{xy}\times (p_x +\mathrm{i} p_y)$ we will keep TRS but break $C_{3z}$ symmetry.
The $d_{xy}\times (p_x+\mathrm{i}p_y)$-wave gap is nodal on only some of the Fermi arcs. This would be an intrinsic realization of the state discussed in Ref.~\cite{Linder2010dwaveTI}, where $d_{xy}$-wave superconductivity is proximity induced on the topological insulator surface state. We conjecture that the gap that breaks TRS will be preferred in our case, because it gives a greater overall gapping of the FS, and so a greater condensation energy \cite{Hutchinson2020Nov, SunMaeland2023Aug, SunMaeland2024Mar}.

The second to last row in Fig.~\ref{fig:gapsymfull} provides an understanding of the nodal $j_x-\mathrm{i}j_y$-wave state. It arises due to the nodes of $i_1$-wave, whose nodes are not at high symmetry lines in this system. Thus they can be classified as accidental nodes. %Make figures of s and i_1 states and compare.

The gaps discussed in the symmetry analyses in Refs.~\cite{Changdar2025iwave, Waje2025PtBi2GL} correspond to $\Tilde{\Delta}_{\boldsymbol{k}}$ \cite{Scheurer2016NCSTSC}. Thus we can see that the states in Fig.~\ref{fig:gapsymfull} correspond to, from top to bottom, irrep $A_1$, $E$, $E$, $A_1$, and $A_2$ of the P31m space group. Hence, the fully gapped $f$- and $h$-wave states are closely related as both belong to the irrep $E$. Similarly, the fully gapped $p$-wave and the nodal $j$-wave states are closely related since both belong to irrep $A_1$.

\subsection{Gaps in spin basis} \label{sec:spinBasis}
We can transform the gaps in the band basis back to the original basis, using $d_{\boldsymbol{k}}^\dagger = \sum_{z_i \ell \sigma} v_{\boldsymbol{k}z_i \ell\sigma} c_{\boldsymbol{k}z_i \ell\sigma}^\dagger$ in the term \cite{Maeland2025Jun}
\begin{align}
    &\sum_{\boldsymbol{k}} \Delta_{\boldsymbol{k}} d_{\boldsymbol{k}}^\dagger d_{-\boldsymbol{k}}^\dagger =\sum_{\boldsymbol{k}}\sum_{\substack{z_i\ell\sigma\\z'_i \ell'\sigma'}} \Delta_{\boldsymbol{k}} v_{\boldsymbol{k}z_i \ell\sigma} v_{-\boldsymbol{k},z'_i\ell'\sigma'}  c_{\boldsymbol{k}z_i \ell\sigma}^\dagger c_{-\boldsymbol{k},z'_i\ell'\sigma'}^\dagger =\sum_{\boldsymbol{k}}\sum_{\substack{z_i\ell\sigma\\z'_i \ell'\sigma'}} \Delta_{\boldsymbol{k}z_i z'_i \ell \ell' \sigma \sigma'} c_{\boldsymbol{k}z_i \ell\sigma}^\dagger c_{-\boldsymbol{k},z'_i\ell'\sigma'}^\dagger.
\end{align}
We focus on the gaps on the bottom surface, with $z_i = z'_i = 1$. They are even in orbital, so we characterize them in terms of spin-singlet and spin-triplet gaps by focusing on the spin degree of freedom. We find singlet-triplet mixing due to SOC in all cases. The up-down spin-triplet gap $\Delta_{\uparrow\downarrow}^{t} = (\Delta_{\uparrow\downarrow}+\Delta_{\downarrow\uparrow})/2$ is zero. 
When comparing to the symmetry analysis in Ref.~\cite{Vocaturo2024PtBi2Effective}, the reason is that we have pure Rashba SOC in the electron model and no Ising SOC.

\begin{table}[]
    \centering
    \caption{Comparison of gap symmetries in band basis and spin basis. The table also marks if the gaps are nodal and if they preserve time-reversal symmetry (TRS).}
    \label{tab:gapspinbasis}
    \begin{ruledtabular}
    \begin{tabular}{c|cccc|ccc}
        Band basis & \multicolumn{4}{c|}{Spin basis} & Nodal & TRS  \\
        \hline
        $\Delta_{\boldsymbol{k}}$ & $\Delta_{\boldsymbol{k}\uparrow\downarrow}^{s}$ & $\Delta_{\boldsymbol{k}\uparrow\downarrow}^{t}$ &  $\Delta_{\boldsymbol{k}\uparrow\uparrow}$ & $\Delta_{\boldsymbol{k}\downarrow\downarrow}$ & &  \\
        \hline
        $p_x +\mathrm{i} p_y$ & $s$ & 0& $p_y+\mathrm{i}p_x$ & $-p_y+\mathrm{i}p_x$ & $\cross$ & \checkmark \\
        $f_x-\mathrm{i}f_y$ & $d_{x^2-y^2}+\mathrm{i}g_2$ & 0 & $-f_y+\mathrm{i}p_x$ & $f_y+\mathrm{i}f_x$ & $\cross$ & $\cross$ \\
        $h_x +\mathrm{i} h_y$ & $g_1+\mathrm{i}g_2$ & 0 &  $-f_y+\mathrm{i}l_x$ & $-h_y+\mathrm{i}h_x$ & $\cross$ & $\cross$ \\
        $i_1 \times (p_x+\mathrm{i}p_y) = j_x-\mathrm{i}j_y$ & $i_1$ & 0 & $-j_y+\mathrm{i}j_x$ & $j_y+\mathrm{i}j_x$ & \checkmark & \checkmark \\
        $i_2\times (p_x +\mathrm{i}p_y) = h_y +\mathrm{i}h_x$ & $i_2$ & 0 & $h_x+\mathrm{i}h_y$ & $-h_x+\mathrm{i}h_y$ & \checkmark & \checkmark
    \end{tabular}
    \end{ruledtabular}
\end{table}

When the gap in the band basis is $p_x +\mathrm{i} p_y$, we find that spin singlet $\Delta_{\uparrow\downarrow}^{s} = (\Delta_{\uparrow\downarrow}-\Delta_{\downarrow\uparrow})/2$ is $s$-wave, up-up $\Delta_{\uparrow\uparrow}$ is $p_y+\mathrm{i}p_x$ and down-down $\Delta_{\downarrow\downarrow}$ is $-p_y+\mathrm{i}p_x$. 

When the gap in the band basis is $f_x-\mathrm{i}f_y$, we find that the spin-singlet gap $\Delta_{\uparrow\downarrow}^{s}$ is  $d_{x^2-y^2}+\mathrm{i}g_2$,
$\Delta_{\uparrow\uparrow}$ is $-f_y+\mathrm{i}p_x$, and $\Delta_{\downarrow\downarrow}$ is $f_y+\mathrm{i}f_x$. This gap breaks TRS \cite{Sigrist, Maeland2023AprPRL}, both because the spin-singlet part is complex and because up-up and down-down are not related as $\Delta_{\boldsymbol{k}\uparrow\uparrow} = -\Delta_{-\boldsymbol{k},\downarrow\downarrow}^*$. It is fascinating that the imaginary part of $\Delta_{\uparrow\uparrow}$ is $p_x$- rather than $f_x$-wave. The singlet part is not far from $d_{x^2-y^2}+\mathrm{i}d_{xy}$, we can view that as the starting state, and then an admixture of $g_2$-wave in the imaginary part. 

When the gap in the band basis is fully gapped $h_x +\mathrm{i} h_y$-wave, we find in the spin basis; $\Delta_{\uparrow\downarrow}^{s}$: $g_1+\mathrm{i}g_2$, $\Delta_{\uparrow\uparrow}$: $-f_y+\mathrm{i}l_x$, and $\Delta_{\downarrow\downarrow}$: $-h_y+\mathrm{i}h_x$. The imaginary part of $\Delta_{\uparrow\uparrow}$ has 18 nodes on the FS, so we name it $l_x$-wave and it goes like $\cos(9\theta)$. Many of the nodes of the imaginary part are in regions with low absolute value indicating mixing of orbitals with many different angular momenta. Similar to above, TRS is spontaneously broken for the linear combination with best gapping of the FS and where the absolute value of the gap is equal on all Fermi arcs. Note that the real part of $\Delta_{\uparrow\uparrow}$ is $f_x$-wave, indicating that this state involves some amount of NN Cooper pairing in real space, since $f$-wave lattice harmonics can be constructed from NN on the triangular lattice \cite{Benestad}.

When the gap in the band basis is $i_1\times (p_x +\mathrm{i}p_y) = j_x -\mathrm{i}j_y$, we have $\Delta_{\uparrow\downarrow}^{s}$: $i_1$, $\Delta_{\uparrow\uparrow}$: $-j_y+\mathrm{i}j_x$, and $\Delta_{\downarrow\downarrow}$: $j_y+\mathrm{i}j_x$. This state preserves TRS and has accidental nodes. As seen in Fig.~3 of the main text, this state's dimensionless coupling $\lambda_j$ depends on $U$ and $V$ unlike what one would expect due to the high angular momentum of the gap symmetries. We interpret that as due to mixing with the fully gapped $p_x +\mathrm{i} p_y$-state. I.e., the orange and grayscale regions are not two separate competing phases, it is rather a gradual change in the mixing of $(c_s s + c_i i_1)\times (p_x +\mathrm{i}p_y)$, where $c_s, c_i$ are constants. The change in symmetry occurs when the $i_1$ part becomes large enough to give additional nodes on the Fermi arcs [see Figs.~\ref{fig:gaps8}(a), (b), and (c) for this change]. Onsite and nearest neighbor Cooper pairing persists, and so $\lambda_j$ decreases with $U$ and $V$.

When the gap in the band basis is $i_2\times (p_x +\mathrm{i}p_y) = h_y +\mathrm{i}h_x$, we have $\Delta_{\uparrow\downarrow}^{s}$: $i_2$, $\Delta_{\uparrow\uparrow}$: $h_x+\mathrm{i}h_y$, and $\Delta_{\downarrow\downarrow}$: $-h_x+\mathrm{i}h_y$, the last two being unusually nodal like an $i_2$-wave. 

Note that the first column in Fig.~\ref{fig:gapsymfull}, namely $\Tilde{\Delta}_{\boldsymbol{k}}$, is similar to the singlet gaps in the spin basis, while the triplet gaps in the spin basis more closely resemble the odd-parity gaps in the band basis. We summarize the comparison between band basis and spin basis in Table \ref{tab:gapspinbasis}.

\end{document}